\documentclass[numberedappendix]{emulateapj}
\usepackage{apjfonts}
\DeclareMathSymbol{v}{\mathord}{cmletters}{"76}
\usepackage{amsmath}
\usepackage{amssymb}
\usepackage{hyperref}

\RequirePackage[log]{snapshot}
\usepackage[usenames,dvipsnames]{xcolor}

\def\sgra{Sgr~A$^{\ast}$}

\newcommand{\msun}{{\rm M_{\odot}}}

\newcommand{\beq}{\begin{equation}}
\newcommand{\eeq}{\end{equation}}
\newcommand{\beqn}{\begin{eqnarray}}
\newcommand{\eeqn}{\end{eqnarray}}

\shortauthors{R. Gold, J.~C. McKinney, M.~D. Johnson, \& S.~S. Doeleman}
\shorttitle{Horizon Scales with Polarized Radiative Transfer Simulations}

\begin{document} %
\label{firstpage}

\title{Probing the magnetic field structure in Sgr~A${}^*$ on Black Hole Horizon Scales with Polarized Radiative Transfer Simulations}

\author{Roman Gold${}^{1}$, Jonathan C.~McKinney${}^{1}$, Michael D.~Johnson${}^{2}$, Sheperd S.~Doeleman${}^{2,3}$}
\affiliation{${}^1$Department of Physics \& Joint Space-Science Institute, University of Maryland, College Park, MD~20742}
\affiliation{${}^2$Harvard-Smithsonian Center for Astrophysics, 60 Garden Street, Cambridge, MA~02138, USA}
\affiliation{${}^3$Massachusetts Institute of Technology, Haystack Observatory, Route 40, Westford, MA~01886, USA}

\begin{abstract} %

Magnetic fields are believed to drive accretion and relativistic jets
in black hole accretion systems, but the magnetic-field structure 
that controls these phenomena remains uncertain.  We perform general
relativistic (GR) polarized radiative transfer of time-dependent
three-dimensional GR magnetohydrodynamical (MHD) simulations to model
thermal synchrotron emission from the Galactic Center source Sagittarius~A$^\ast$ (\sgra).
We compare our results to new polarimetry measurements by the Event
Horizon Telescope (EHT) and show how polarization in the visibility
(Fourier) domain distinguishes and constrains accretion flow models
with different magnetic field structures.  These include models with
small-scale fields in disks driven by the magnetorotational
instability (MRI) as well as models with large-scale ordered fields in
magnetically-arrested disks (MAD).  We also consider different
electron temperature and jet mass-loading prescriptions that control
the brightness of the disk, funnel-wall jet, and Blandford-Znajek-driven
funnel jet.  Our comparisons between the simulations and observations
favor models with ordered magnetic fields near the black hole event
horizon in \sgra, although both disk- and jet-dominated emission can
satisfactorily explain most of the current EHT data.  We show that
stronger model constraints should be possible with upcoming circular
polarization and higher frequency ($349~{\rm GHz}$) measurements.

\end{abstract}   %


\section{Introduction}
\label{sec:intro}
The origin of the radio emission of \sgra\ has been the subject
of intense observational studies
\citep{1998ApJ...499..731F,2009IJMPD..18..889R,2008Natur.455...78D,2009ApJ...698..676D}
and theoretical modeling
\citep{1994ApJ...428L..13N,2002AA...383..854Y,2009ApJ...703L.142D,2009ApJ...706..497M,2010MNRAS.408..752P,2012ApJ...752L...1M,2013AA...559L...3M,2014AA...570A...7M}.
Near-infrared observations \citep{2010RvMP...82.3121G} of stars
orbiting an unseen central mass so-far provide the most direct
evidence for the existence of a black hole (BH) and yield a BH mass of
$M=4.5\pm 0.4 \times
10^6\msun$ \citep{2008ApJ...689.1044G}. Observations covering a wide
range of the electromagnetic spectrum rule out a standard thin disk
model and clearly reveal that \sgra\ is highly underluminous (compared
to its Eddington limit), presumably due to a highly sub-Eddington
accreting
BH \citep{1998ApJ...499..731F,2009IJMPD..18..889R,2009ApJ...698..676D}. This
regime of the accretion disk has been studied extensively
\citep{2014ARAA..52..529Y}
and features a hot, magnetized accretion flow composed of a weakly
collisional plasma. Synchrotron emission is the main contribution to
the near-mm flux due the low density and dynamically important
magnetic fields.  The unresolved (``zero-baseline'') flux has been measured
in radio \citep{1998ApJ...499..731F,Bower2015},
infrared \citep{Schodel2011}, and X-ray \citep{2003ApJ...591..891B}
and exhibits diverse phenomena, such as flaring in the
near-infrared \citep{2003Natur.425..934G,2004ApJ...606..894Y,2006AA...455....1E}
and X-rays \citep{2006AA...450..535E}.

Very-long baseline interferometric (VLBI) radio measurements, such as
those with the Event Horizon Telescope 
(EHT) \citep{2009astro2010S..68D}, offer an unprecedented capability
to identify the physics near a rotating BH in \sgra\ due to
the EHT's high observing frequency (230~GHz), resolving power, and sensitivity.  Recently,
VLBI observations with the EHT have determined the correlated flux density of
\sgra\ (and M87) on VLBI baselines, thereby partially 
resolving the emission structure and constraining the size of the
emitting 
region \citep{2008Natur.455...78D,2011ApJ...727L..36F,2012Sci...338..355D,Akiyama2015,JohnsonEtAl2015Science}.
The EHT probes the strong field regime of GR and may be capable of
detecting the BH's shadow
\citep{Bardeen1973,Luminet1979,2000ApJ...528L..13F,2010ApJ...717.1092D,2014ApJ...795..134F,2015ApJ...814..115P}.

The EHT can also resolve polarized structure on event horizon scales,
which may allow one to distinguish between competing models of
accretion disks and jets.  Whether a jet is launched depends upon the
BH spin, structure of the magnetic field threading the disk and BH,
and the mass-loading of the polar magnetic
field \citep{1977MNRAS.179..433B,2007MNRAS.377L..49K}.  If the
magnetic field structure consists of small-scale MHD turbulence driven
by the magnetorotational instability (MRI) \citep{MRI1998}, then the
production of a jet is either not possible due to rapid reconnection
of a disorganized magnetic
field \citep{2008ApJ...678.1180B,2009MNRAS.394L.126M} or is at least
difficult without a collisional ideal MHD
plasma \citep{mu12,2012MNRAS.423.3083M}.  Some researchers call such
MRI-driven disks a type of standard-and-normal-evolution (SANE)
accretion flow \citep{2012MNRAS.426.3241N}.  In another limit, a
plentiful supply of ordered vertical magnetic flux builds-up near the
BH until reaching saturation, in which case the MRI is marginally
suppressed and the disk enters the so-called magnetically arrested
disk (MAD) state driven by magnetic Rayleigh-Taylor
instabilities \citep{2003PASJ...55L..69N,2003ApJ...592.1042I,2011MNRAS.418L..79T,2012MNRAS.423.3083M}.

Time-dependent global general relativistic magnetohydrodynamical
(GRMHD) simulations of a variety of BH accretion flow types are
essential to understand the possible range of disk and jet states and
their underlying dynamics. A significant theoretical uncertainty in
modeling \sgra\ is that such weakly collisional flows involve kinetic
physics with undetermined heating
rates \citep{1999ApJ...520..248Q,2006ApJ...637..952S,2007ApJ...660.1273J,2008PhRvL.100f5004H,2012ApJ...755...50R,2015ApJ...800...27R}
for a population of thermal or non-thermal
electrons \citep{1997ApJ...490..605M,Ozel2000,2003ApJ...598..301Y,2014ApJ...791...71L}
not fully accounted for in GRMHD simulations.  For example, the jet
might light-up only if actually launched under favorable physical
conditions, or the jet could be always present but the particle
heating could control whether the jet lights up. Relativistic jets are
commonly invoked to interpret the emission from compact radio sources
\citep{1979ApJ...232...34B}. In particular, for \sgra\ and
M87, jet models have been successfully applied to explain the spectral
energy distribution (SED) with certain assumptions about the electron
temperatures \citep{2000AA...362..113F,2012Sci...338..355D}. While no
clear unambiguous spectral or imaging signature of such a jet has been
found in \sgra (frequency-dependent light curves show time-lagged
correlations indicative of expansion or
outflows \citep{Marrone2008,Yusef-Zadeh2008,2015arXiv150203423B}),
evidence for or against jets could come from polarimetric observations
that depend upon the nature of the magnetic field.

There is also theoretical uncertainty in the amount of particles that
should be present inside \citet{1977MNRAS.179..433B} (hereafter BZ)
driven jets.  GRMHD numerical simulations with BZ-driven jets must
inject matter in some way to keep the numerical scheme
stable \citep{2003ApJ...589..444G}. In real astrophysical systems, the
nature of mass-loading of jets remains uncertain and could be due to
photon annihilation or pair cascades to some degree
for \sgra\ \citep{2011ApJ...735....9M}, but this creates only a
low-level of mass-loading (much lower than GRMHD numerical schemes can
handle).  In some cases, like for \sgra, the level of mass-loading
might even be insufficient to enable force-free or MHD conditions in
the highly-magnetized
funnel \citep{2005ApJ...631..456L,2015ApJ...809...97B}.  MHD dynamical
mass-loading due to magnetic Rayleigh-Taylor instabilities at the
disk-jet interface might lead to significant mass-loading, which could
depend upon the disk type, with MADs generating more mass-loading due
to large-scale magnetic oscillations that connect the disk and
jet \citep{2012MNRAS.423.3083M}.  However, no work has yet quantified
such an MHD-based mass-loading mechanism.

Accounting for these various uncertainties, GRMHD simulations can then
be used as dynamical models in a radiative transfer calculation in
order to compare with
observations \citep{2009ApJ...703L.142D,2009ApJ...706..497M,2010ApJ...717.1092D,2012ApJ...755..133S,2014AA...570A...7M,2015ApJ...812..103C,2015ApJ...799....1C}.
In particular, polarized radiative transfer offers up to four times
the information of unpolarized
studies \citep{2011MNRAS.410.1052S,2012ApJ...755..133S}, potentially
leading to much better constraints on models and theories of accretion
flows and jets than studies that only use
intensity \citep{2000ApJ...545..842Q,2007ApJ...671.1696S,2010ApJ...725..750B}. The
EHT 2013 campaign has shown how linear polarization begins to
distinguish between generic ordered and turbulent field
configurations \citep{JohnsonEtAl2015Science}. As part of the EHT
collaboration, we analyzed a single polarized radiative transfer GRMHD
model that was broadly consistent with the linear polarization of \sgra\ measured by the EHT and its
correlation with total intensity.

In this work, we follow-up 
\citet{JohnsonEtAl2015Science} by considering a larger array of
GRMHD simulations of both SANE and MAD types for rapidly rotating BHs,
the role of electron heating prescriptions, and the role of the funnel
mass-loading that lead to more disk-dominated or jet-dominated
emission.  We analyze in more depth this extended synthetic data set, assess the level of agreement with current EHT observations, and explore expectations and implications for future EHT campaigns.  The
main goal is to use sparse (primarily linear) polarization
measurements in the visibility plane to place constraints upon the
horizon-scale magnetic field structure, electron heating physics
implied by varying the electron temperature prescription, and
mass-loading of the BZ-driven funnel jet.

The structure of the paper is as follows: In \S~\ref{sec:methods}, we
describe the models used and explain our methods including the
generation of synthetic VLBI data. In \S~\ref{sec:results}, we
describe our results that include fitting to observations, showing
differences due to the underlying dynamical GRMHD model chosen,
constraining the magnetic field structure, variability of linear
polarization, shadow features, changes due to electron temperature and
mass-loading prescriptions, and prospects for future EHT efforts
focused on higher frequencies and circular
polarization. In \S~\ref{sec:future}, we discuss future plans.
In \S~\ref{sec:conclusions}, we summarize and conclude.

\section{Methods}
\label{sec:methods}
In this section, we describe our GRMHD simulations, electron heating
prescriptions, jet mass-loading prescriptions, polarized radiative
transfer scheme, scattering kernel, Stokes parameters computed, model
fitting procedures, and the generation of synthetic VLBI data in the
visibility domain.

\subsection{GRMHD simulations}
\label{sec:models}

\begin{figure}
  \centering
  \begin{tabular}{@{}cc@{}}
    \includegraphics[width=0.23\textwidth]{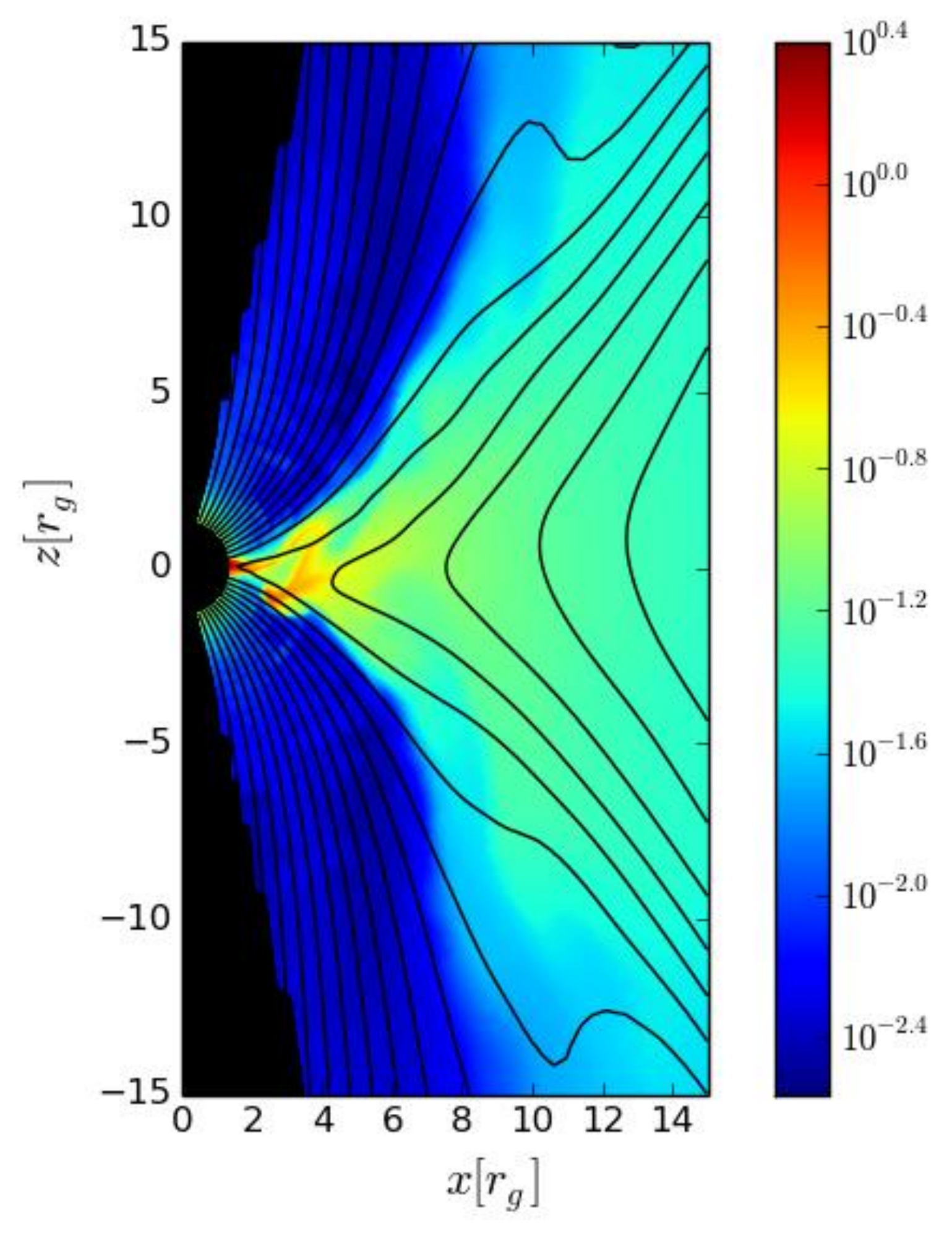} &
    \includegraphics[width=0.23\textwidth]{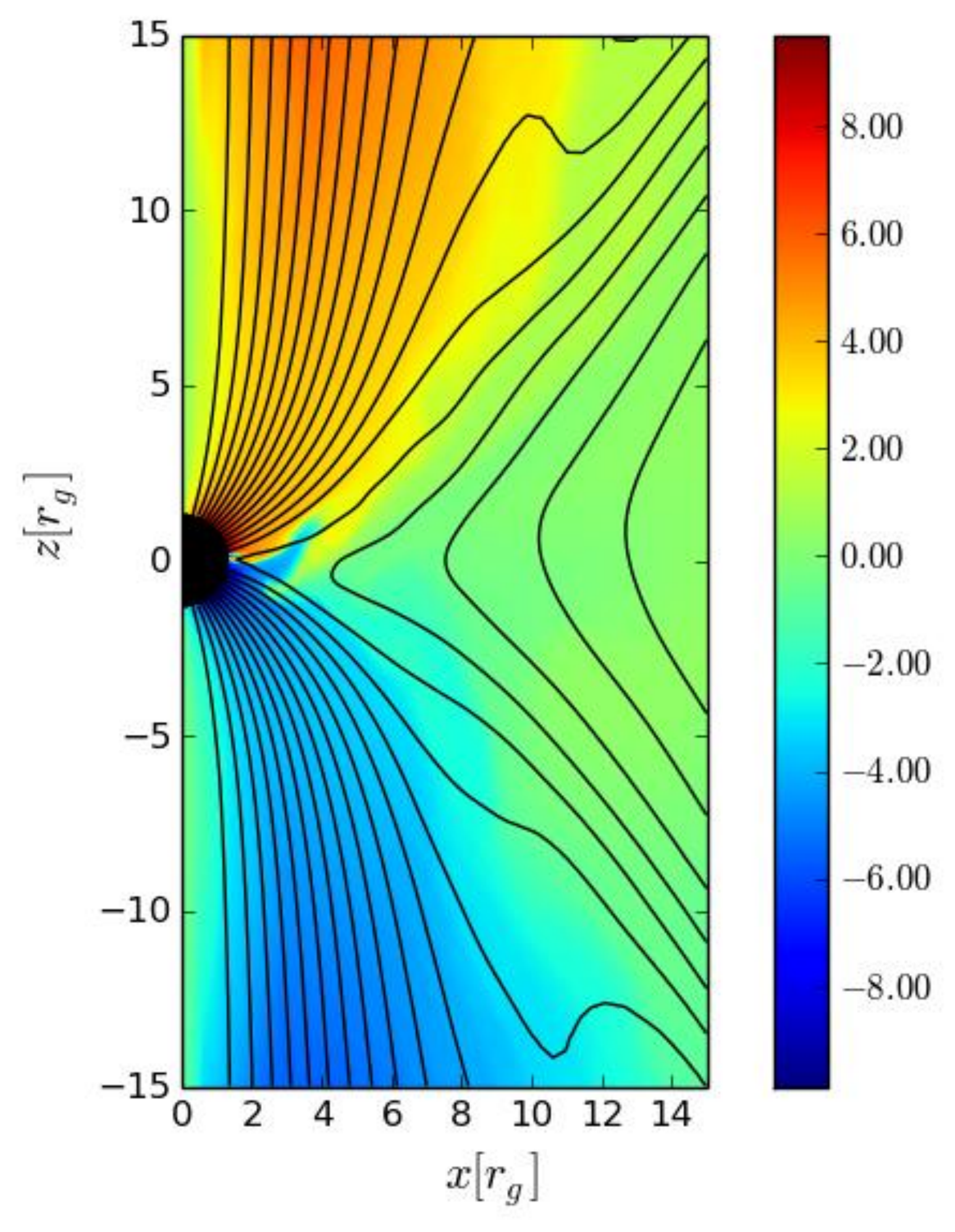} \\
    \includegraphics[width=0.23\textwidth]{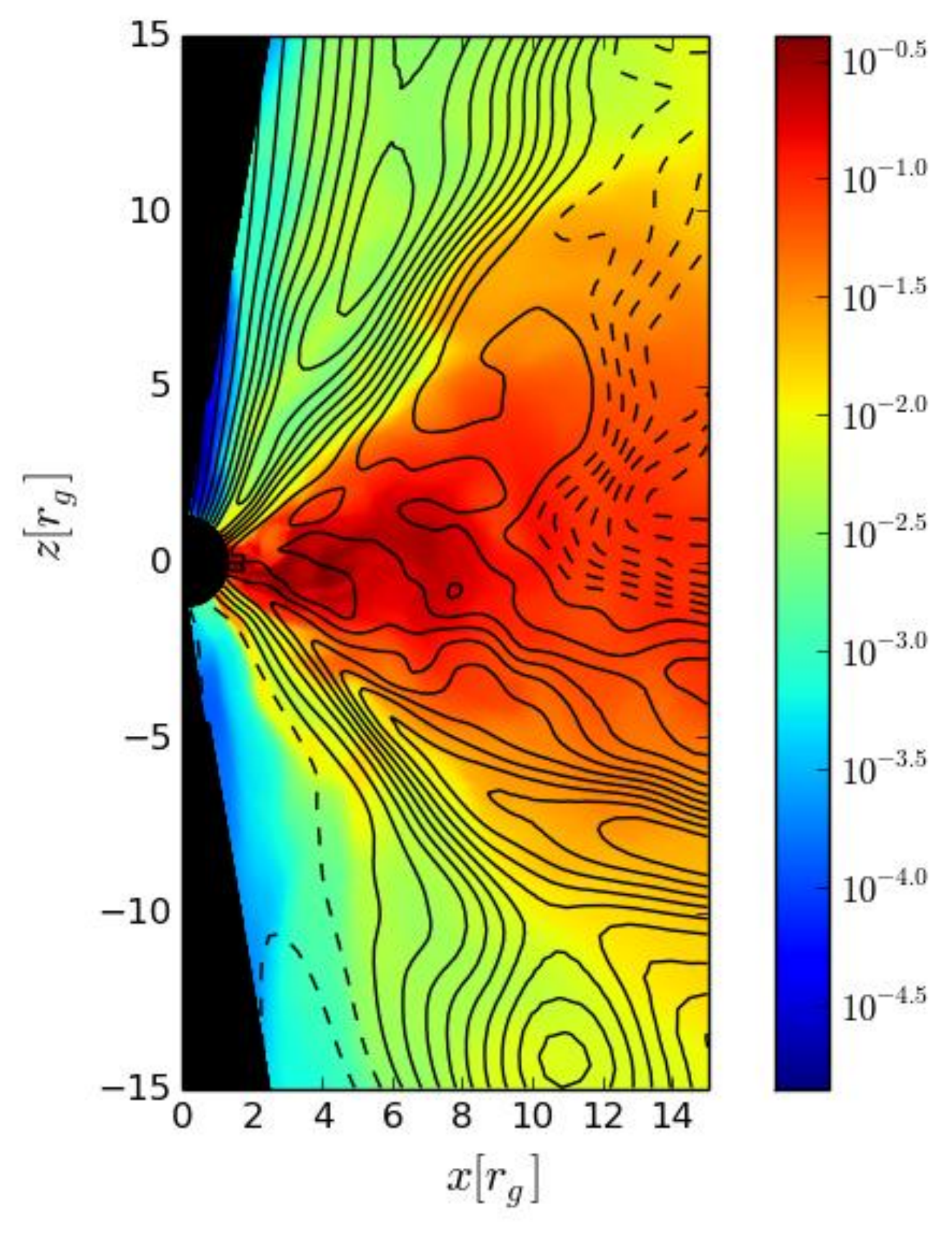} &
    \includegraphics[width=0.23\textwidth]{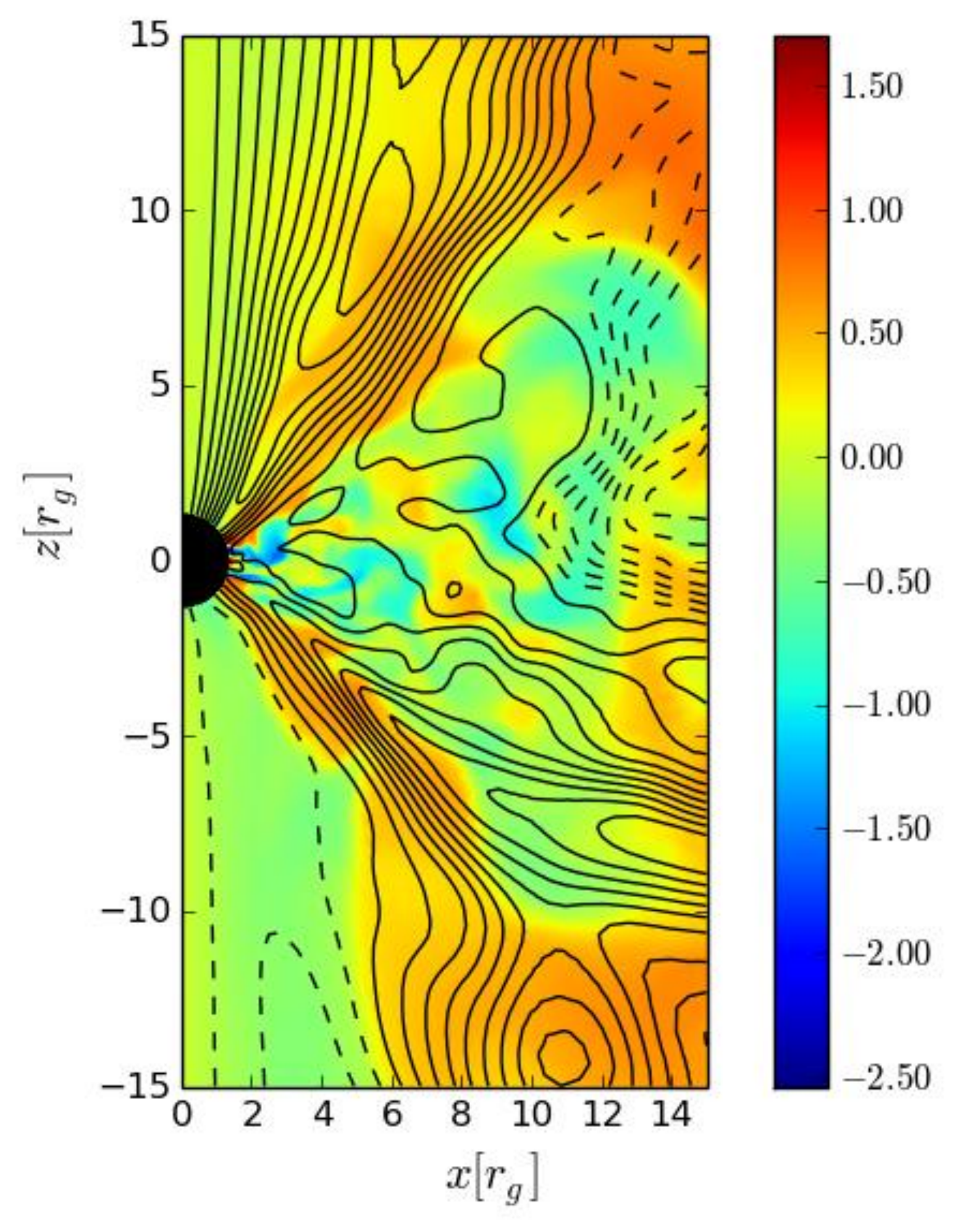} \\
    \includegraphics[width=0.23\textwidth]{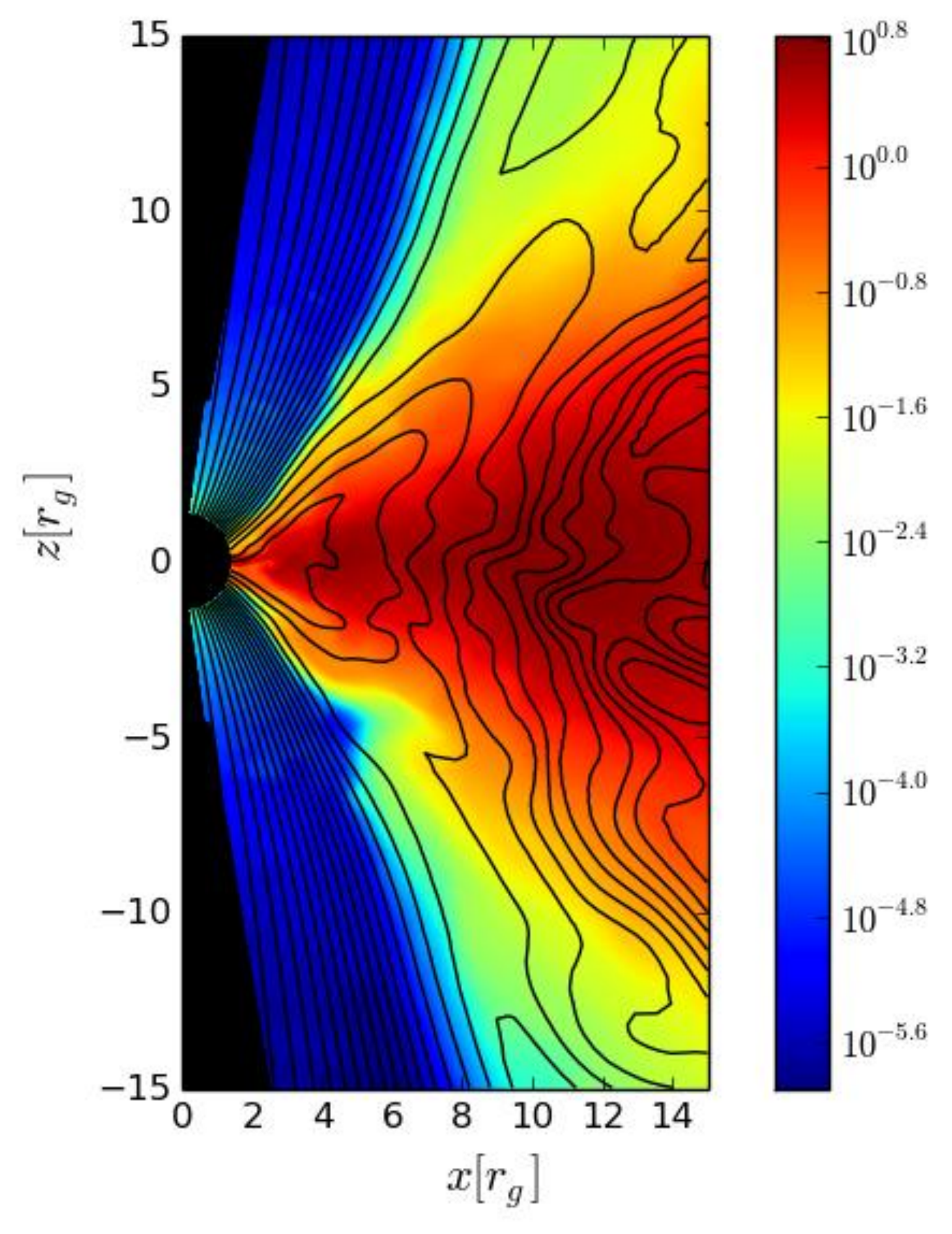} &
    \includegraphics[width=0.23\textwidth]{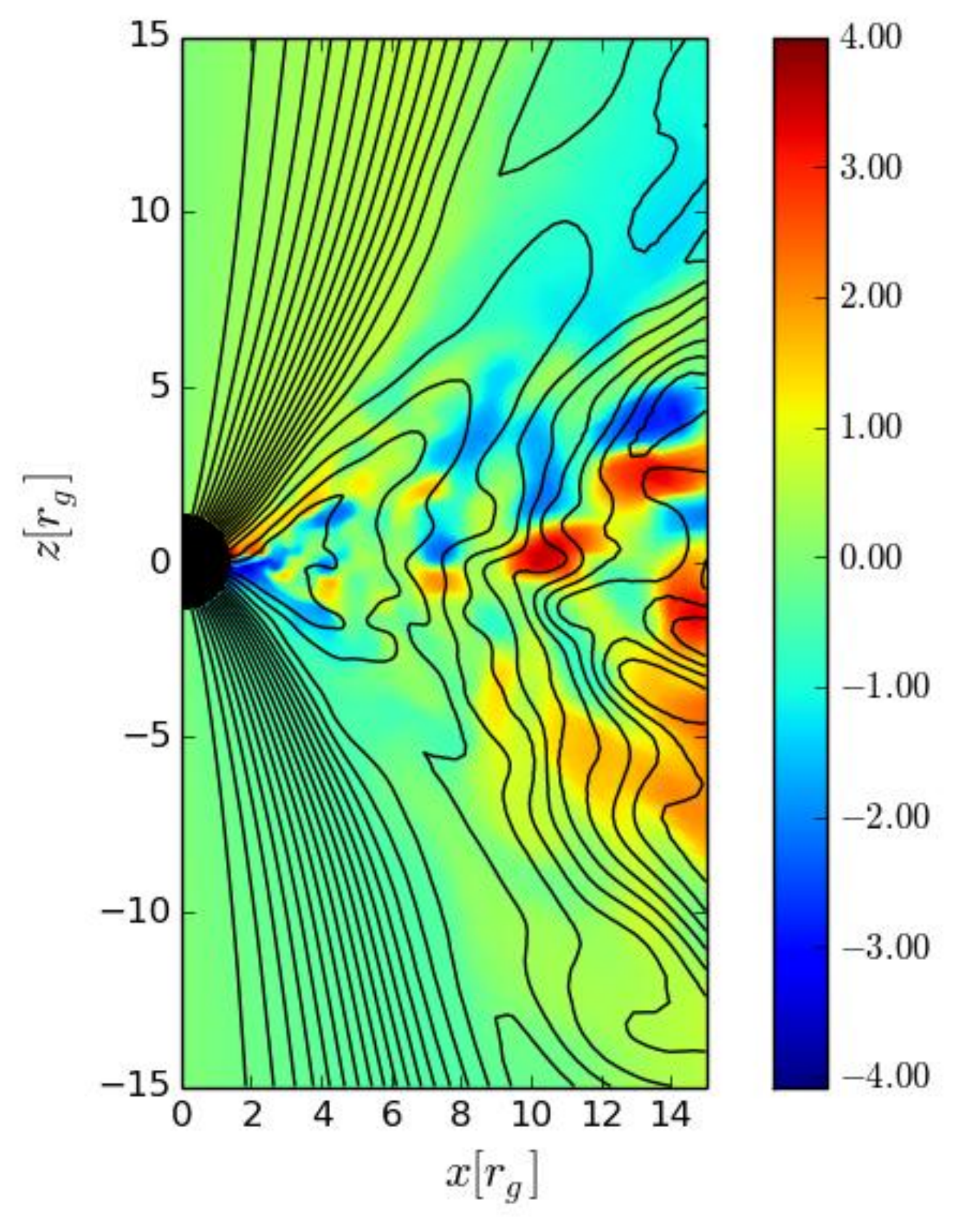} \\
  \end{tabular}
  \caption{Poloidal plane slices ($z$ vs.\ $x$) for snapshots in time
    of the dimensionless rest-mass density ($\rho r_g^2c/\dot{M}$,
    left panels) and dimensionless toroidal magnetic field scaled by
    radius ($(r/r_{\rm H}) B_\phi r_g/\sqrt{c\dot{M}}$ with $B_\phi$
    in Gaussian units, right panels) for the {\tt MAD\_thick} models
    (upper two panels), {\tt SANE\_quadrupole-disk} model
    (vertically-middle two panels), and {\tt SANE\_dipole-jet} model
    (lowest two panels).  Magnetic field lines (from the
    $\phi$-integrated vector potential for clarity of the field
    behavior) are shown as black contour lines (negative contours are
    dashed, positive are solid, as originating from the initial values
    of the vector potential) with arbitrarily-sized uniform spacing to
    give $20$ contours within plotted domain.  These are our
    simulation models with the default polar axis angle-based density
    cut-out (and no $b^2/\rho$-based removal procedure).  The {\tt
      MAD\_thick} models have a relatively strong ordered poloidal and
    toroidal field in the jet, and the disk has an ordered magnetic
    field.  The {\tt SANE\_quadrupole-disk} has no jet but does
    contain a toroidal magnetic field in a wind that has comparable
    strength in the disk. The {\tt SANE\_dipole-jet} model has a weak
    jet with an ordered toroidal and poloidal magnetic field, while
    the disk has a disordered toroidal field.}
  \label{fig:sims}
\end{figure}

GRMHD simulations are typically based upon the ideal MHD equations of
motion that assume no explicit viscosity, resistivity, or
collisionless physics.  Our ideal GRMHD simulations are based upon the
ideal GRMHD code called HARM \citep{2003ApJ...589..444G}, which has
been used to perform several simulations to explore the role of BH
spin
\citep{2004ApJ...611..977M,2004ApJ...602..312G,2005ApJ...630L...5M,2010ApJ...711...50T,2012MNRAS.423L..55T},
magnetic field type \citep{2009MNRAS.394L.126M,2012MNRAS.423.3083M},
large-scale jet propagation \citep{2006MNRAS.368.1561M}, disk
thickness
\citep{2008ApJ...687L..25S,2010MNRAS.408..752P,2015arXiv150805323A},
how disks and jets pressure balance \citep{2007MNRAS.375..513M},
relative tilt between the BH and disk \citep{2013Sci...339...49M}, and
dynamically-important radiation
\citep{2014MNRAS.441.3177M,2015MNRAS.454L...6M}.  Despite being ideal
MHD solutions, they still provide the most state-of-the-art way of
describing the fully three-dimensional global plasma behavior around a
BH.

We use several previously published GRMHD models as input for the
polarized radiative transfer calculations. The models used in this
paper are labeled to designate the type of GRMHD simulation ({\tt MAD}
and {\tt SANE}) followed by an identifier for the choices made for the
radiative transfer scheme ({\tt -disk} and {\tt -jet}) for our primary
models, for which we perform full fits (described in \S\ref{modelfit}). In the following, we list the
underlying GRMHD dynamical models used.

\begin{enumerate}

  \item {\tt MAD\_thick}: A rapidly spinning $a/M=0.9375$
    geometrically thick MAD model \citep{2012MNRAS.423.3083M} with a
    large-scale dipolar field with plentiful supply of magnetic flux.
    Dynamically produces a powerful jet with magnetic Rayleigh-Taylor
    instabilities.  Includes rare magnetic field polarity inversions
    that drive transient jets \citep{2010ApJ...725..750B}.

  \item {\tt SANE\_quadrupole}: A rapidly spinning $a/M=0.9375$ MRI
    disk model \citep{2009MNRAS.394L.126M} with an initially
    large-scale quadrupolar magnetic field.  Dynamically leads to no
    jet and contains an MRI-driven MHD-turbulent disk.

  \item {\tt SANE\_dipole}: A rapidly spinning $a/M=0.92$ MRI disk
    model \citep{2009MNRAS.394L.126M} with an initially dipolar
    magnetic field consisting of a single set of nested field loops
    following rest-mass density contours.  Dynamically leads to a weak
    jet and MRI-driven MHD-turbulent disk.

\end{enumerate}

Fig.~\ref{fig:sims} shows the set of three GRMHD simulations that form
the basis of our various models.  These models are all of relatively
rapidly rotating BHs but span a range of types of magnetic fields in
the disk (ordered and disordered) and jet types (powerful, weak, and
no jet).  Jet radiation could be an important contribution to observed
emission even for \sgra\ and might help explain the synchrotron
self-absorption emission at low frequencies
\citep{2002AA...383..854Y,2013AA...559L...3M}.  We only show the
poloidal ($z$ vs.\ $x$) plane. All simulations have a toroidal
direction with a turbulent toroidally-dominated disk at large radii
and a more mixed laminar-turbulent (with equal toroidal and poloidal
field strengths) disk at smaller radii near the photon orbit.  The jet
present in {\tt MAD\_thick} and {\tt SANE\_dipole} consists of a
helical field with comparable poloidal and toroidal magnetic field
strengths near the horizon (i.e., the light cylinder, or Alfven
surface, where the toroidal field strength must become comparable to
the poloidal field strength, is near the horizon for these high spin
models).  The {\tt SANE\_quadrupole} model has no BZ-driven jet or
persistent low-density funnel region, but there is still a
well-defined toroidally-dominated wind.

As discussed in detail in \citet{2012MNRAS.423.3083M}, the {\tt
  MAD\_thick} model very well-resolves the MRI and turbulent modes,
the {\tt SANE\_quadrupole} model well-resolves the MRI and turbulent
modes, and the {\tt SANE\_dipole} model marginally resolves the MRI
and turbulent modes.  A quasi steady-state inflow equilibrium is
reached out to $r\sim 100r_g$ (gravitational radii) over a run-time of
$30,000r_g/c$ ($180$ hours for \sgra) for the {\tt MAD\_thick} model,
$r\sim 20r_g$ for the {\tt SANE\_quadrupole} model, and $r\sim 12r_g$
for the {\tt SANE\_dipole} model with both SANE models having run-time
of $\sim 5,000r_g/c$ ($30$ hours for \sgra).  Fig.~\ref{fig:sims} uses
the snapshot at time $t=20612r_g/c$ for the {\tt MAD\_thick} model,
$t=4280r_g/c$ for the {\tt SANE\_quadrupole} model, and $t=3200r_g/c$
for the {\tt SANE\_dipole} model.

\subsection{Electron temperature prescription}
\label{sec:te}

\begin{figure}
  \centering
  \begin{tabular}{@{}cc@{}}
    \includegraphics[width=0.22\textwidth]{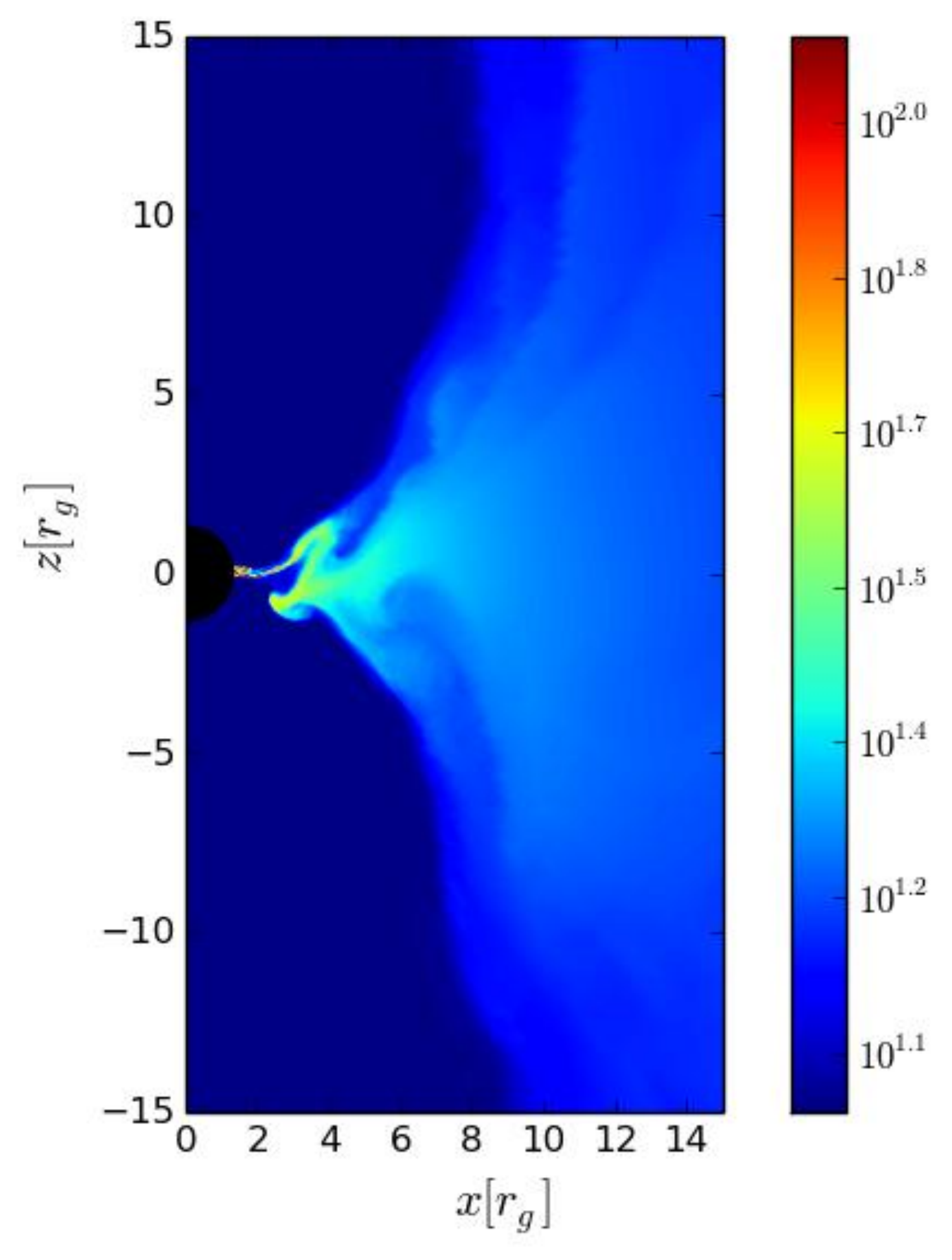} &
    \includegraphics[width=0.22\textwidth]{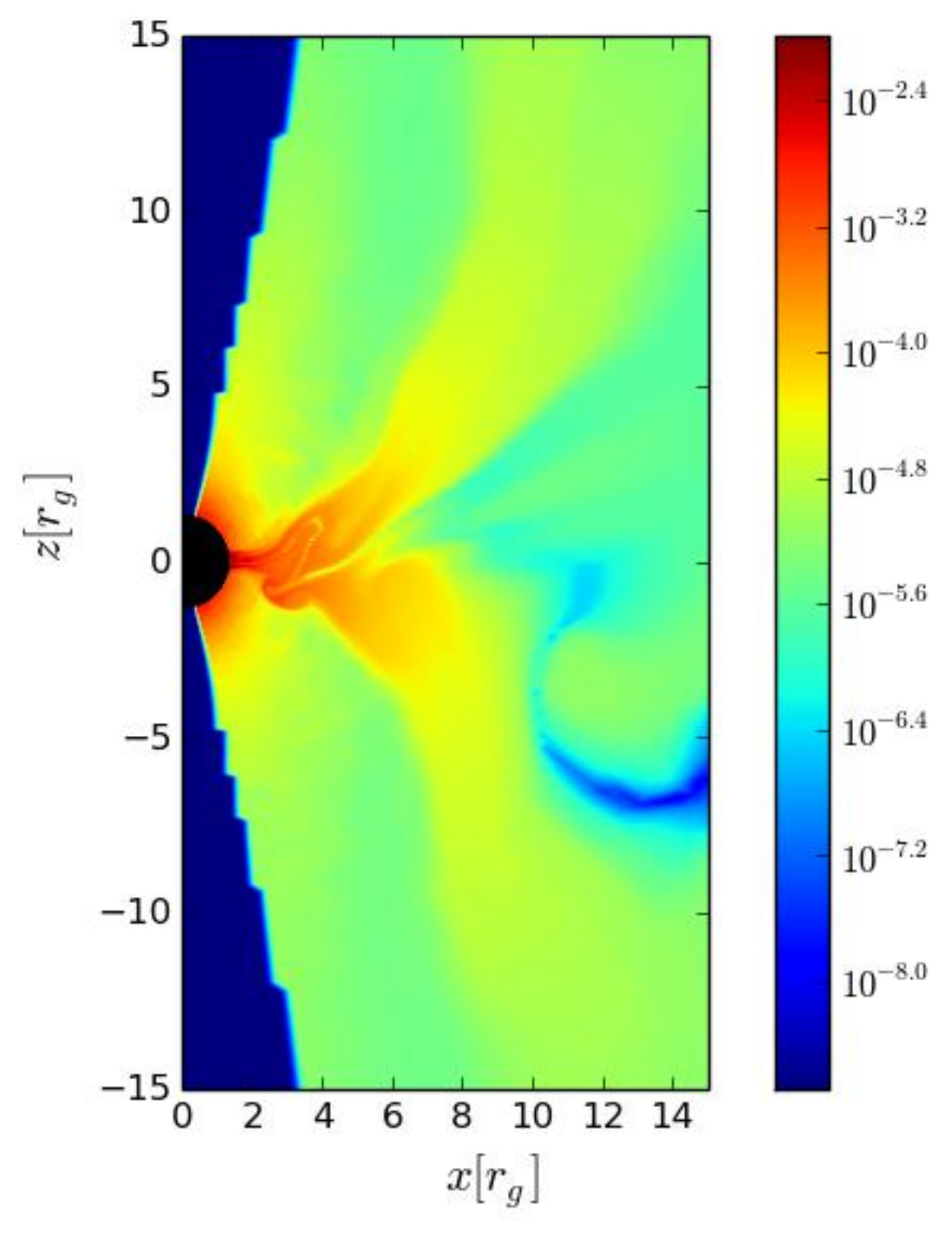} \\
    \includegraphics[width=0.22\textwidth]{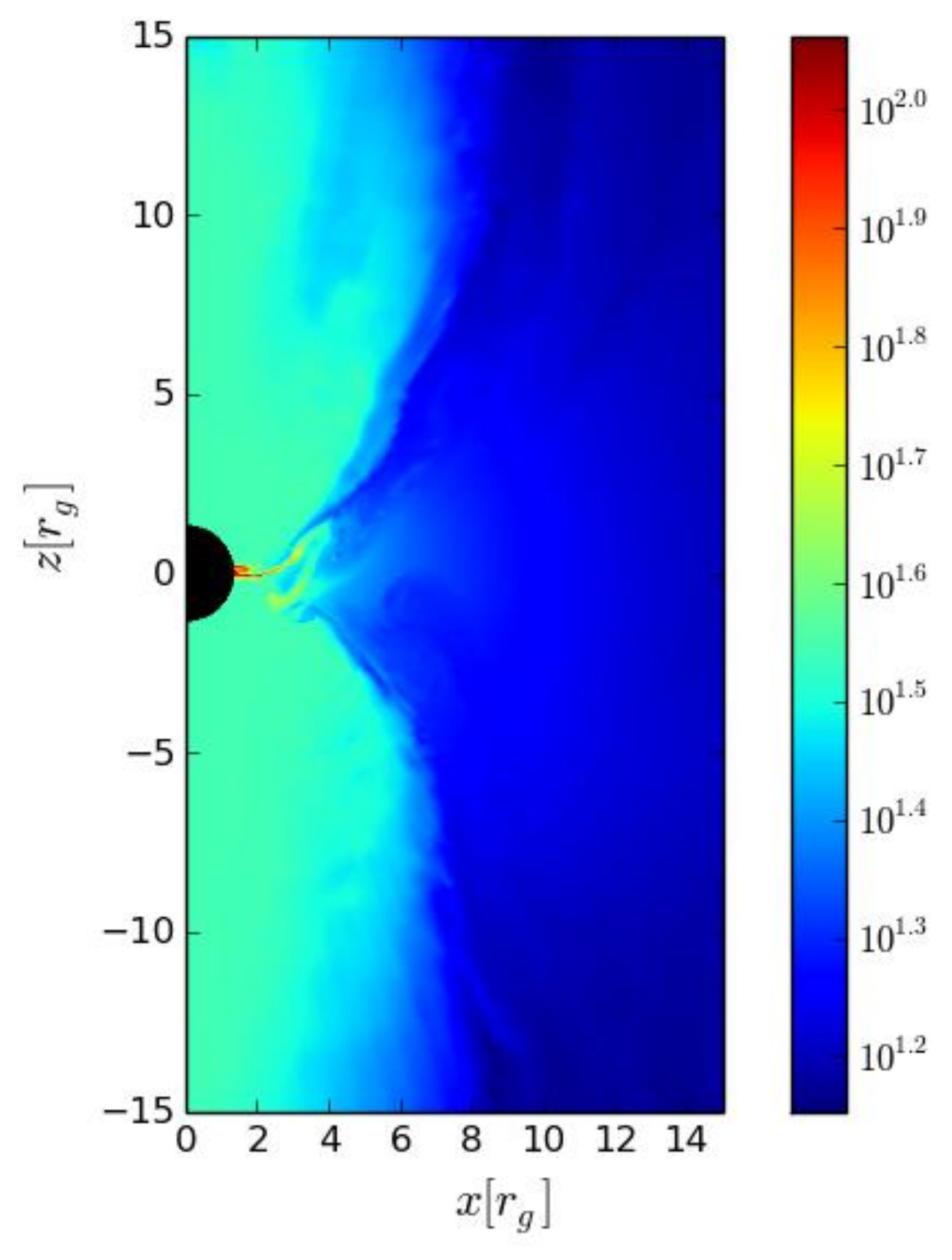} &
    \includegraphics[width=0.22\textwidth]{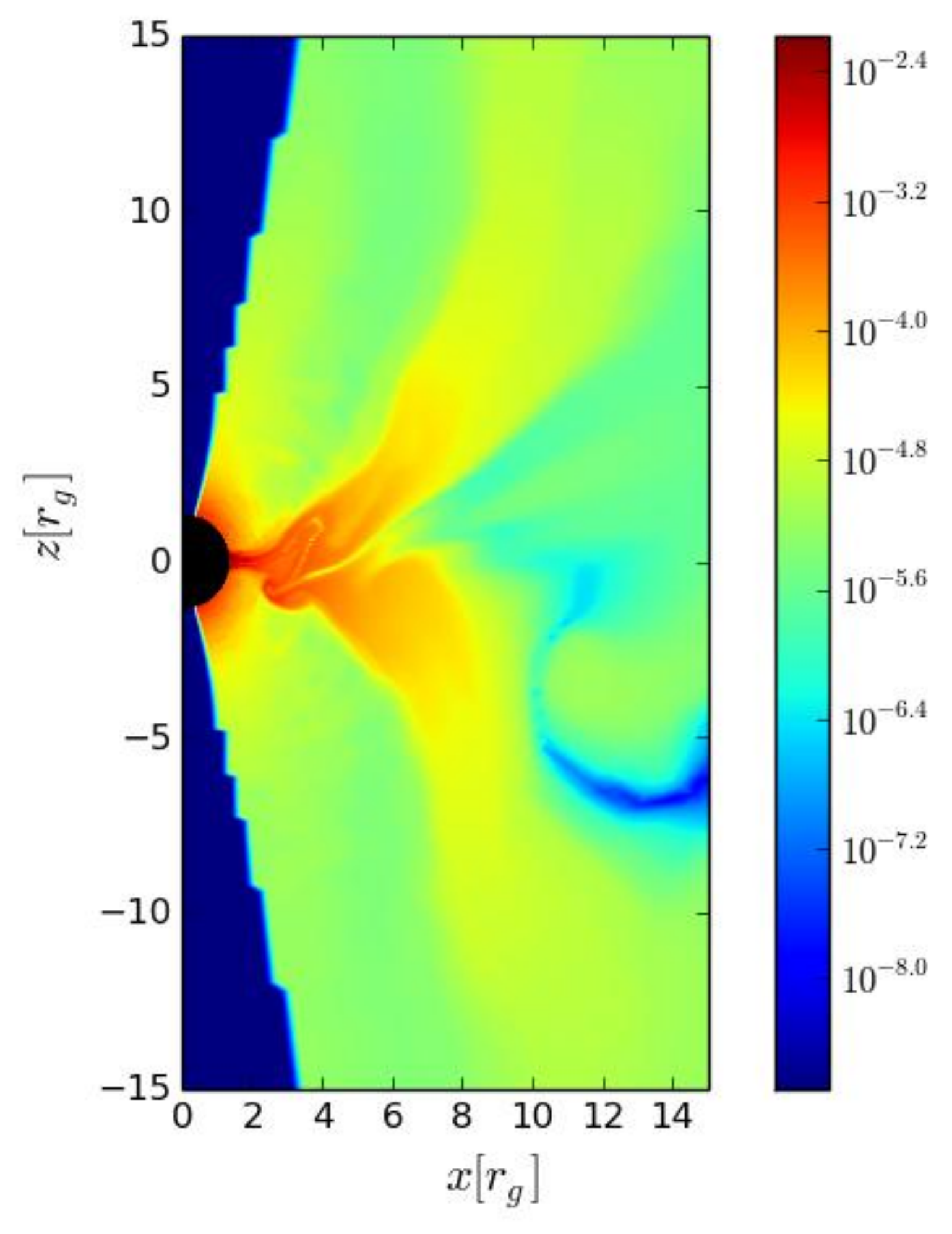} \\
    \includegraphics[width=0.22\textwidth]{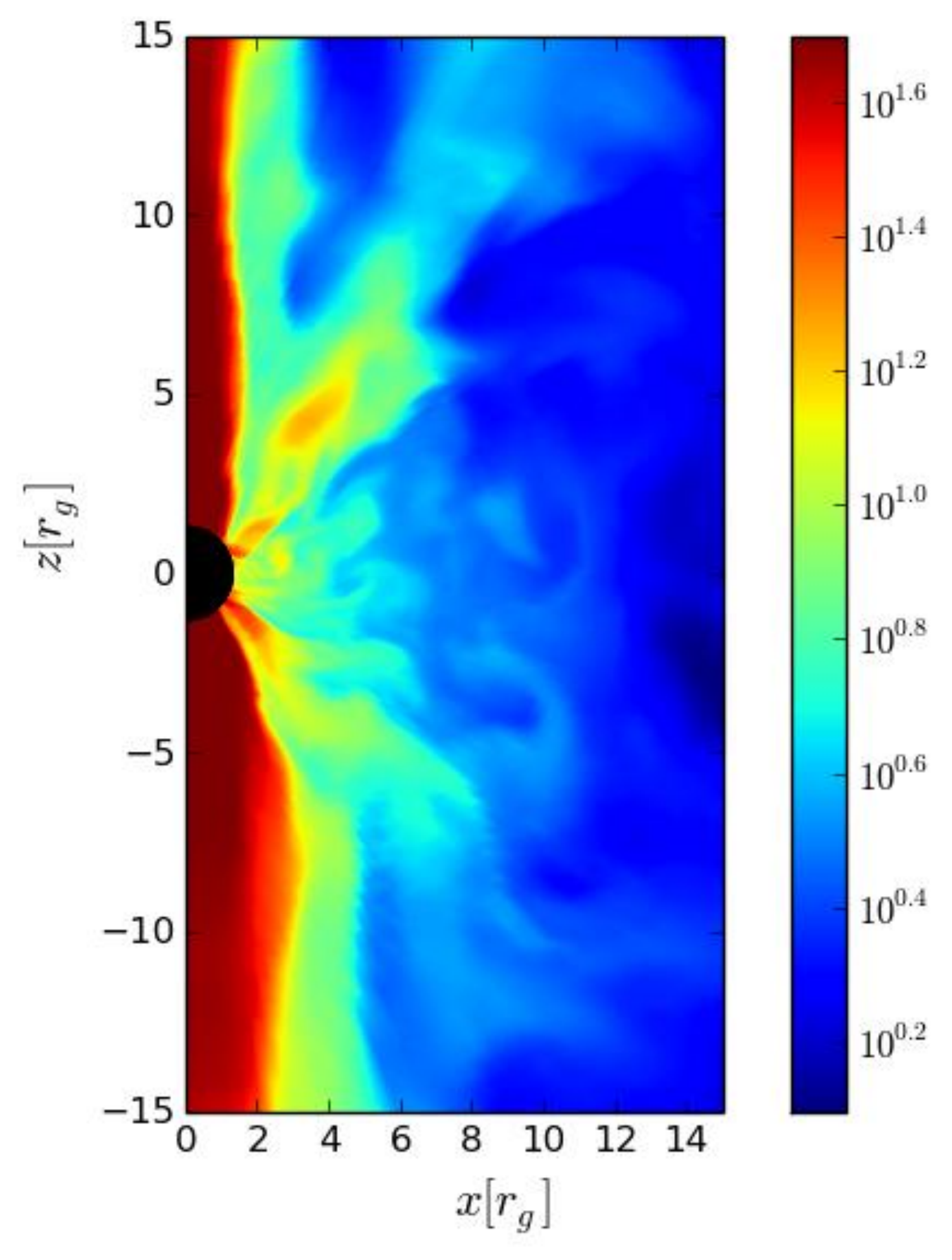} &
    \includegraphics[width=0.22\textwidth]{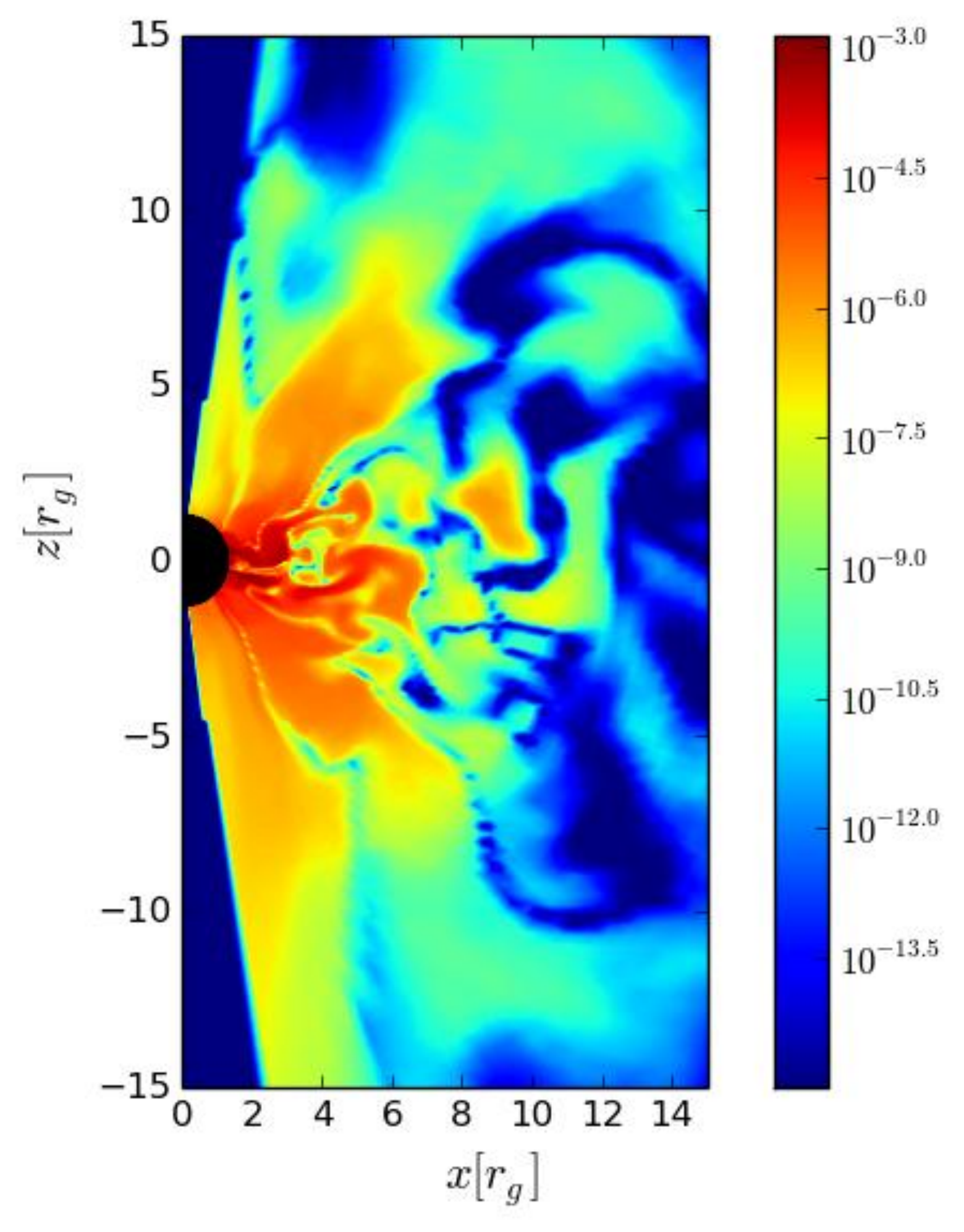} \\
    \includegraphics[width=0.22\textwidth]{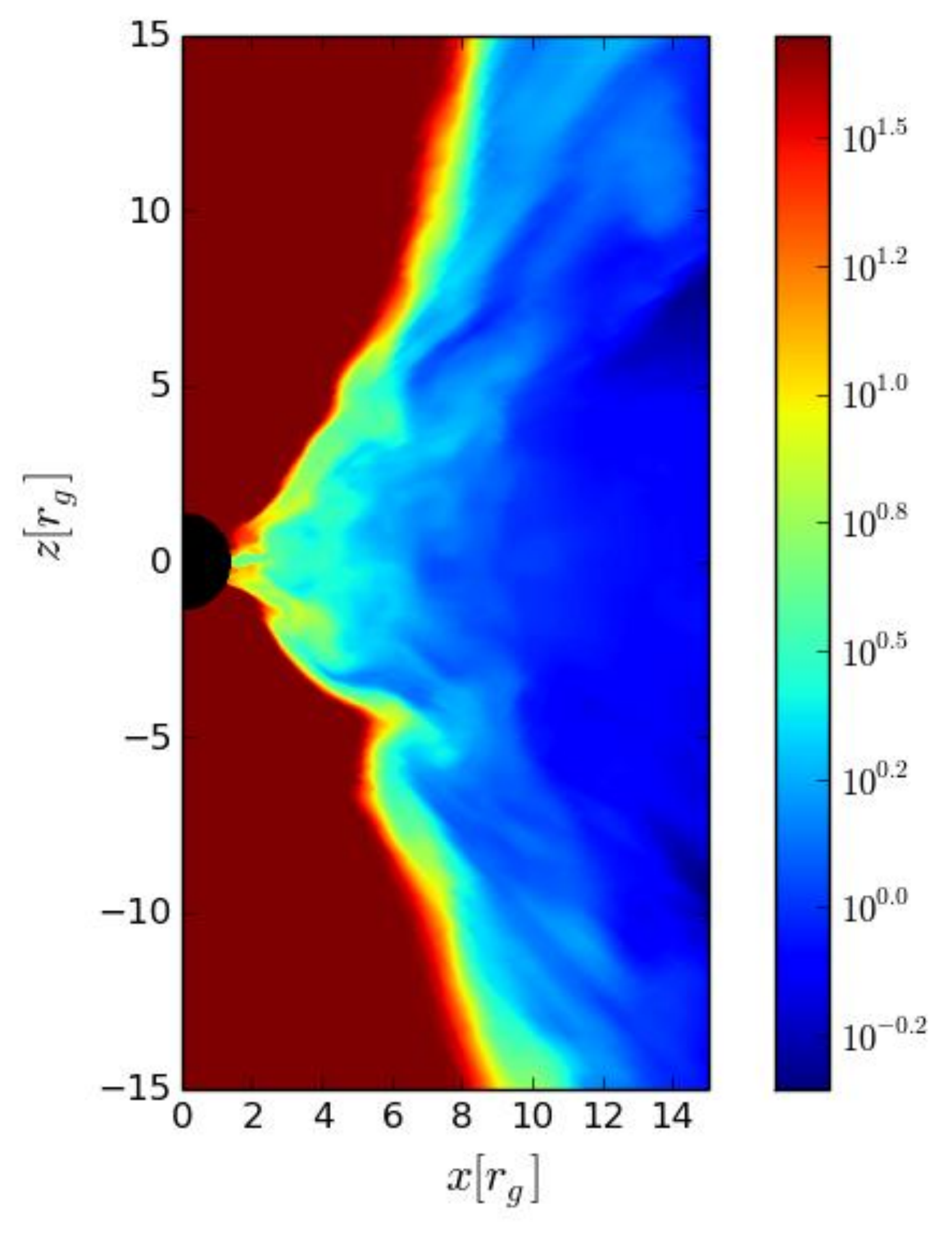} &
    \includegraphics[width=0.22\textwidth]{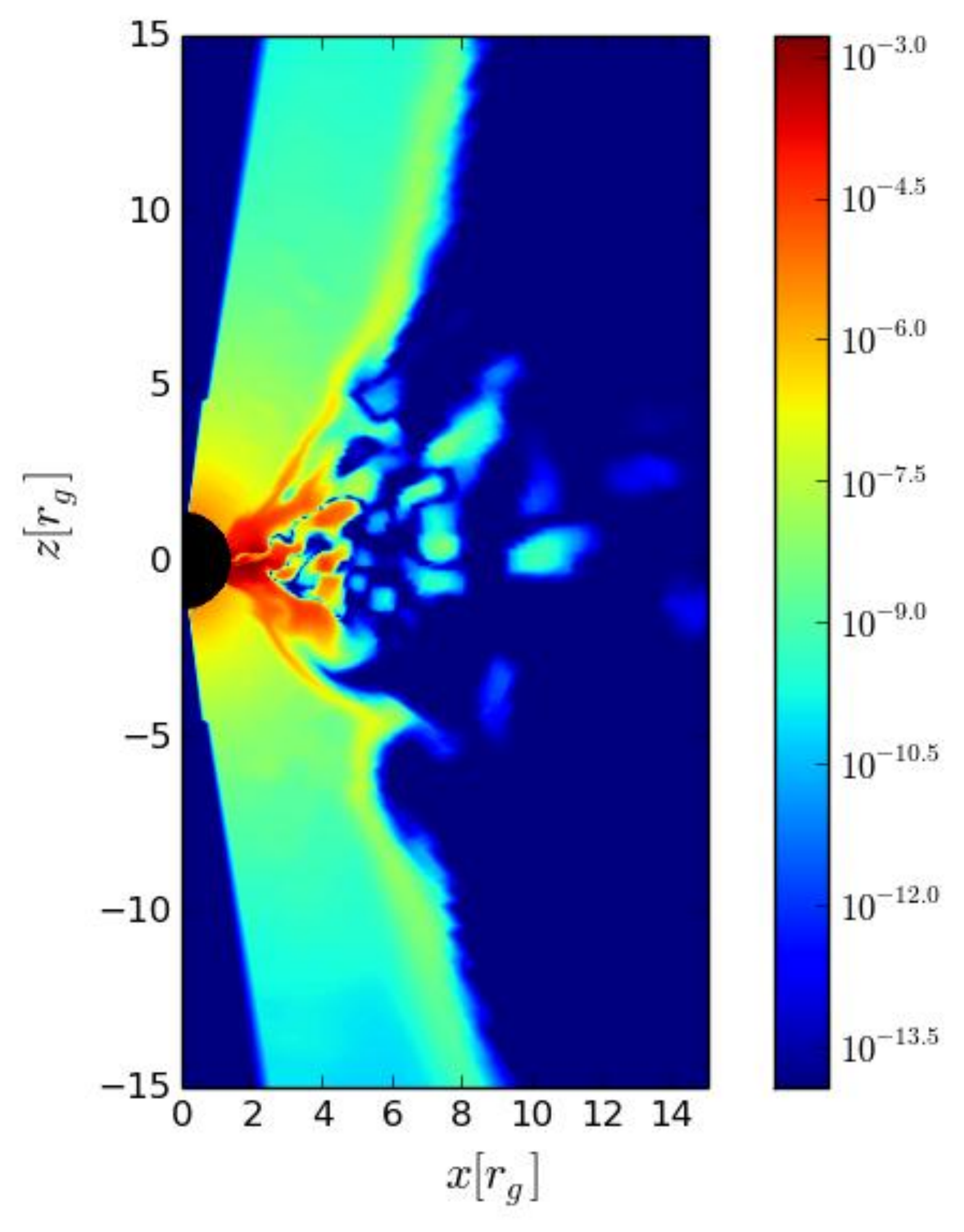} \\
  \end{tabular}
  \caption{Shows snapshot of electron temperature as
    $\log_{10}(\theta_e)$ where $\theta_e=k_{\rm B} T_{\rm e}/(m_{\rm e} c^2)$
    (left panels) and arbitrarily scaled 230GHz synchrotron emissivity
    per unit mass accretion rate $\log_{10}(j_\nu/\dot{M})$ (right
    panels) for the {\tt MAD\_thick-disk} model (upper two panels),
    {\tt MAD\_thick-jet} model (next vertically-middle panels), {\tt
      SANE\_quadrupole-disk} model (next lower two panels), and {\tt
      SANE\_dipole-jet} model (next lowest two panels). These are our
    default models we consider for parameter fitting.}
  \label{fig:Tejnu}
\end{figure}

\begin{figure}
  \centering
  \begin{tabular}{@{}cc@{}}
    \includegraphics[width=0.23\textwidth]{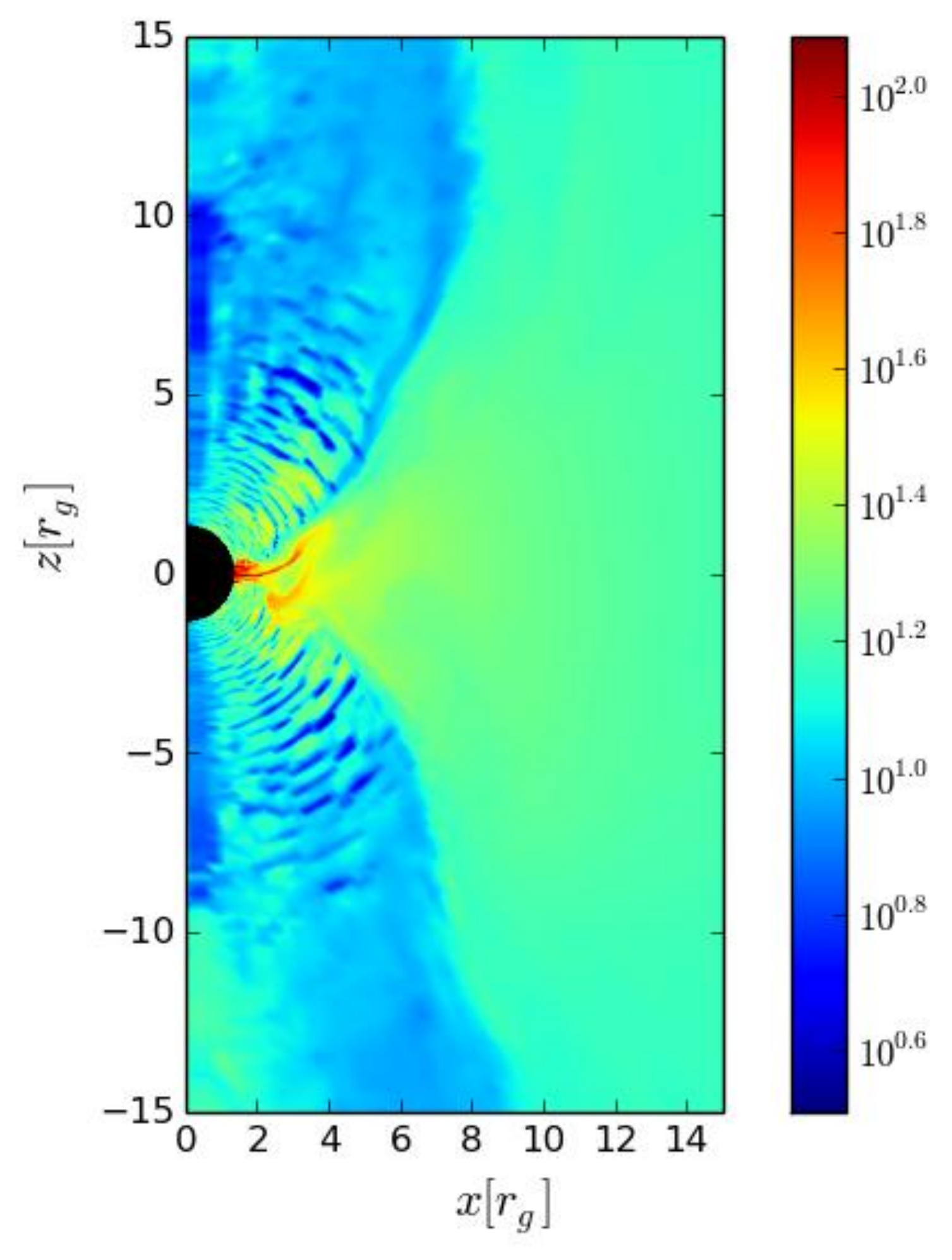} &
    \includegraphics[width=0.23\textwidth]{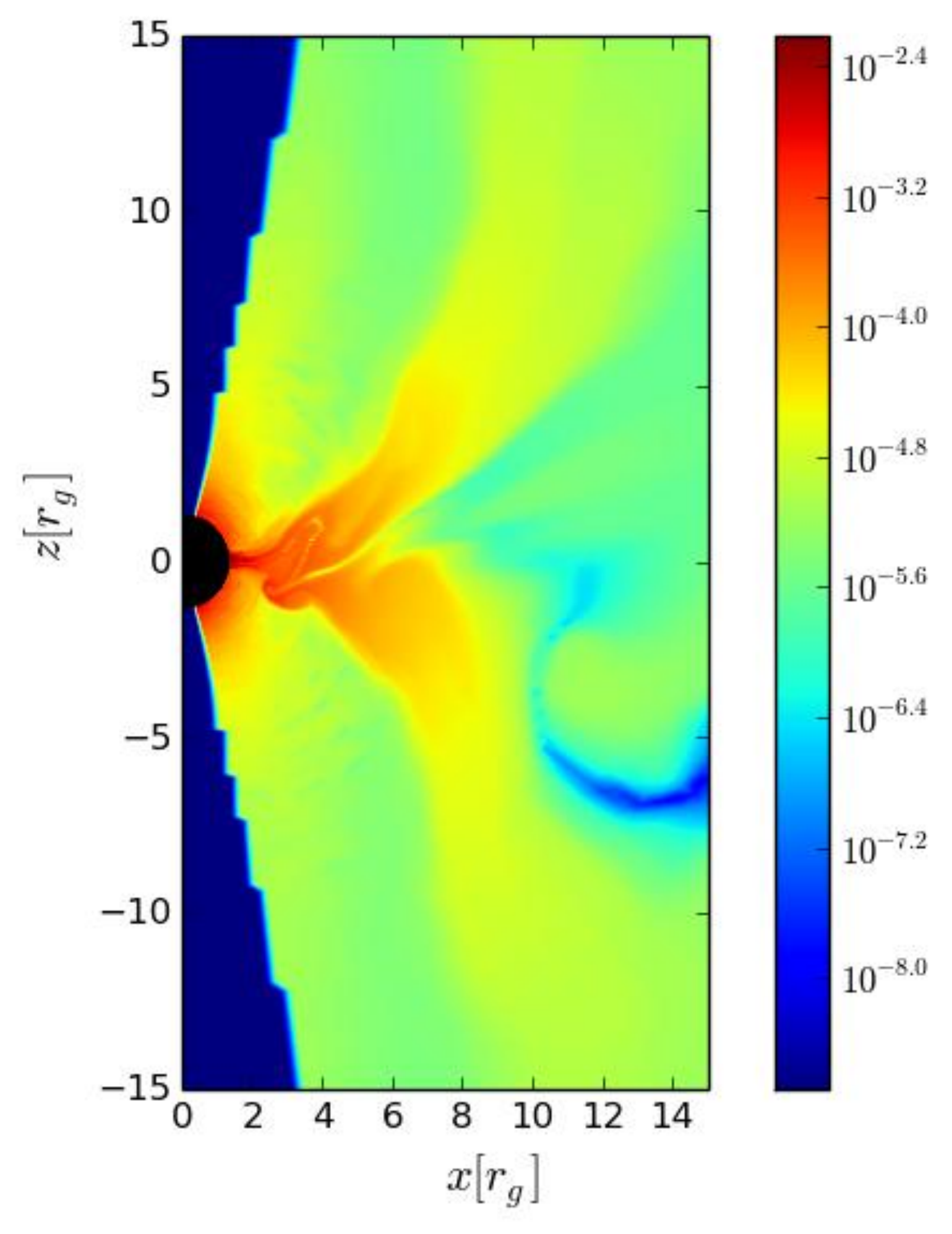} \\
    \includegraphics[width=0.23\textwidth]{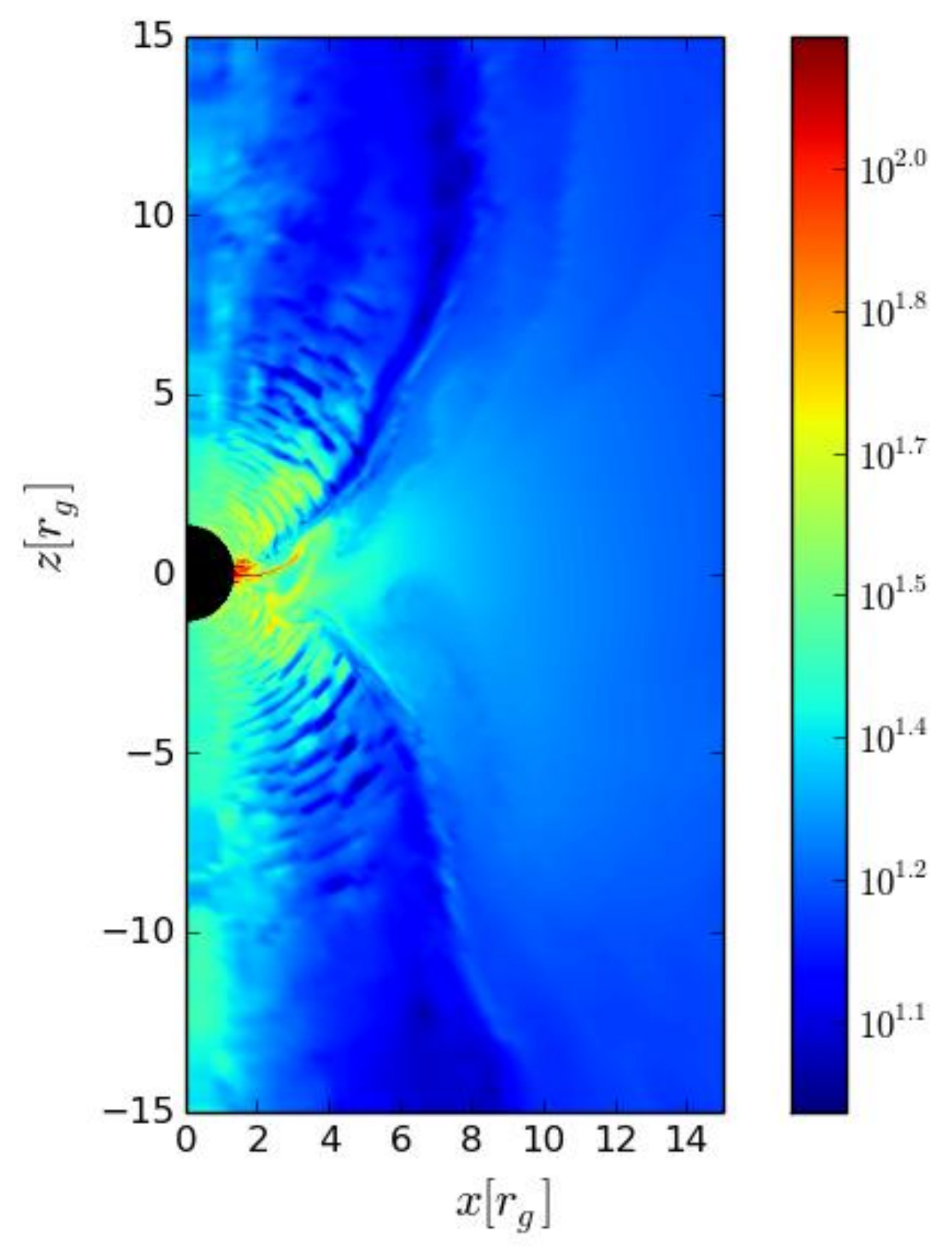} &
    \includegraphics[width=0.23\textwidth]{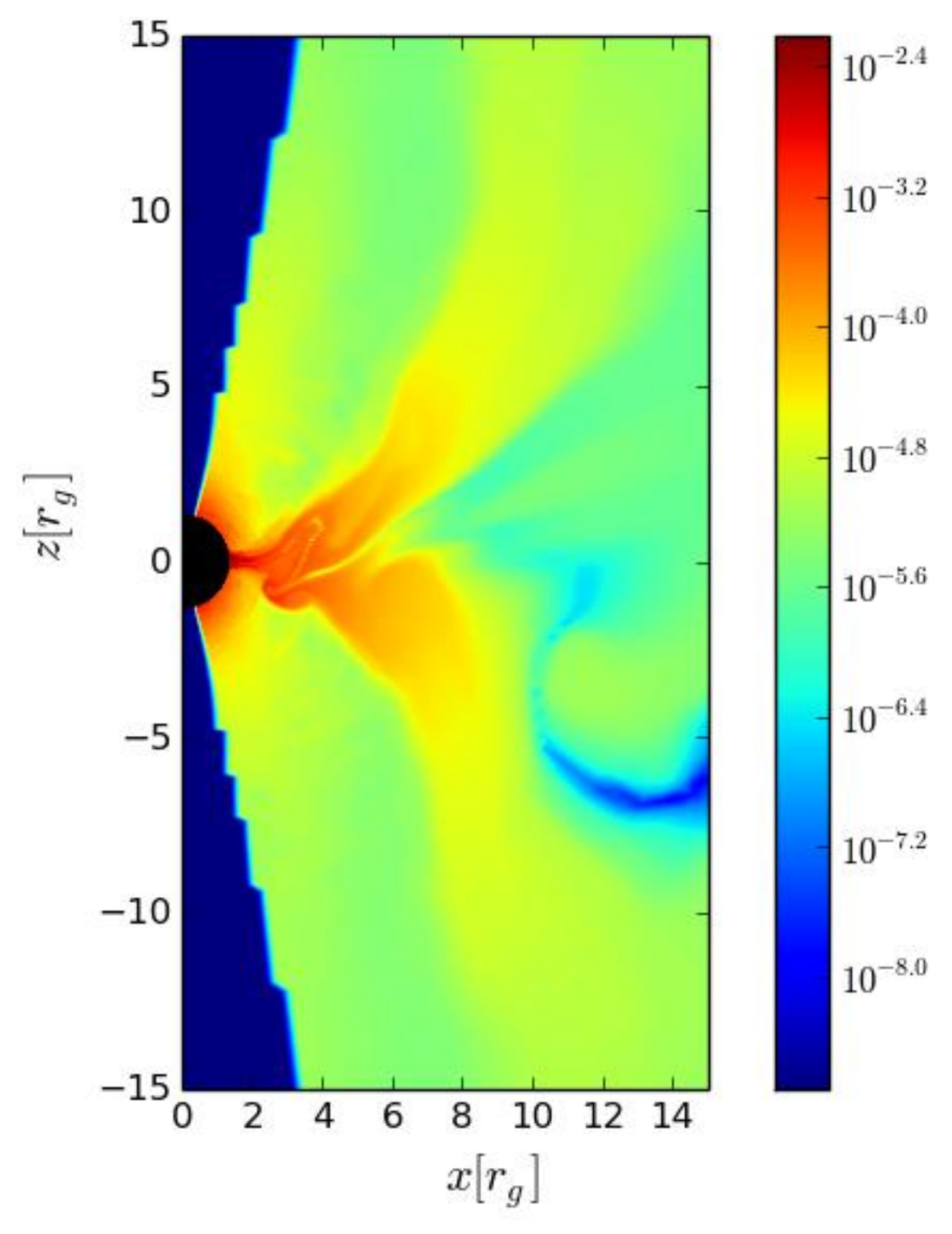} \\
    \includegraphics[width=0.23\textwidth]{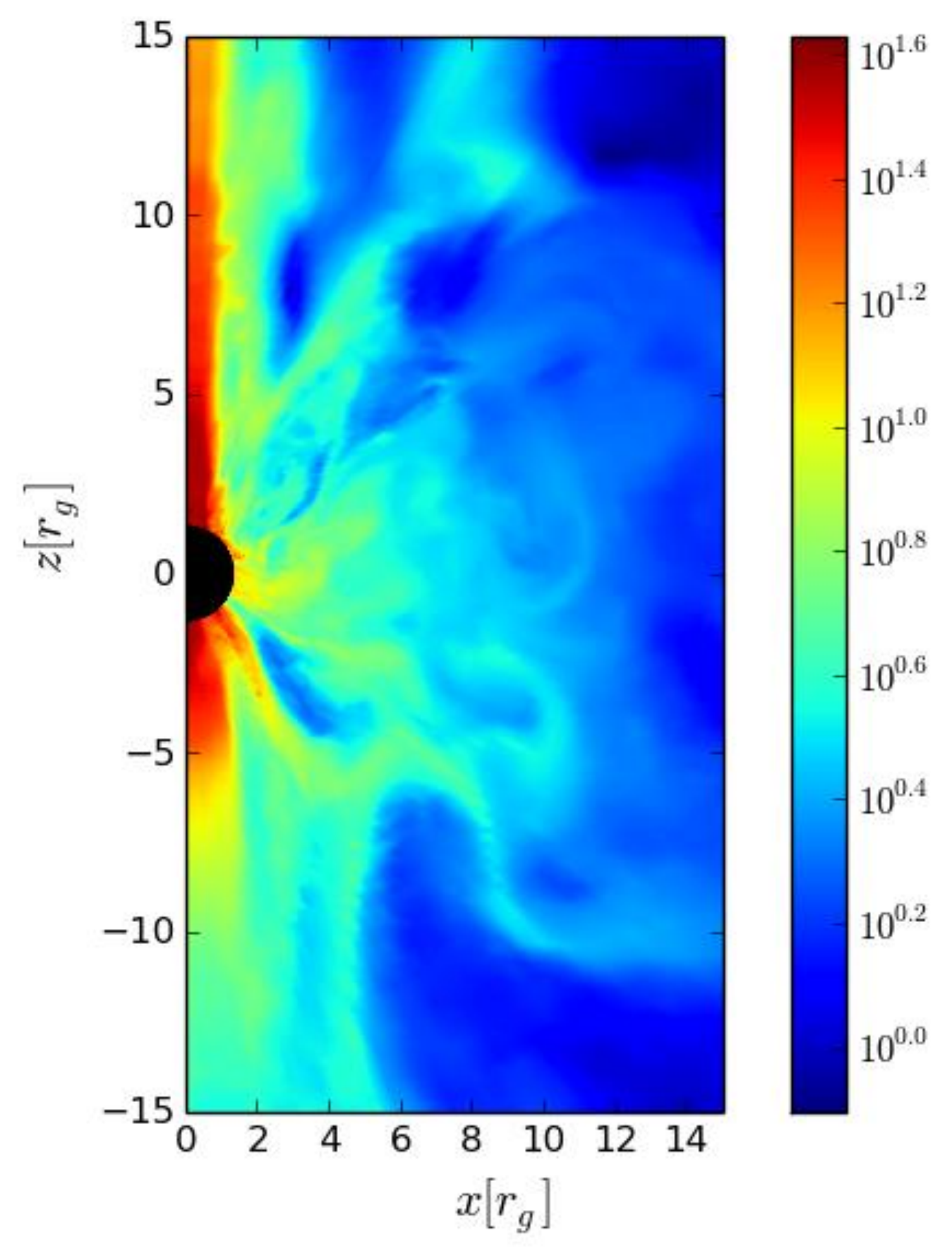} &
    \includegraphics[width=0.23\textwidth]{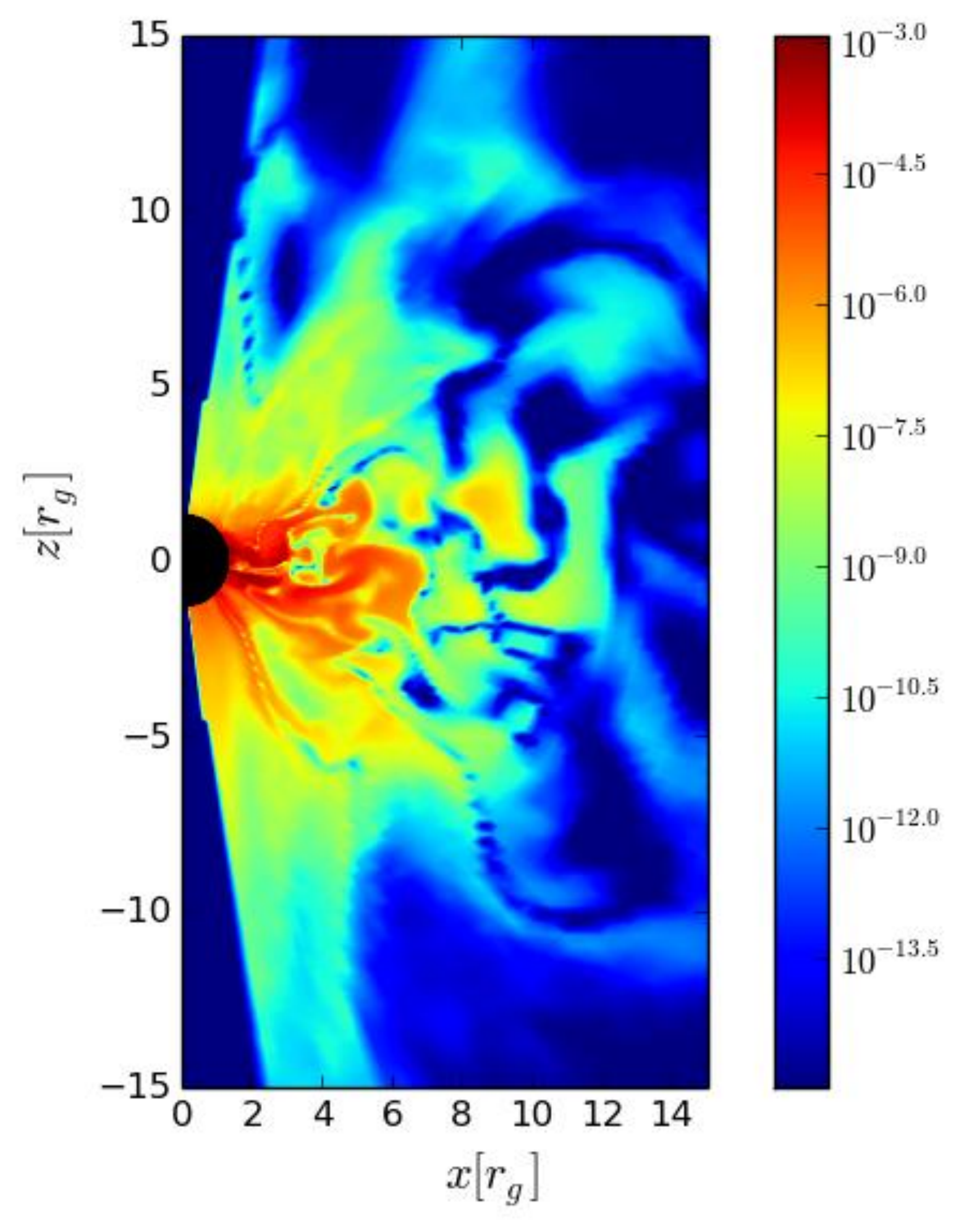} \\
  \end{tabular}
  \caption{Same as Fig.~\ref{fig:Tejnu}, but here using the
    alternative value of $\sigma_{T_{\rm e}}=40$ for the {\tt
      MAD\_thick-disk} model (upper two panels), {\tt MAD\_thick-jet}
    model (vertically-middle two panels), and {\tt
      SANE\_quadrupole-disk} model (lower two panels).  Later
    discussions do not focus on how these parameters affect the {\tt
      SANE\_dipole-jet} model, so such variations in prescriptions are
    not shown for that model.  As compared to Fig.~\ref{fig:Tejnu},
    the higher $\sigma_{T_{\rm e}}=40$ pushes higher electron
    temperatures into the region with higher $b^/2\rho$, so that for
    the {\tt MAD\_thick} model higher electron temperatures are
    achieved in the funnel where $b^2/\rho\sim 50$.  The higher
    $\sigma_{T_{\rm e}}$ leads to {\rm lower} temperatures for much of
    the {\tt SANE\_quadrupole-disk} model, because polar material has
    only up to $b^2/\rho\sim 2$ and has no BZ-driven jet.  Hence,
    $\sigma_{T_{\rm e}}=40$ allows us to focus higher isothermal
    temperatures in the BZ-driven funnel jet polar region if it exists
    in a model.}
  \label{fig:Tejnu-magn40}
\end{figure}

A major uncertainty in what leads to emission of \sgra\ is the
electron physics in weakly collisional plasmas.  Often ad hoc
prescriptions are adopted for electron temperatures.  A
``first-principle'' approach as in pair plasma pulsar wind studies
\citep{2015arXiv151001734P} is computationally unfeasible at present,
but some progress has been made.

We use the procedure described in \citet{2012ApJ...755..133S} to
modify the simulation proton and electron temperature based upon an
evolution of the temperatures within the disk based upon the
collisionless physics described by \citet{2007ApJ...671.1696S}.  This
involves extending the simulation data from some radius ($r\sim 50r_g$
for MAD and $r\sim 30r_g$ for SANE) with a power-law extension out to
the Bondi radius in \sgra.  We use a density power-law of $\rho\propto
r^{-0.85}$ and magnetic field strength scaling of $|b|\propto r^{-1}$,
which are consistent with GRMHD simulations and the densities implied
by Chandra X-ray observations \citep{2012ApJ...755..133S}. This outer
material primarily affects circular polarization and Faraday rotation
measures as the polarized radiation passes through a significant
column of toroidal field in the disk or corona.

The electron and proton temperatures are determined at all radii by a
combination of the GRMHD simulations and an evolution equation for
temperature.  The temperature is evolved radially inwards from an
initial temperature of $T_{\rm e}=T_{\rm p}=1.5\times 10^7$K at
$r=3\times 10^5r_g$ implied by Chandra X-ray emission, while the
density derives from the power-law extension mentioned above until
small radii when the $\theta$- and $\phi$- averaged GRMHD simulation
data is used.  The evolution of $T_{\rm e}$ and $T_{\rm p}$ is
controlled by proton-electron collisions (which dominate for $r\gtrsim
10^4r_g$), electrons being either non-relativistic or relativistic
(matters for $r\lesssim 10r_g$ in the disk), and the electron to
proton heating ratio $f_{\rm e}/f_{\rm p} = C_{\rm heat} \sqrt{T_{\rm e}/T_{\rm
    p}}$ for electron temperature $T_{\rm e}$ and proton temperature
$T_{\rm p}$.  This gives a result for the equatorial $T_{\rm e}$ and $T_{\rm p}$
vs.\ radius that uses both the GRMHD simulation data and radial
extension model.  As in our prior work, then, the original
simulation's temperature is modified by a mapping (at any $\theta$ or
$\phi$) from the simulation value of the internal energy per unit
rest-mass ($u/\rho$) to the evolution equation's result for $T_{\rm
  e}$ and $T_{\rm p}$ (for more details see
\citealt{2012ApJ...755..133S}).  This introduces another free model
parameter $C_{\rm heat}$ that is nominally of order $C_{\rm heat}\sim
0.3$ \citep{2007ApJ...671.1696S}, but it may be quite small in the
disk \citep{2015MNRAS.454.1848R}.  We identify this modified electron
temperature computed for all points in space and time as $T_{\rm
  e,gas}$.

Most recently, the combined efforts of following electrons
\citep{2015MNRAS.454.1848R} and protons \citep{2015arXiv151104445F}
suggest the proton temperature $T_{\rm p}$ is more isothermal than expected
from ideal MHD for regions like the funnel-wall jet, where the electron
temperature $T_{\rm e}$ rises up to $T_{\rm p}/2$ due to a high electron heating
rate at low plasma $\beta=p_{\rm g}/p_{\rm b}$ (gas pressure $p_{\rm g}$ and magnetic
pressure $p_{\rm b}$).  The funnel-wall jet is defined by the boundary
between the BZ-driven jet and the coronal wind where the magnetic
field, density, and pressure change dramatically.  This temperature
prescription leads to a fairly isothermal electron temperature within
the funnel-wall jet region, and this helps to produce the observed
flat radio spectra \citep{2014AA...570A...7M,2015arXiv151007243M}.

In order to mimic the isothermal approximation for the funnel-wall
jet, we prescribe the electron temperature as a smooth transition from
$T_{\rm e,gas}$ to a chosen value in the funnel-wall jet of $T_{\rm
  e,jet}$, using
\begin{equation}
  \label{eq:Te}
  T_{\rm e} = T_{\rm e,gas}  e^{-b^2/\rho \sigma_{T_{\rm e}}} + T_{\rm e,jet} (1 - e^{-b^2/\rho \sigma_{T_{\rm e}}}) ,
\end{equation}
where $b^2/2$ is the magnetic energy density in Heaviside-Lorentz
units and $\rho$ is the rest-mass density.  This mimics what would be
the combined results of \citet{2015MNRAS.454.1848R} and
\citet{2015arXiv151104445F} and is similar to what others have done
when wanting to mimic the effects of emitting non-thermal particles in
a jet \citep{2014AA...570A...7M,2015arXiv151007243M}.  The reference
magnetization is given by $\sigma_{T_{\rm e}}$ that defines the jet
funnel-wall boundary, where $\sigma_{T_{\rm e}}=1$ for {\tt
  MAD\_thick-disk}, {\tt SANE\_quadrupole-disk}, and {\tt
  SANE\_dipole-jet} models while $\sigma_{T_{\rm e}}=4$ for the {\tt
  MAD\_thick-jet} model.  The funnel-wall
jet electron temperature $T_{\rm e,jet}$ (per unit 
$m_{\rm e} c^2/ k_{\rm B}$, hereafter we drop the $m_{\rm e} c^2/ k_{\rm B}$ factor) is
$T_{\rm e,jet}=10$ for model {\tt MAD\_thick-disk},
$T_{\rm e,jet}=35$ for model {\tt MAD\_thick-jet},
$T_{\rm e,jet}=50$ for model {\tt SANE\_quadrupole-disk},
and $T_{\rm e,jet}=100$ for model {\tt SANE\_dipole-jet}.
For models (except {\tt SANE\_dipole-jet}), we also
consider the alternative value of $T_{\rm e,jet}=100$.

The electron temperature prescription was chosen such that emission
from the jet material was suppressed in the {\tt MAD\_thick-disk}
model, whereas somewhat more jet emission was allowed in {\tt
  MAD\_thick-jet}.  Parameters for the electron temperature
prescription were chosen to represent a jet-dominated emission model
for {\tt SANE\_dipole-jet} that contains a BZ-driven jet, and the
electron temperatures were chosen to be disk-dominated for {\tt
  SANE\_quadrupole-disk} that contains no BZ-driven jet.

We use a smooth interpolation to prescribe electron temperatures and
BZ-jet mass-loading, which avoids sharp features that can suddenly
appear and disappear due to using hard cuts on a specific single value
of a physical quantity, like plasma $\beta$ or the unboundedness of
the fluid \citep{2013AA...559L...3M,2015ApJ...812..103C}.  In future
work, we will consider more advanced evolution equations for the
electron temperatures \citep{2015MNRAS.454.1848R} and proton
temperatures \citep{2015arXiv151104445F}.

\subsection{Jet mass-loading prescription}
\label{sec:rhojet}

\begin{figure}
  \centering
  \begin{tabular}{@{}cc@{}}
    \includegraphics[width=0.23\textwidth]{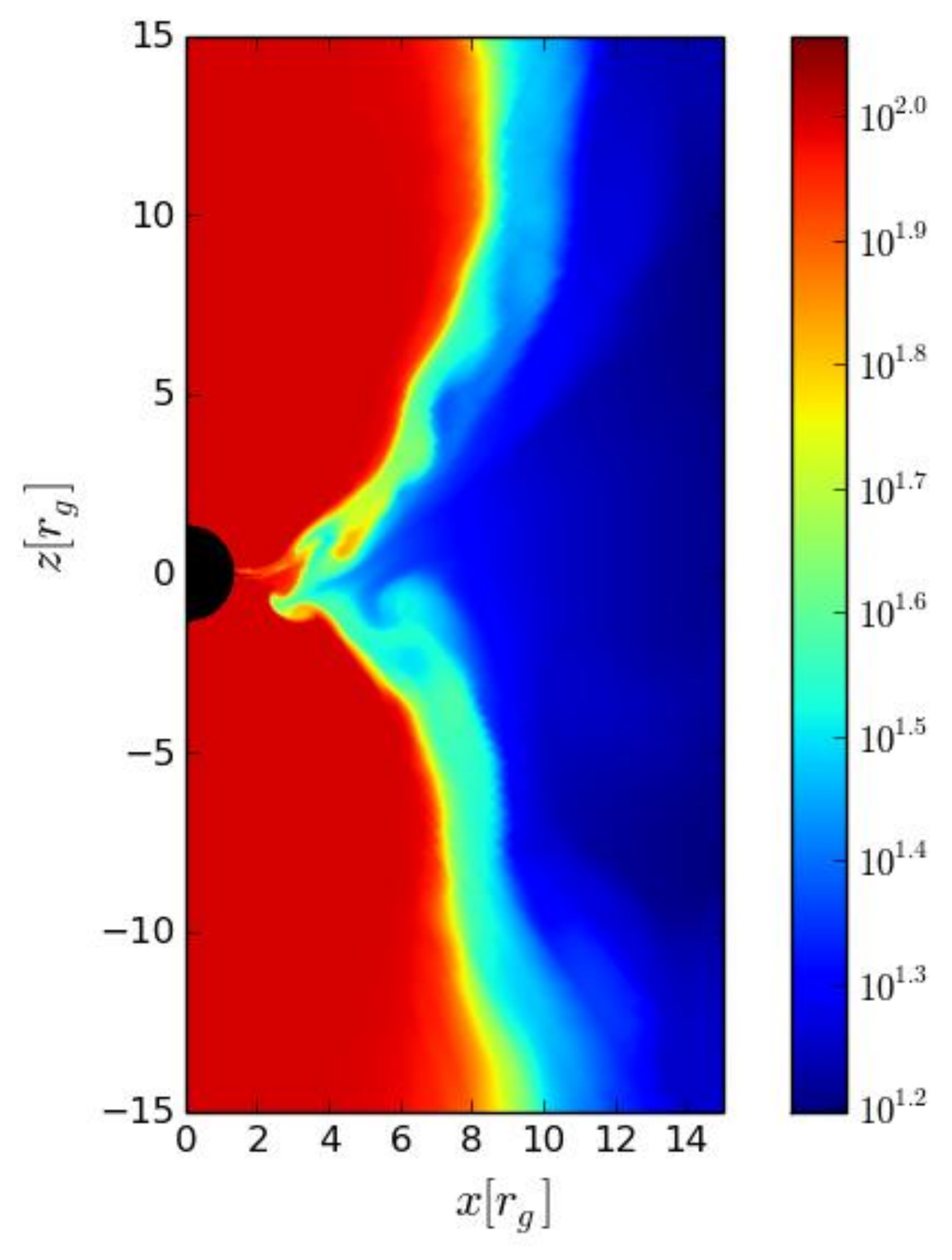} &
    \includegraphics[width=0.23\textwidth]{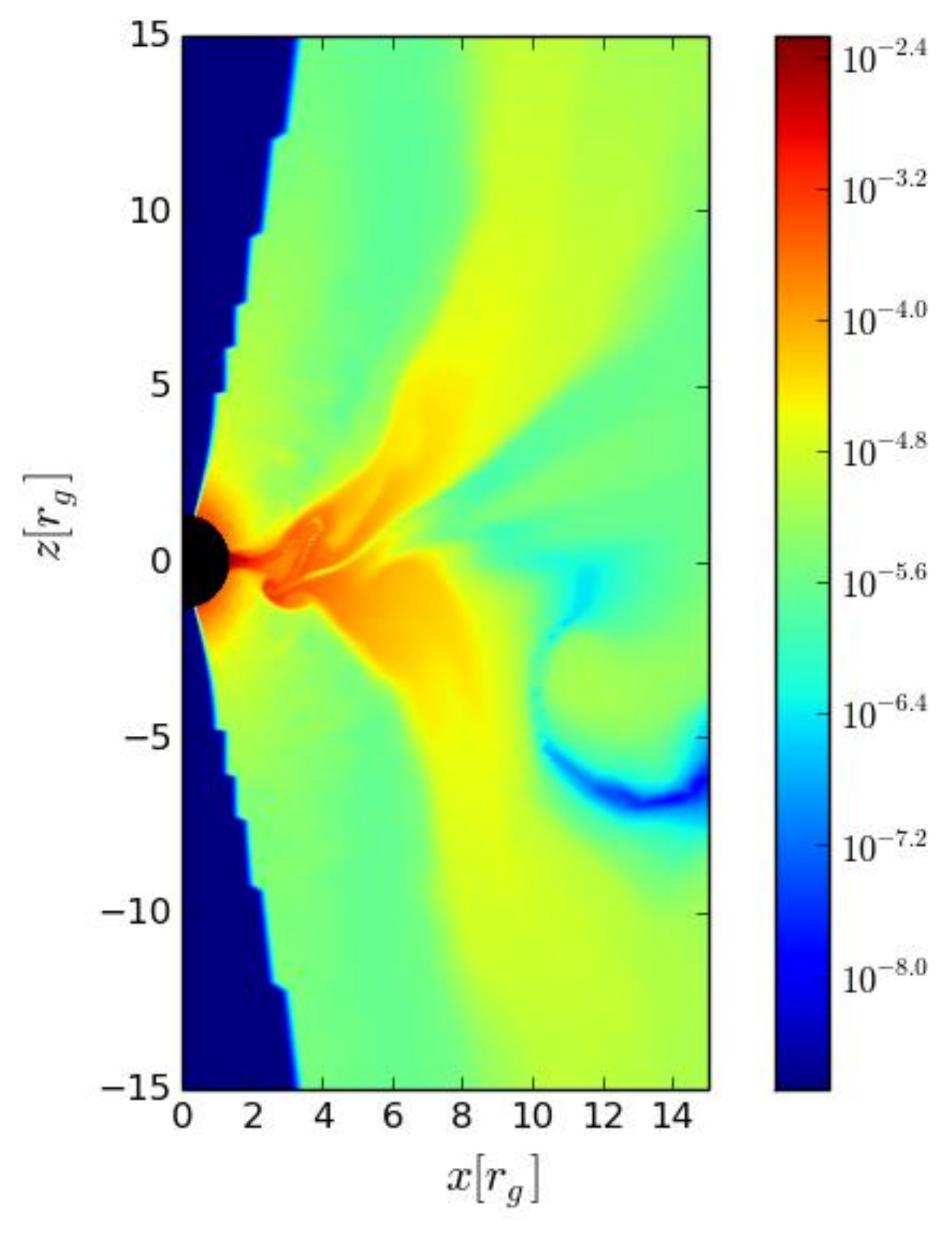} \\
    \includegraphics[width=0.23\textwidth]{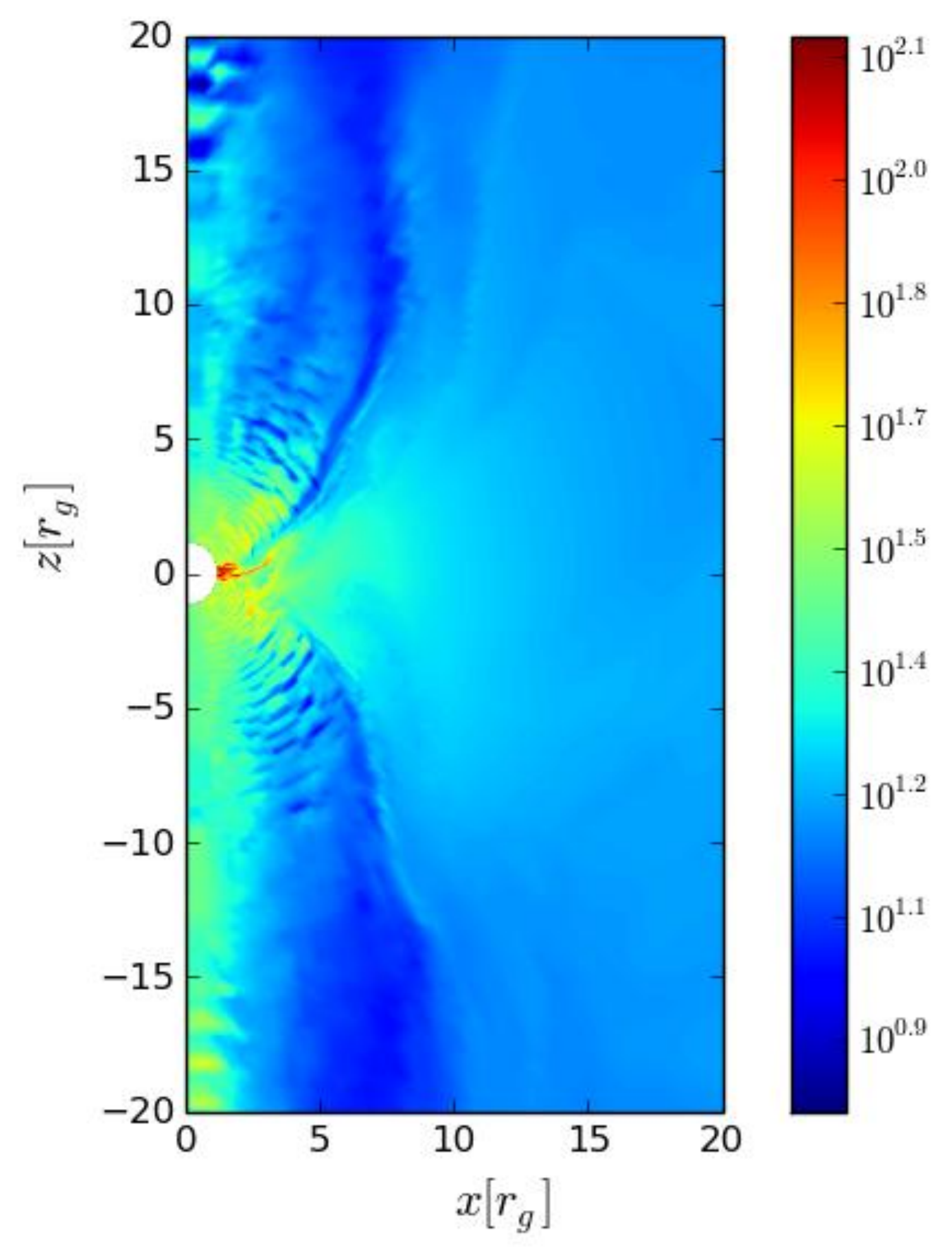} &
    \includegraphics[width=0.23\textwidth]{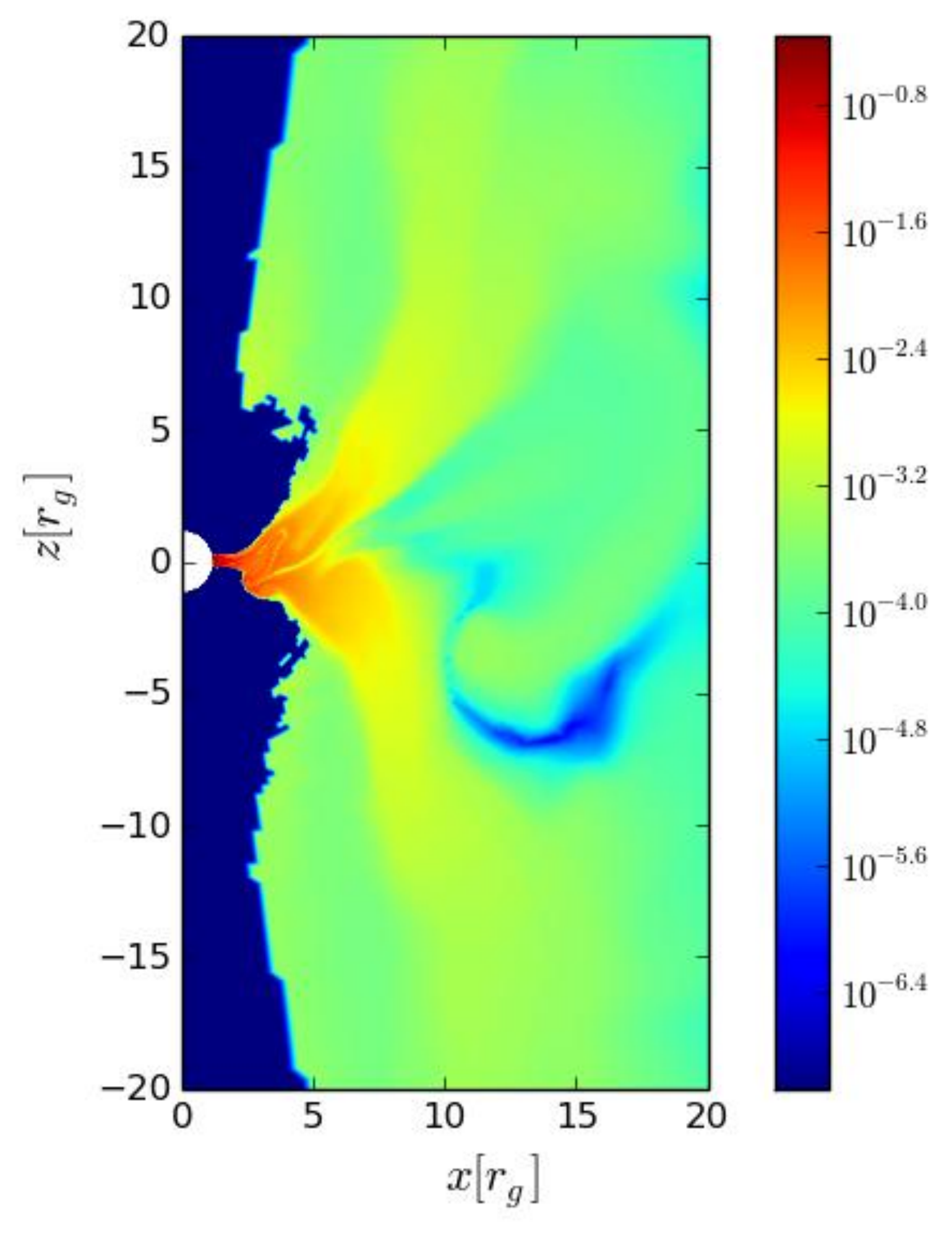} \\
    \includegraphics[width=0.23\textwidth]{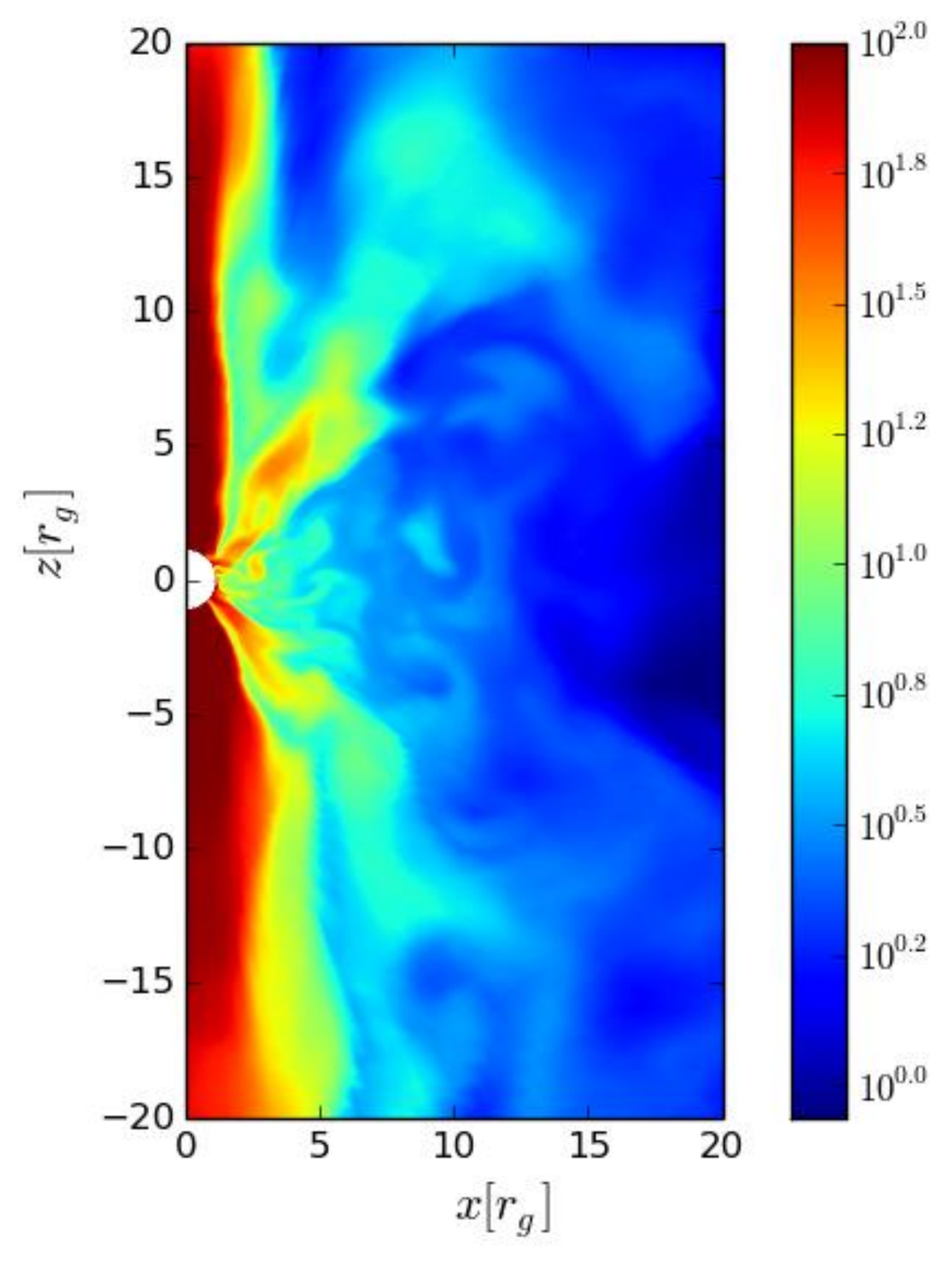} &
    \includegraphics[width=0.23\textwidth]{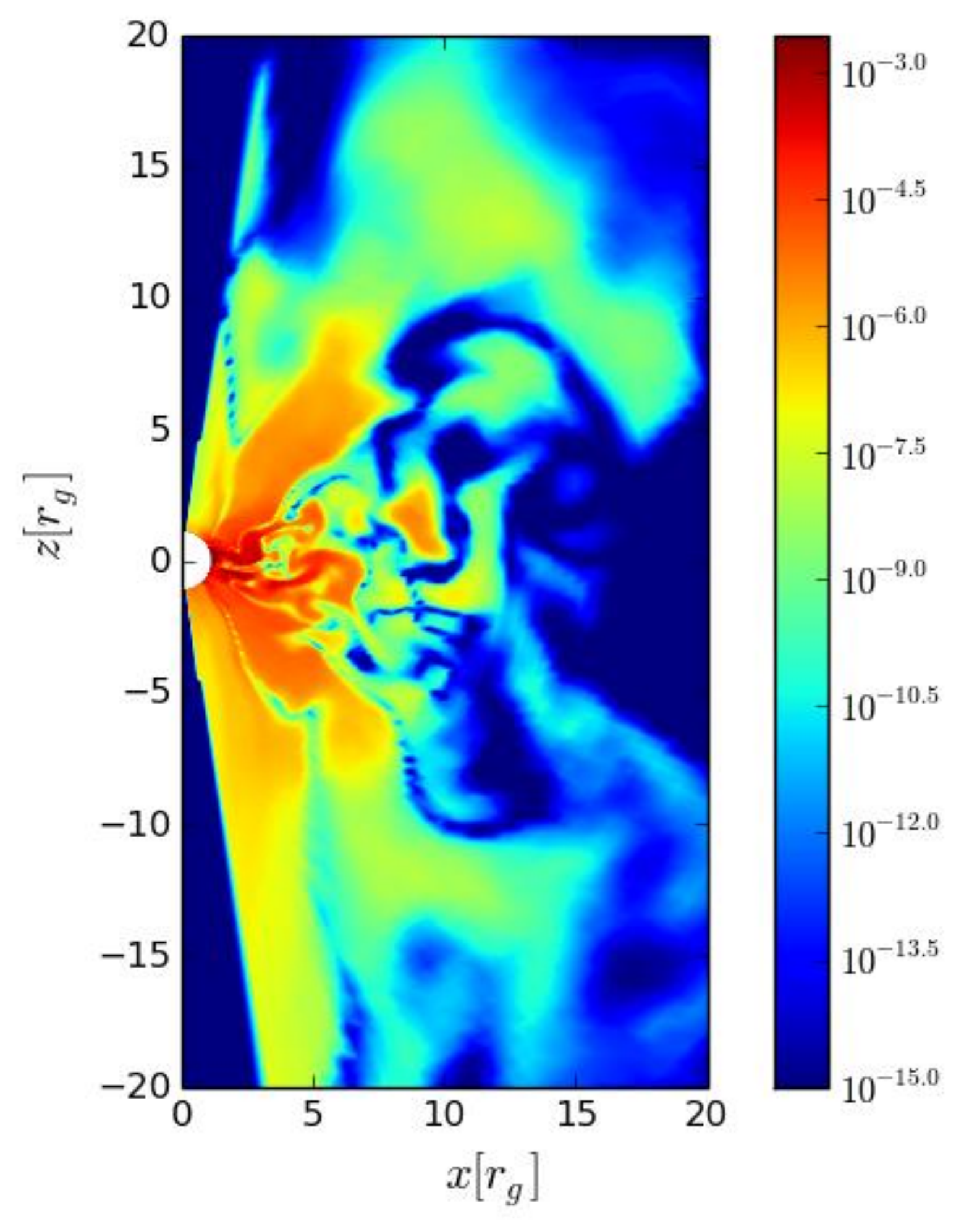} \\
  \end{tabular}
  \caption{Same as Fig.~\ref{fig:Tejnu}, but for the {\tt
      MAD\_thick-jet} model with $T_{\rm e,jet}=100$ (upper two
    panels), {\tt MAD\_thick-jet} model with $\sigma_{T_{\rm e}}=40$ and
    $\sigma_{\rho}=10$ (vertically-middle two panels), and {\tt
      SANE\_quadrupole-disk} model with $T_{\rm e,jet}=100$ (lower two
    panels).  These panels show how the wind or jet emission is
    enhanced in different ways.  Lighting up only self-consistent
    material that obeys mass conservation (middle two panels) leads to
    less BZ-jet funnel emission as compared to the otherwise similar
    model shown in Fig.~\ref{fig:Tejnu-magn40}.  This way we can
    identify the difference between funnel-wall jet emission and
    BZ-driven funnel jet emission.}
  \label{fig:Tejnu-te100}
\end{figure}

We control the rest-mass density in the BZ-jet funnel region where,
nominally, matter is injected to keep the numerical scheme stable as
necessary for any GRMHD code that models BZ jets
\citep{2003ApJ...589..444G}.  The simulations we consider inject mass
once $b^2/\rho>50$ (our MAD models) or $b^2/\rho>200$ (our SANE
models), but this leads to much more mass accumulating near the
stagnation radius (where the flow either moves in due to gravity or
out due to the jet) \citep{2013PhRvD..88h4046G}, leading to
$b^2/\rho\gtrsim 10$ there.  As long as the relativistic jet has not
had much radial range to accelerate significantly, large values of
$b^2/\rho$ do nothing to modify the GRMHD solution except to rescale
the funnel density.  So, the density could be chosen to be much higher
(up to $b^2/\rho\sim 5$) or much lower without actually affecting the
dynamics because the jet Lorentz factor never goes beyond $\gamma\sim
5$ (only occurring at large radii $r\sim 10^3r_g$, beyond the range of
interest for this paper focused on horizon-scale structures).

By default, we control the density by only applying a polar axis
cut-out, which removes material very near the polar axis that is
numerically inaccurate and causes the densities to be momentarily
artificially high.  A similar approach is taken by others
\citep{2015ApJ...799....1C}.  For MAD models, we remove material
within $0.025$ radians, while for our SANE models we remove material
within $0.01$ radians.  This polar axis cut still leads to material
being present near the horizon in the polar region that would
represent some mass-loading of the jet.  For the MAD model, this leads
to a density in the funnel comparable to the density in the disk, and
so the default MAD model can have (depending upon $T_{\rm e,jet}$)
competing emission from the disk or jet.  For the dipole model, the
numerical floor-injected density in the funnel is much lower than in
the MAD case, so only a very high temperature would (at some higher
frequency) cause the funnel jet light up.

We also consider an alternative density removal procedure (still
including the polar angle cut-out), where we remove the polar material
using
\begin{align}
  \label{eq:rho}
  \rho = \rho_{\rm gas} e^{-b^2/\rho \sigma_{\rho}} + \rho_{\rm
    jet} (1 - e^{-b^2/\rho \sigma_{\rho}}) ,
\end{align}
with $\sigma_{\rho}=10$ and $\rho_{\rm jet}=0$, which for all models
does a good job at removing material that is numerically injected near
the BH.  Here $\rho_{\rm gas}$ is the original simulation rest-mass
density that includes the floor injection material.  After removing of
the density using $\sigma_{\rho}=10$ and $\rho_{\rm jet}=0$, only
self-consistent material that obeys conservation of mass is left.  The
default polar axis angle-based density removal procedure keeps funnel
jet material that is dependent upon uncertain mass-loading physics.
For the {\tt SANE\_quadrupole-disk} model, there is never any region
with $b^2/\rho\gtrsim 2$, so only the polar angle cut matters.  In
future work, we will consider simulations that directly track the
injected mass.

Fig.~\ref{fig:Tejnu} shows the coordinate x-z plane for our default
models (i.e., models with default choices for electron temperature and
mass-loading prescriptions). The figures show the electron temperature
and the arbitrarily scaled 230GHz thermal synchrotron emissivity,
which helps to identify the origin of emission seen in the final
radiative transfer results.  In some cases, like the {\tt
  SANE\_quadrupole-disk} model, low-level emissivity is truncated a
bit by the density removal or polar cuts, but in other models the
polar axis density cut-out has little effect on the emissivity.

Fig.~\ref{fig:Tejnu-magn40} shows some alternative electron
temperature prescriptions using a higher $\sigma_{T_{\rm e}}=40$.
Fig.~\ref{fig:Tejnu-te100} show cases where $T_{\rm e,jet}=100$ or our
alternative mass-loading choice. In the {\tt MAD\_thick} model, our
default polar angle cut-off only cuts-out matter that was injected by
the numerical floors but does a poor job of removing all injected
material, because its primary purpose was to only remove material
near the polar axis.  Our alternative additional cut-off using
$\sigma_{\rho}=10$ somewhat accurately removes the
numerically-injected floor material \citep{2012MNRAS.423.3083M}.

\subsection{Polarized radiative transfer scheme}

The data from GRMHD simulations presented in the previous section
serve as input for a general relativistic polarized radiative transfer
scheme \citep{2014ascl.soft07007S,2011MNRAS.410.1052S}, an extended
version of {\tt ASTRORAY} \citep{SPM12}. In its present stage, the
code assumes a thermal, isotropic distribution function for the
electrons, and it includes Faraday rotation and conversion. The code
has been used to model polarized synchrotron emission and absorption
of \sgra\ in the past \citep{SPM12,SM13}.

The {\tt ASTRORAY} code performs direct transfer of light through the
entire GRMHD simulation's dependence vs.\ space and time. The GRMHD
data is sampled every $4r_g/c$ (MAD model, $1.4$ minutes for \sgra) or
$2r_g/c$ (SANE models, $0.7$ minutes for \sgra).  We do not use the
so-called fast-light approximation, which assumes an infinite speed of
light.  The fast-light approximation has been found to be inaccurate
on timescales less than $10r_g/c$ ($3.5$ minutes for \sgra)
\citep{SPM12} and can considerably change the character of the
behavior in time \citep{2010ApJ...717.1092D}.  The fast-light
approximation may lead to stronger lensing features, which would
otherwise be washed out, because nearby emitting regions would not as
easily correlate their emission.  This also means we do not have to
choose between a time-averaged flow (to try to obtain a more realistic
distribution of density and other plasma properties for each snapshot)
vs.\ the average of snapshots (see discussion in
\citealt{2015ApJ...799....1C}).  For our models, errors introduced by
the fast-light approximation at $f \geq 230{\rm GHz}$ exceed $\Delta
F/F \gtrsim 15\%$, although the effects are insignificant at lower
frequencies.  Instead, we compute snapshots (or time-averages of
snapshots) in the observer's reference frame using transfer through
the full time-dependent simulation data.

As compared to prior similar work
\citep{2009ApJ...703L.142D,2009ApJ...706..497M,2010ApJ...717.1092D,2014AA...570A...7M,2015ApJ...812..103C,2015ApJ...799....1C},
no other work has considered the role of polarization with GRMHD
simulations except our own prior works \citep{SPM12,SM13} and
preliminary work by \citet{2014IAUS..303..298D}.  The {\tt
  MAD\_thick-disk} model is similar to models used in prior
polarimetric work by us \citep{2012ApJ...755..133S,SM13}.  The {\tt
  SANE\_quadrupole-disk} model has not been used before in polarized radiative
transfer studies.  The {\tt SANE\_dipole-jet} model has been
used in prior studies of both \sgra\ \citep{2010ApJ...717.1092D} and
M87 \citep{2012MNRAS.421.1517D} without polarization,
while the {\tt SANE\_dipole-jet} model has been applied to parsec-scale active galactic nuclei
jets with polarization and Faraday rotation \citep{2010ApJ...725..750B}.

\subsection{Scattering}

For image plane quantities, we apply a simple circular Gaussian
blurring to simulate the effects of scattering off inhomogeneities in
the ionized interstellar medium. The width of the Gaussian is lowered
by a factor of two to account for the possible (partial) mitigation
technique presented in \citet{2014ApJ...795..134F}.

However, no scattering kernel 
is applied to the visibility plane quantities (see \S\ref{sec::VLBI_Visibility_Plane}), so our results should be compared with data that have been ``deblurred'' using the estimated scattering kernel extrapolated from longer wavelengths \citep[see][]{2014ApJ...795..134F}. 

\subsection{Image Plane Quantities}

We generate synthetic images for each Stokes parameter $\{ I, Q, U, V
\}$ for all models and as a function of time.  $I$, representing intensity, is a 
non-negative quantity, while positive or negative $V$ corresponds to
right and left circular polarization, respectively, following IAU/IEEE
definition of the sign of circular polarization. The linear
polarization intensity is given by $LP=\sqrt{|Q|^2+|U|^2}$, where
linear polarization fraction is given as per unit intensity in
percent. The linear polarization direction, or electric vector position angle (EVPA) is determined by the argument of the complex polarization field: ${\rm EVPA}={\rm arg}(Q+iU)/2$, corresponding
to the angle of the electric polarization EAST of NORTH.  Circular polarization fraction is given as $V$ per unit $I$
in percent. These additional diagnostics are used to assess further
the viability of each model beyond the spectrum of image and
time-averaged quantities.

\subsection{Model Fitting}\label{modelfit}

The free parameters in our radiative transfer model are:
\begin{itemize}
  \item inclination: $i$
  \item heating ratio between electrons and protons: $C_{\rm heat}$
  \item Mass accretion rate normalization: $\dot{M}$
\end{itemize}

We determine the mass accretion rate, inclination ($i=0$ is face-on,
while $i>0$ moves toward NORTH), and heating ratio by fitting the
measured fluxes and image-integrated linear and circular polarization
as in \citet{SPM12}.  Specifically, we fit (unpolarized) fluxes at 7
frequencies from $f=87{\rm GHz}$ to $f=857{\rm GHz}$, 3 linear
polarizations at $f=\{87{\rm GHz},230{\rm GHz},349{\rm GHz}\}$, and 2
circular polarizations at $f=\{230{\rm GHz},349{\rm GHz}\}$ resulting
in 9 degrees of freedom (3 free parameters).

We employ a steepest descent method to minimize $\chi^2$ as in
\cite{SPM12} where $\chi_I^2 = \sum_i (F_{\nu,i}-F^{\rm
  obs}_{\nu,i})^2/\Delta F^2_{\nu,i}$ with $F_{\nu,i}$ are the
computed fluxes, $F^{\rm obs}_{\nu,i}$ are the observed fluxes
(averaged as described and tabulated in \cite{SPM12} and $\Delta
F_{\nu,i}$ are errors, see \cite{SPM12}. We compute the analogous
quantity $\chi^2_P$ for polarization using
$LP@\{87{\rm GHz},230{\rm GHz},349{\rm GHz}\}=\{1.4\pm0.5\%,7.4\pm0.7\%,6.5\pm0.6\%\}$
and $CP@\{230{\rm GHz},349{\rm GHz}\}=\{-1.2\pm0.3\%,-1.5\pm0.3\%\}$. The total
residual of the fits is then $\chi^2=\chi^2_I+\chi^2_P$. In the table
we quote $\chi^2$ and the resulting $\chi^2_I$ (even though the latter
was not separately fitted for).  Note, that this procedure does not
optimize the fit for $230{\rm GHz}$ (the main focus of this work) in any
way. We exclude from the fits any lower frequencies for which non-thermal
particles could be required, with the goal of not biasing our
models at $230{\rm GHz}$. EVPA is not included in the fits just like
prior work using {\tt ASTRORAY} \cite{SPM12,SM13}, because it is influenced
by the uncertainties in the radial extension. These important aspects
will be tackled in future work.

Similar to previous work there are nuisance model parameters, including $\sigma_{T_{\rm e}}$, $T_{\rm e,jet}$, $\sigma_{\rho}$, which we
only vary as part of a specific model and do not minimize $\chi^2$
over.  Some prior works directly include image size in the fitting
procedure as additional observational data \citep{2015ApJ...799....1C},
but we do not.  In principle, we could tune our fitting procedure to primarily fit the 230GHz emission (the main focus of this work) instead of just by the error of each independent observation.  This was useful for the {\tt SANE\_dipole-jet} model, for which we slightly adjusted the original fit parameters in order to get better agreement with linear polarization at 230GHz. We plan further development of fitting procedures.

\subsection{VLBI Visibility Plane Quantities}
\label{sec::VLBI_Visibility_Plane}

After fixing the free parameters as determined from the fits to
image-integrated flux, linear polarization, and circular polarization,
we generate the corresponding Fourier transformed visibility data $\{
\tilde{I}, \tilde{Q}, \tilde{U}, \tilde{V} \}$.  We focus on
measurements of fractional linear and circular polarization in the
visibility domain: $\breve{m} \equiv (\tilde{Q}+i\tilde{U})/\tilde{I}$
and $\breve{v} \equiv \tilde{V}/\tilde{I}$, respectively. Note that each of these quantities is complex. We also
compute the visibility domain ${\rm EVPA}={\rm
  arg}((\tilde{Q}+i\tilde{U})/\tilde{I})/2$. The amplitudes and phases
of these visibility domain ratios are immune to a wide range of
station-based calibration errors and uncertainties and thus provide
excellent VLBI observables \citep{RWB94}. Fractional polarization in
the visibility domain is also insensitive to the ensemble-average
``blurring'' effect of scattering \citep{Johnson2014}, which increases the thermal noise in measurements but does not bias them. Because we do not include thermal noise in the current comparisons with observations, when applicable (e.g., for $|\tilde{I}|$), we show quantities that would be obtained after ``deblurring'' \citep{2015ApJ...799....1C}.  
In practice, long baselines may show slight additional variations from ``refractive
substructure,'' which will vary stochastically with a timescale of
${\sim}1~{\rm day}$ \citep{JohnsonGwinn2015}.

\section{Results}
\label{sec:results}

In this section, we present the results of our polarized radiative
transfer calculations.  We describe our findings for each GRMHD model,
quantify the level of agreement with several observational constraints
and point out remaining issues, and elaborate on the different trends
seen between different models.

\subsection{Results of Zero-Baseline Model Fitting}

\begin{figure}

  \centering
  \includegraphics[width=0.49\textwidth]{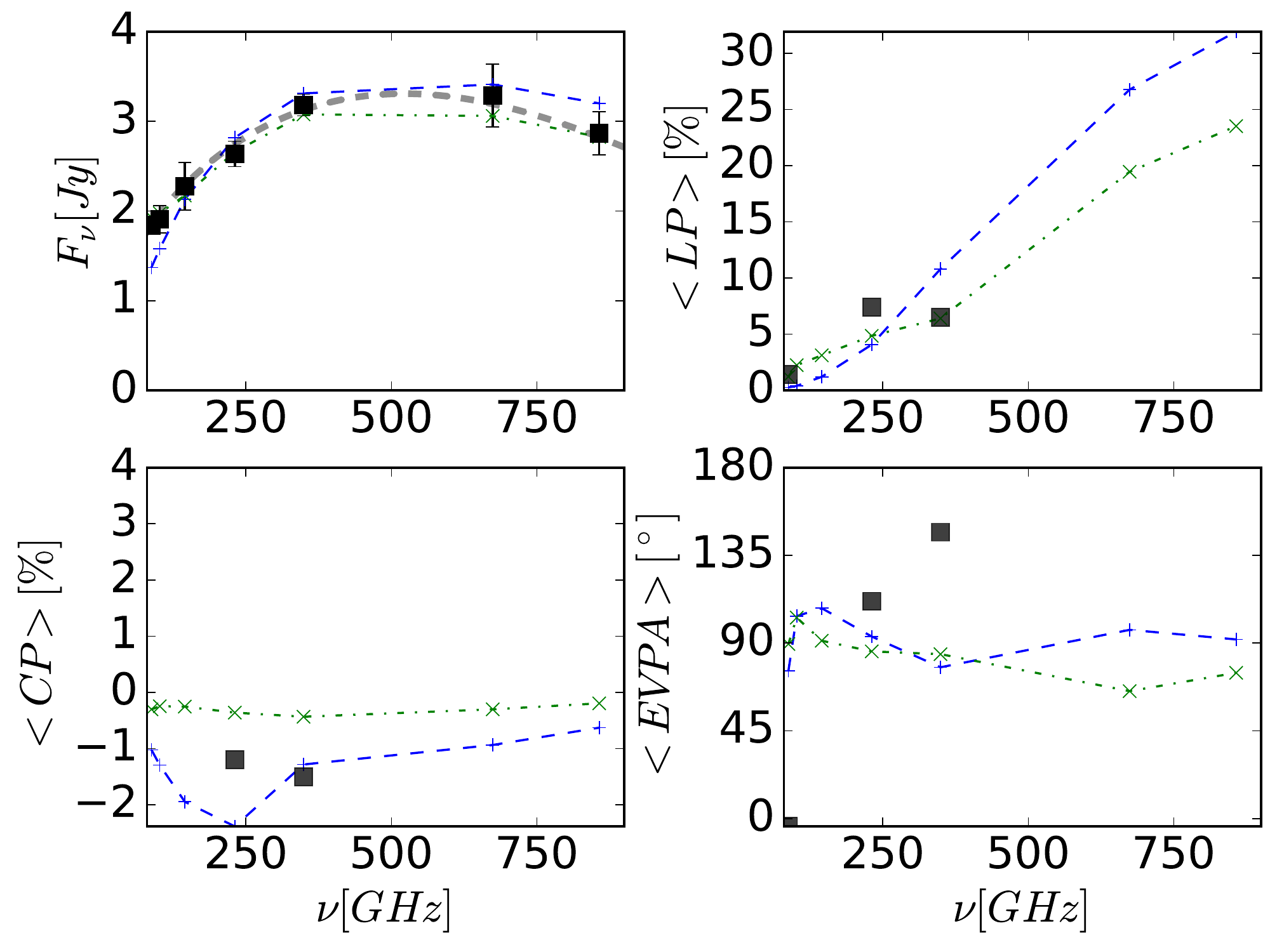}
  \caption{Spectral energy distribution (SED, $F_\nu$ in Jy),
    fractional linear polarization $<\!\!\!LP\!\!\!>$, fractional circular
    polarization $<\!\!\!CP\!\!\!>$, and electric vector position
    angle $<\!\!\!EVPA\!\!\!>$ (all image-integrated) vs.\ frequency in
    GHz for model {\tt MAD\_thick-disk} (green dotted line and cross) and {\tt
      SANE\_quadrupole-disk} (blue dashed line and plus).  Positive or negative
    $<\!\!\!CP\!\!\!>$ corresponds to right and left circular
    polarization, respectively.  These model data are compared to the
    observed data (black squares) and a fit through observed data
    (grey dashed line). See Tab.~\ref{tab:models} for results from
    fitting.  More work is needed on our model that extends the
    simulation data to large radii in order to fit the
    frequency-dependence of CP and EVPA. The frequency dependence of
    image-integrated flux and linear polarization measurements are in
    good agreement with our models, while CP and EVPA are roughly the
    right magnitude (except for one observation of low EVPA, but we do not fit for EVPA).\label{fig:SED}}
\end{figure}

\begin{table}[t]
  \caption{List of models considered including parameters: $i$
    inclination ($i=0$: face-on), $C_{\rm heat}$ (related to the
    electron-to-proton heating ratio for the disk), $\dot{M}$
    rest-mass accretion rate, and $\chi^2/{\rm dof}$ (fitted to I,LP, and CP)
    and $\chi^2_I/{\rm dof}$ (fitted only to unpolarized I) (dof: degrees of
    freedom) quantifying the goodness of fit.
    \label{tab:models}}


  \begin{tabular}{ccccccc} \hline\hline

Model          & $i$ & $C_{\rm heat}$ & $\dot{M}$ [$\frac{M_\odot}{{\rm yr}}$] & $\frac{\chi^2}{{\rm dof}}$ & $\frac{\chi^2_I}{{\rm dof}}$ \\ \hline 

{\tt MAD\_thick-disk}& $99^\circ$     & $0.025$      & $5.5 \times 10^{-9}$    & $4$                        & $0.5$ \\ 
{\tt MAD\_thick-jet} & $140^\circ$    & $0.05$       & $5.4 \times 10^{-9}$    & $5$                        & $8$   \\ 
{\tt SANE\_quadrupole-disk} & $98^\circ$ & $0.47$    & $4.0 \times 10^{-8}$    & $7$                        & $7$   \\ 
{\tt SANE\_dipole-jet} & $126^\circ$  & $0.55$       & $2.6 \times 10^{-8}$    & $13$                       & $12$  \\ 

\hline\hline 

  \end{tabular}
\end{table}

We first consider the frequency-dependent zero baseline observations
as compared to our models. Fig.~\ref{fig:SED} shows our model fits
using the default electron temperature and mass-loading prescriptions
using the fitting procedure discussed in section~\ref{modelfit} with
an assumed BH mass of $M=4.3\times 10^6\msun$.  We time-averaged
spectra over an interval $4000r_g/c-5600r_g/c$ for {\tt
  SANE\_quadrupole-disk}, $2500r_g/c-3300r_g/c$ for {\tt
  SANE\_dipole-jet}, and $20212r_g/c-22212r_g/c$ for our MAD models.

The fits to $I$, $LP$ percentage, and $CP$ percentage give an
intensity vs.\ frequency with a relatively good fit for any
model. Linear polarization adds an additional constraint on the
inclination angle due to how different inclination angles lead to
varying amounts of cancellation in polarized emission from different
parts of the disk or jet.  We do not fit EVPA, which is not fit well
vs.\ frequency, although their values are roughly correct.  We do not
focus on fitting EVPA because it is controlled by the flow at larger
radii than the simulations reach a steady-state out to.  In principle,
fitting to intensity alone will produce a different fit than our
fitting to $I$, $LP$, and $CP$, which may affect prior intensity-only
fits, but we did not consider intensity-only fits in this paper.  For
{\tt SANE\_dipole-jet}, we slightly adjusted the original fit to
obtain better agreement for the linear polarization fraction at
$230$GHz (The original overall lowest $\chi^2/{\rm dof}$ was for
$i=124^\circ$, $C_{\rm heat}=0.47$ and $\dot{M}=3.6\times
10^{-8}\msun/{\rm yr}$, extremely similar to our slightly tuned fit.).

The fits prefer different inclination angles for the jet and disk
cases. For the disk cases, the inclination must be close to edge-on
(${\approx}10^\circ$ above edge-on), whereas for the {\tt jet} models it must
be much higher (${\approx}45^\circ$ below edge-on).

The {\tt MAD\_thick-disk} model did not need an isothermal jet or non-thermal
particles to account for the low frequency flux seen in
Fig.~\ref{fig:SED}.  The MAD model constrains a hot funnel-wall visible
as high emissivity in
Figs.~(\ref{fig:Tejnu},\ref{fig:Tejnu-magn40},\ref{fig:Tejnu-te100})
for any electron temperature prescription.  At much larger radii, the
prescription for the electron temperature still ensures that enough
material is close to isothermal, which is sufficient to fit the
low-frequency synchrotron self-absorption part of the spectrum.

\begin{figure*}[t]
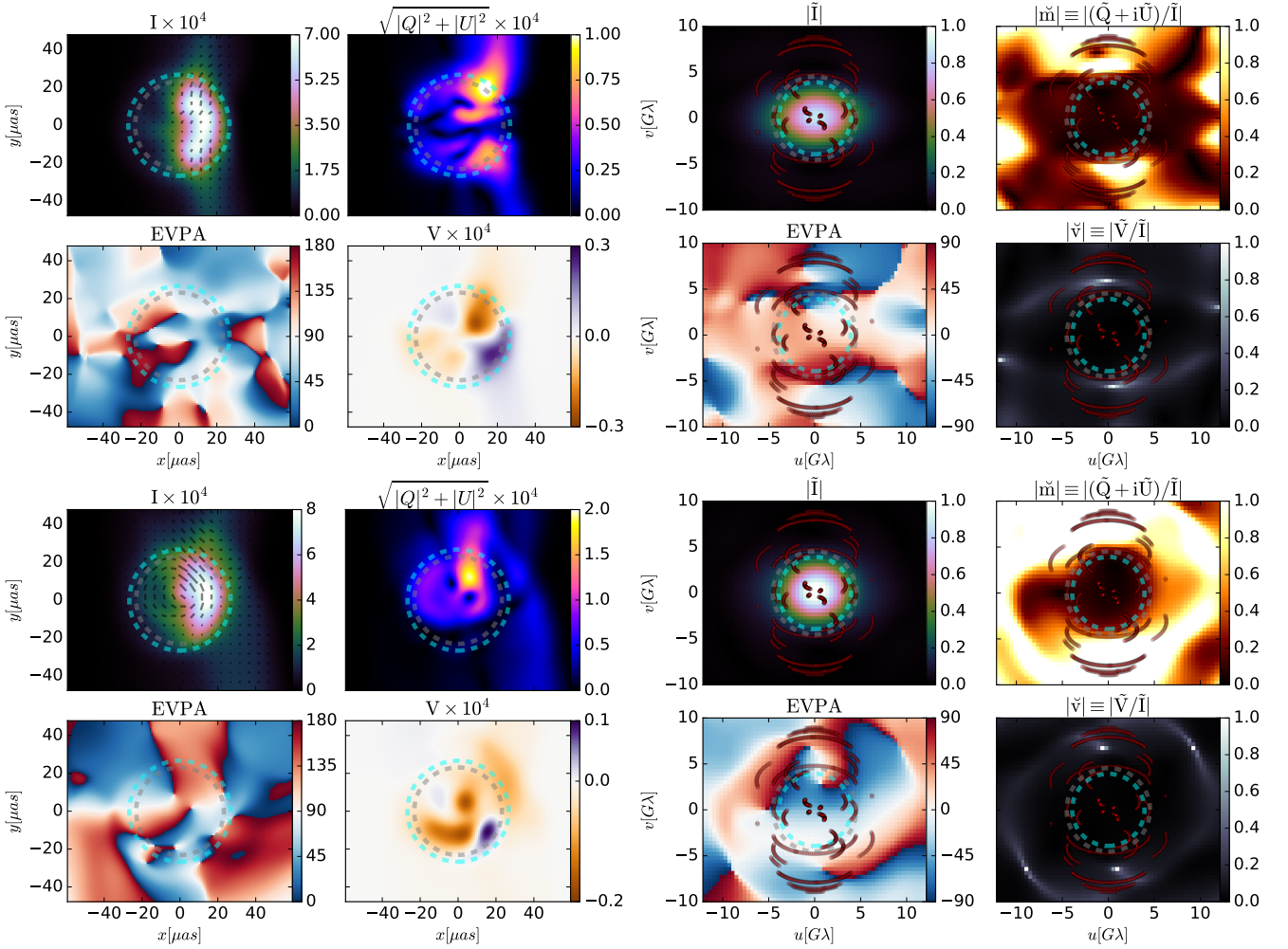

  \centering

  \includegraphics[width=0.49\textwidth]{{{thickdisk7-disk_shotimag93.75th173f230fn5550case70816924_151_I-LP-EVPA-CP_xy}}}
  \includegraphics[width=0.49\textwidth]{{{thickdisk7-disk_shotimag93.75th173f230fn5550case70816924_151_miniversion_4panel_uv}}} \\

  \includegraphics[width=0.49\textwidth]{{{thickdisk7-jet_shotimag93.75th245f230fn5550case78114_151_I-LP-EVPA-CP_xy}}}
  \includegraphics[width=0.49\textwidth]{{{thickdisk7-jet_shotimag93.75th245f230fn5550case78114_151_miniversion_4panel_uv}}}

  \caption{All Stokes parameters and several derived quantities in the
    image plane (intensity $I$ in Jy/pixel with overlaid polarization ticks that scale in length with $LP=\sqrt{|Q|^2+|U|^2}$ and are oriented by the EVPA, linear polarization
    intensity $\sqrt{|Q^2|+|U|^2}$ in Jy/pixel, EVPA showing angle of
    electric field polarization, and $|V|$ showing magnitude of
    circular polarization) and visibility plane (Fourier intensity
    $|\tilde{I}|$, fractional linear polarization $|\breve{m}|$, EVPA
    showing polarization angle, and $|\breve{v}|$
    showing fractional circular polarization) at $f=230{\rm GHz}$ for
    model {\tt MAD\_thick-disk} (upper panel) and {\tt MAD\_thick-jet} (lower panel)
    from snapshots in time.  All such plots have $151\times 151$ pixels for quantities given in per pixel. Dashed circles indicate the expected
    shadow size for a back-lit BH that has no spin (cyan, $\approx
    10.4r_g$) or has maximal spin (grey, $\approx 9r_g$, roughly averaged over viewing plane
    angle and inclination with respect to the BH spin axis, see \citealt{2004ApJ...611..996T}). Tracks
    are shown in the $uv$ plane that will be probed by the EHT in 2017 (see \citealt{2015ApJ...813..132J} for site details).  These
    figures give a general impression of the relationship between
    image and visibility domain. The visibility plots are further
    useful to judge the importance of baseline coverage and source
    orientation.  \label{fig:16panels-thickdisk7}}

\end{figure*}

\subsection{Image and Visibility Plane}

\begin{figure*}[t]
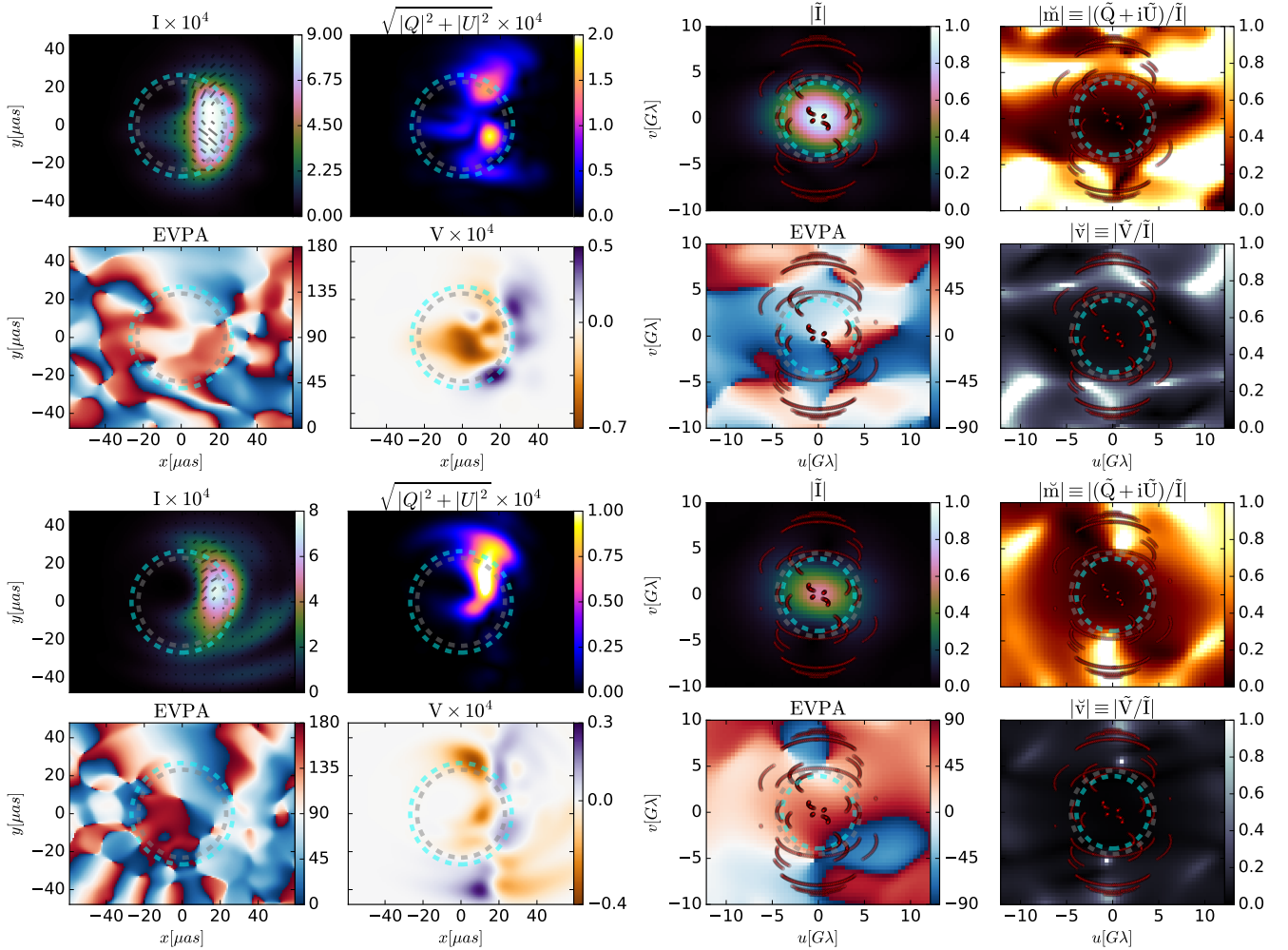

  \centering

  \includegraphics[width=0.49\textwidth]{{{shotimag93.75th170f230fn2140case4442140_151_I-LP-EVPA-CP_xy}}}
  \includegraphics[width=0.49\textwidth]{{{shotimag93.75th170f230fn2140case4442140_151_miniversion_4panel_uv}}} \\

  \includegraphics[width=0.49\textwidth]{{{shotimag92th215f230fn1600case2511_151_I-LP-EVPA-CP_xy}}}
  \includegraphics[width=0.49\textwidth]{{{shotimag92th215f230fn1600case2511_151_miniversion_4panel_uv}}}

  \caption{Same as Fig.~\ref{fig:16panels-thickdisk7}, but for models
    {\tt SANE\_quadrupole-disk} (upper panels) and {\tt
      SANE\_dipole-jet} (lower panels).  The BH shadow in both
    intensity and polarization are considerably different than the MAD
    models, suggesting detectability of the BH shadow feature may have
    to account for such model variations.
    \label{fig:16panels-quadrupole}}

\end{figure*}

For the default models that we fit zero baseline observations to in
the previous section, we next consider what the full image and
visibility planes reveal.  Figs.~\ref{fig:16panels-thickdisk7} and
\ref{fig:16panels-quadrupole} show all Stokes parameters for our
default set of four models at the same time show for
Figs.~(\ref{fig:sims},\ref{fig:Tejnu},\ref{fig:Tejnu-magn40},\ref{fig:Tejnu-te100})
in section~\ref{sec:models}.  No time-averaging is performed.

In these and similar plots, the image plane has the BH spin axis
pointing along the vertical (NORTH) direction, where the
left-direction is WEST and right-direction is EAST (as if seeing
projected image on surface of Earth and the EHT), while sometimes the
opposite choice is made for EAST-WEST as if one is viewing the source.
The image plane linear ($Q$ and $U$) and circular polarization ($V$)
are not shown as fractional values, so that the primary polarized
emitting regions can be identified.  Visibility plane linear
polarization ($\breve{m}$) and circular polarization ($\breve{v}$) are
shown as fractions, consistent with what the EHT can most
robustly measure.

Across all these models, polarization persists to longer baselines
(smaller scale structure) than total intensity.  In temporal
evolutions of such plots, stronger variability is seen generally on
longer baselines as expected. These findings suggest that the emission
fine-scale structure is best constrained by high resolution (in time
and space) polarization studies, and they highlight the growing importance
of polarization as the EHT expands to longer baselines
\citep{2009ApJ...706.1353F,2013AAS...22114304F,2015MNRAS.446.1973R}.

In all our models, the image plane polarization is highest outside
from where the image plane intensity is highest. Regions with highest
total intensity have somewhat lower (but still significant)
polarization. Therefore, the visibility plane intensity will show the
different scale of intensity and polarization as a high visibility
fractional polarization due to regions with relatively lower
intensity.  Hence, accurate high fractional polarization measurements
can only be achieved with sensitive measurements to detect and
characterize points with little correlated total-intensity flux
$\tilde{I}$ \citep{JohnsonEtAl2015Science}.

\subsection{Observational Constraint on Magnetic Field Structure}

In \citet{JohnsonEtAl2015Science}, it is shown how EHT data constrain
the degree of order in the magnetic field in the emission region by
comparing the relative amplitudes of two dimensionless quantities: the visibility plane fractional polarization with the
visibility domain normalized intensity.  The price one has to pay by using
$\breve{m}$ is that the interpretation is not as straightforward as
with the image plane fractional polarization. The quantity $\breve{m}
\equiv (\tilde{Q}+i\tilde{U})/\tilde{I}$ is a measure of linear
polarization in the visibility (Fourier) domain, but its inverse
Fourier transform is {\it not} the fractional linear polarization in
the image domain $m_{LP} \equiv (Q+iU)/I$.  However, direct
comparisons can still be made between observations and models
\citep{JohnsonEtAl2015Science}.

\begin{figure}

  \centering
  \includegraphics[width=0.47\textwidth]{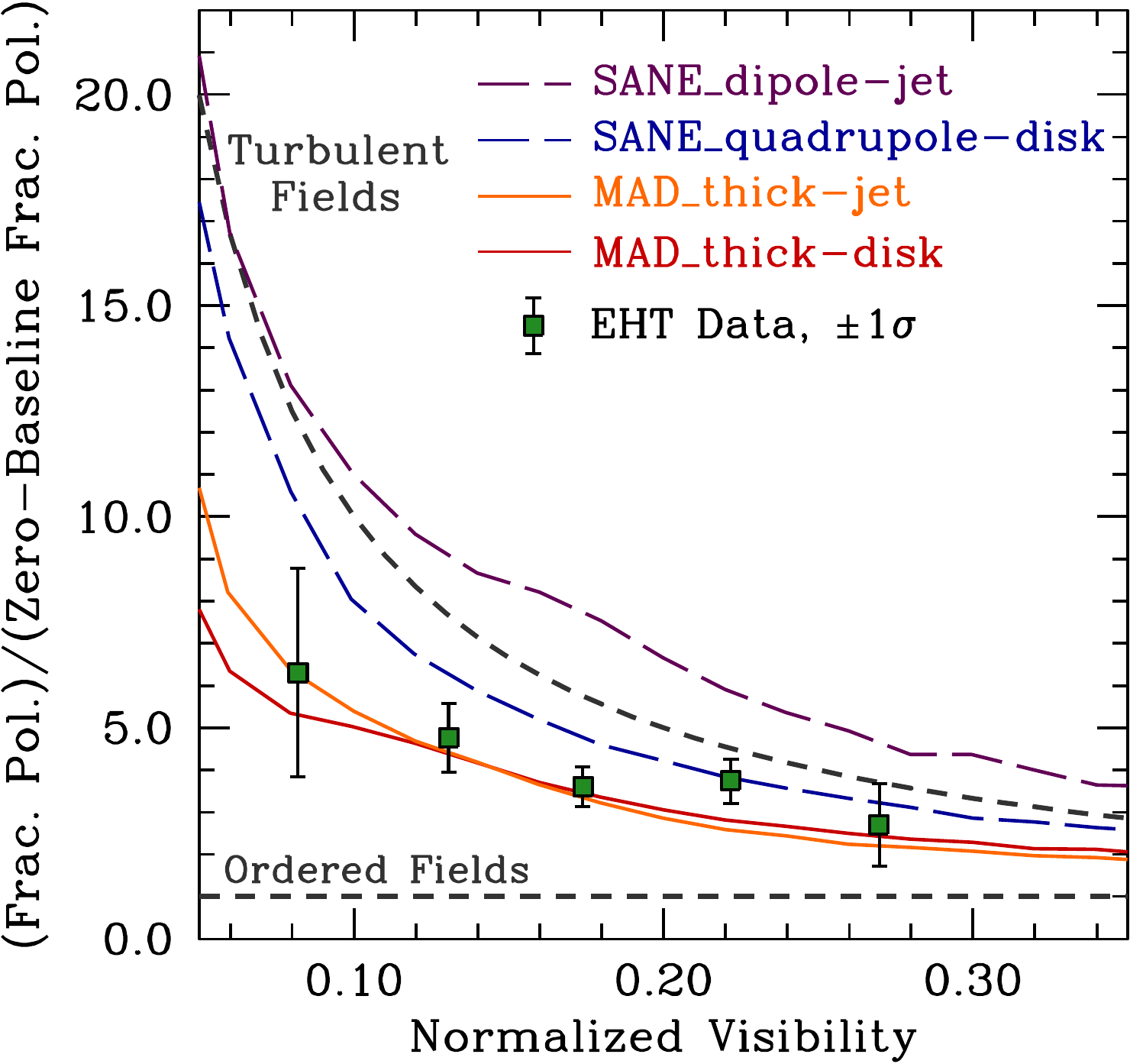}
  \caption{Fractional linear polarization $|\breve{m}|$ (as measured in
    the visibility plane) per unit zero baseline value vs.\ the
    normalized visibility $|\tilde{I}/\tilde{I}_0|$ for all
    models. Observational data from the EHT are shown as green
    squares.  Asymptotic results for perfectly ordered fields are shown as a horizontal
    dashed black line and for completely turbulent fields as a
    curved dashed black line. SANE models are shown as colored dashed
    lines and MAD models as colored solid lines. Our MAD models are
    preferred over our SANE models. {\tt SANE\_quadrupole-disk} with a
    large-scale quadrupolar field in the disk is only marginally
    inconsistent with observations. \label{fig:I-vs-mbreve} }

\end{figure}

Following \citet{JohnsonEtAl2015Science}, Fig.~\ref{fig:I-vs-mbreve}
shows $|\breve{m}|$ as a function of normalized total intensity
$|\tilde{I}/\tilde{I}_0|$ for all of our default models and the asymptotic cases described in \citet{JohnsonEtAl2015Science}, where
$\tilde{I}_0$ is the correlated flux density on the zero-baseline.
For uniform polarization across the image one expects
$|\breve{m}(|\tilde{I}/\tilde{I}_0|)|={\rm const}{.}$, whereas for maximally
disordered fields 
$|\breve{m}(|\tilde{I}/\tilde{I}_0|)| \propto \tilde{I}_0/\tilde{I}$ on average. Our simulations confirm that EHT
measurements of $\breve{m}$ indicate the magnetic field's degree of
order vs.\ disorder, preferring both MAD models over both SANE models.

\begin{figure}
  \centering
  \includegraphics[width=0.47\textwidth]{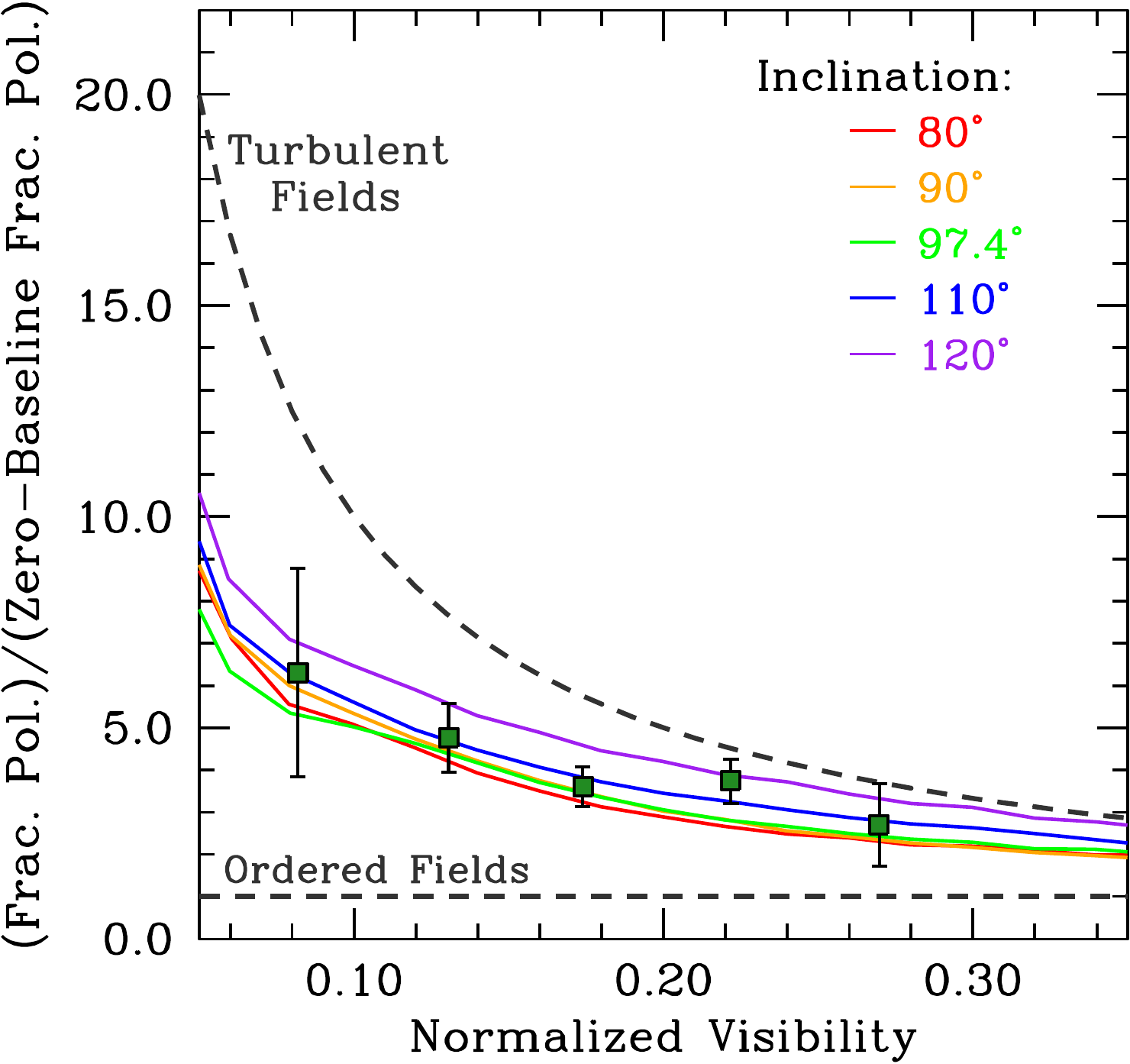}
  \caption{Same as Fig.~\ref{fig:I-vs-mbreve}, but showing five
    different viewing inclinations of the {\tt MAD\_thick-disk} model while
    holding all other parameters fixed. The middle value
    ($i=97.4^\circ$) is the fiducial case used throughout the
    paper. Unlike other observables, such as the total flux and image
    size, $|\breve{m}|$ vs.\ $|\tilde{I}/\tilde{I}_0|$ is relatively insensitive to the
    changing inclination and robustly quantifies the order of the
    magnetic field throughout the emission
    region. \label{fig:I-vs-mbreve_VaryingInclination} }

\end{figure}

\begin{figure}

  \centering
  \includegraphics[width=0.47\textwidth]{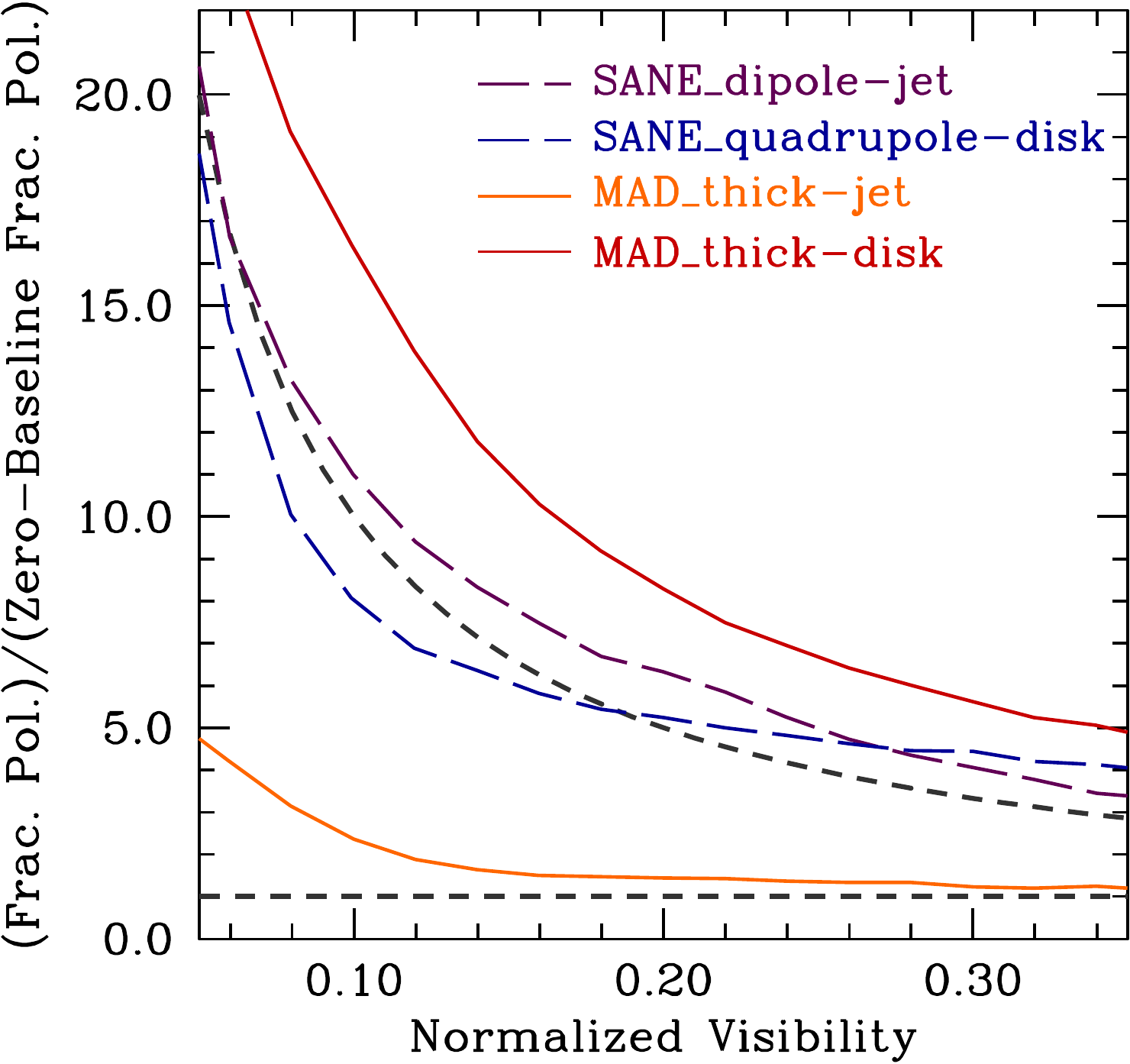}
  \caption{Similar to Fig.~\ref{fig:I-vs-mbreve}, but showing
    fractional circular polarization $|\breve{v}|$ per unit zero
    baseline value (as measured in the visibility plane)
    vs.\ normalized visibility $|\tilde{I}/\tilde{I}_0|$ for all models.
    No EHT data yet exists, but apparently jet models tend to have
    higher $|\breve{v}|$ than disk models, so $\breve{v}$ might help
    differentiate between disk and jet models and provide a unique
    constraint compared to linear polarization via
    $\breve{m}$. \label{fig:I-vs-vbreve} }

\end{figure}

Unlike other image and polarization characteristics, these conclusions are relatively insensitive to the viewing inclination. Fig.~\ref{fig:I-vs-mbreve_VaryingInclination} shows the average of $|\breve{m}(|\tilde{I}/\tilde{I}_0|)|$ for a fixed fiducial 
model ({\tt MAD\_disk}) at varying inclination. The results are similar over a range of $40^\circ$, showing that this metric of field order is fairly robust.  Figure~\ref{fig:I-vs-vbreve} shows the varying models
but for $|\breve{v}|$, the visibility plane circular polarization
fraction (note that EHT data for $\breve{v}$ are not yet available). Here we see that the MAD models are not clustered
together, and instead the jet models tend to have high $|\breve{v}|$ and
the disk models tend to have lower $|\breve{v}|$.  More extensive comparisons are necessary to conclude whether $|\breve{v}|$ can
help distinguish disk vs.\ jet models in general.

\subsection{Variability}

\begin{figure*}[ht]

  \centering
  \includegraphics[width=0.49\textwidth]{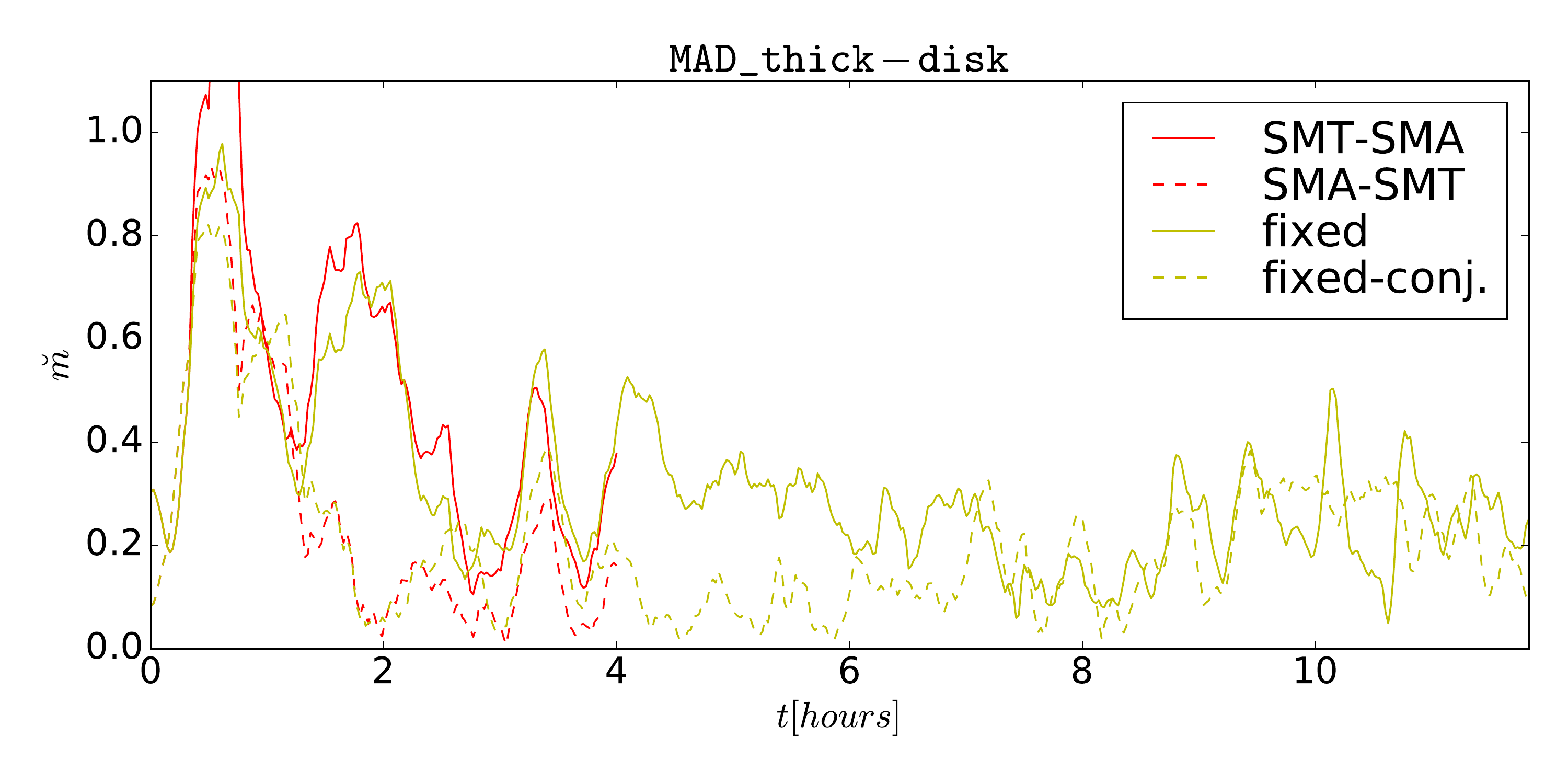} 
  \includegraphics[width=0.49\textwidth]{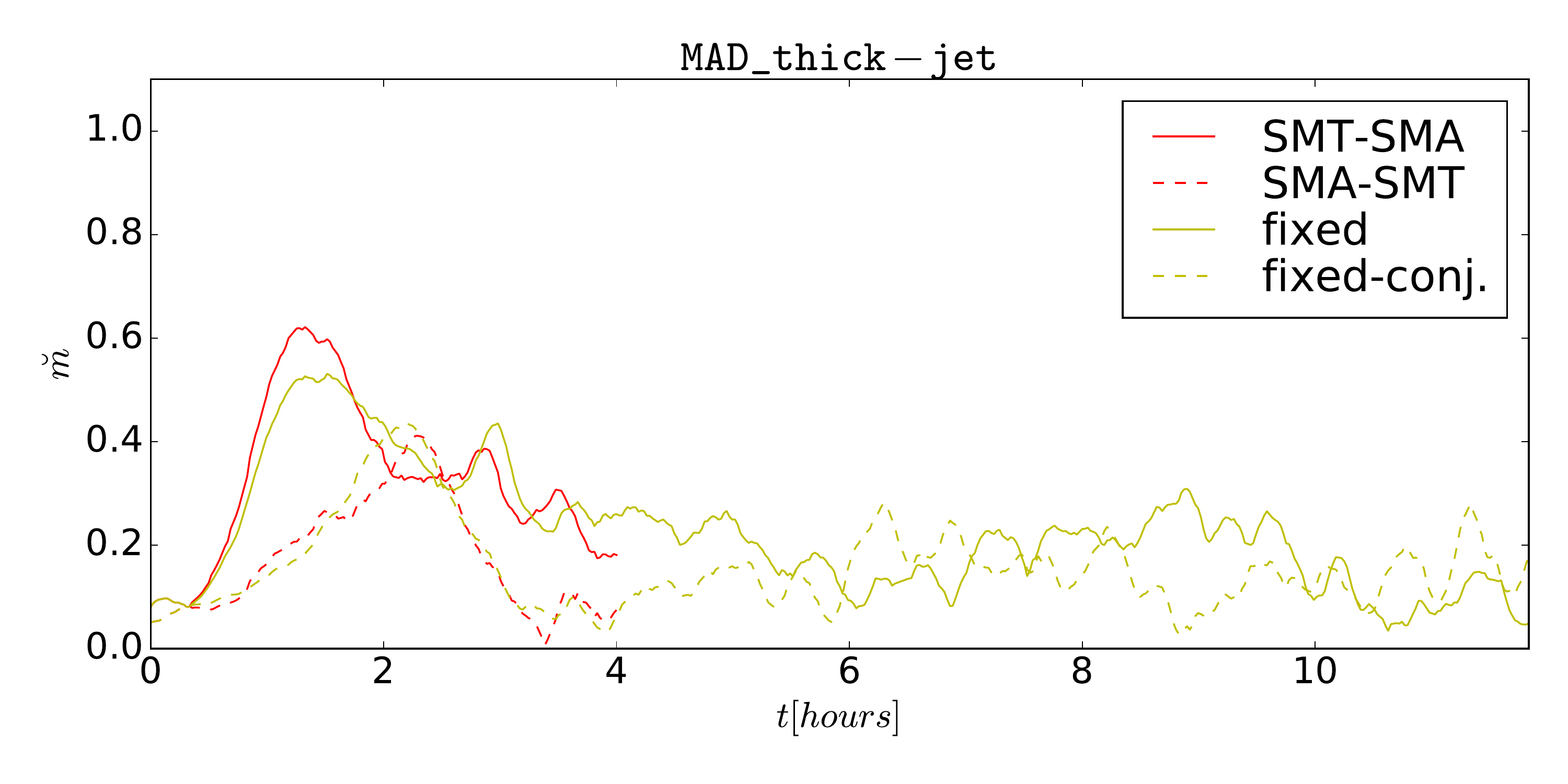}  
  \includegraphics[width=0.49\textwidth]{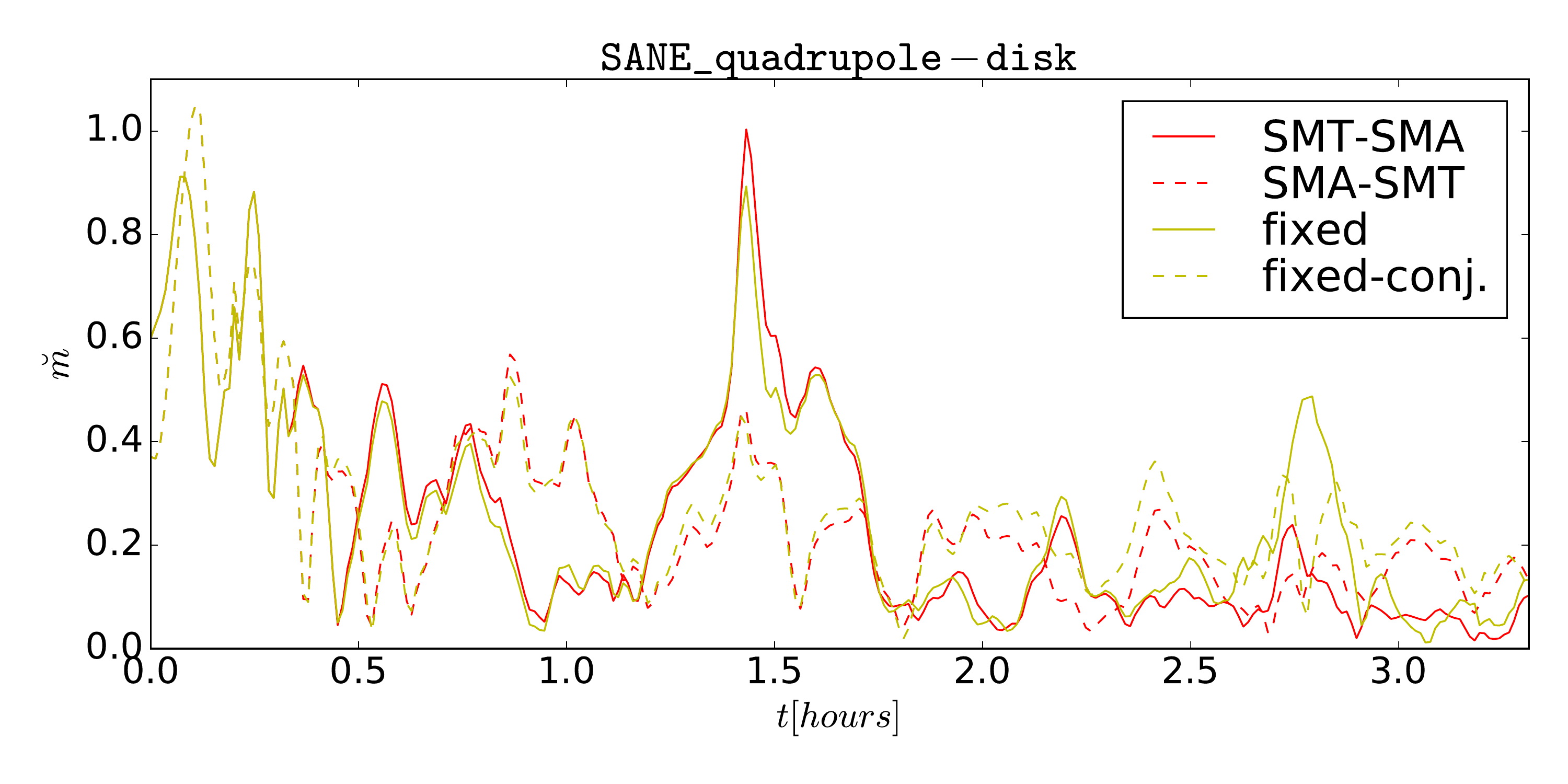}
  \includegraphics[width=0.49\textwidth]{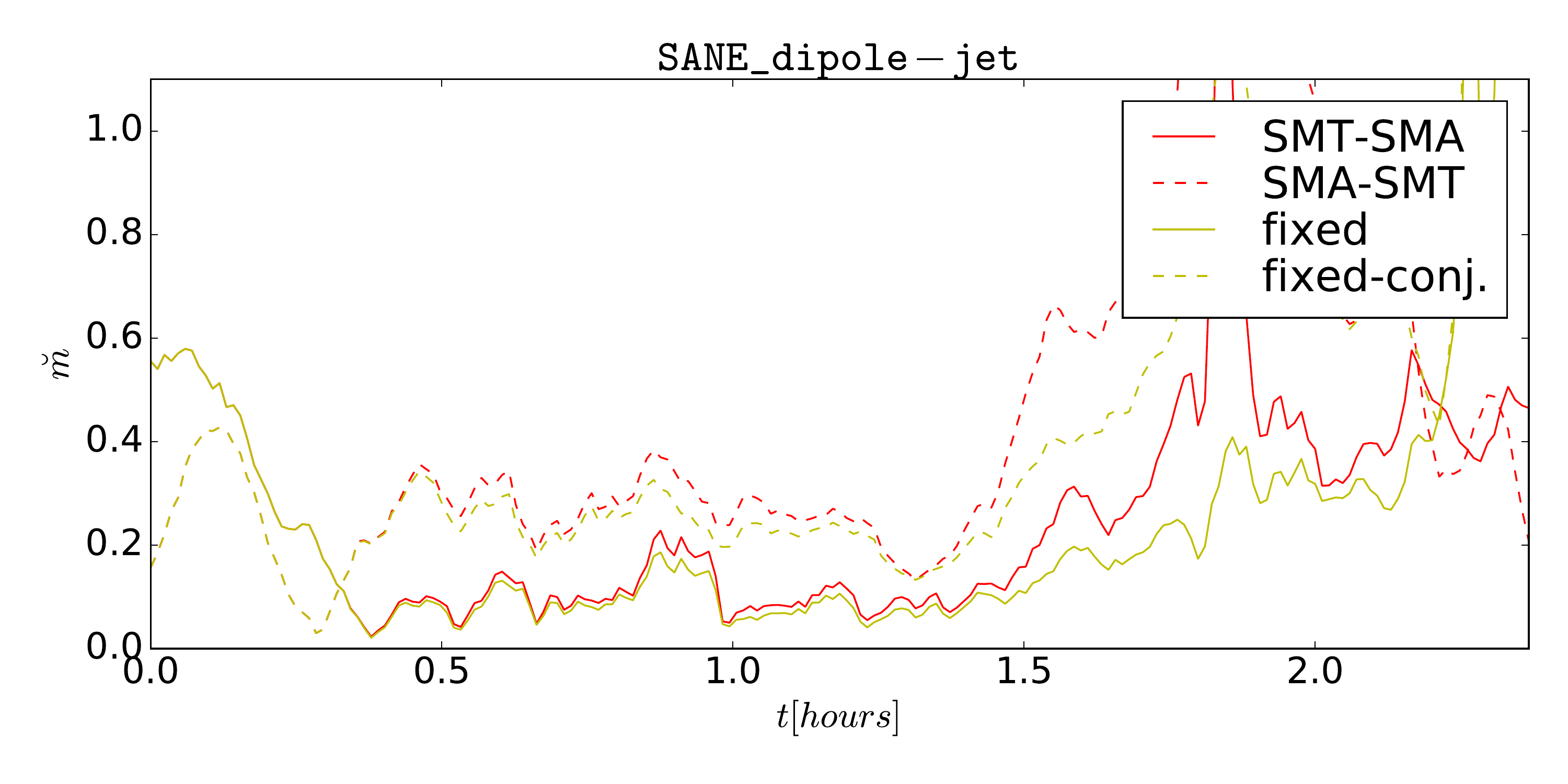} 
  \caption{Value of $|\breve{m}|\equiv
    |\tilde{Q}+i\tilde{U}|/|\tilde{I}|$ vs.\ time along the SMA-SMT
    baseline (red/dark) and at fixed (stationary) baselines
    (yellow/light) and as a function of time $t$ for models {\tt
      MAD\_thick-disk} (upper left panel), {\tt MAD\_thick-jet} (upper
    right panel), {\tt SANE\_quadrupole-disk} (lower left panel), and
    {\tt SANE\_dipole-jet} (lower right panel).  All data sets are
    variable, but only the MAD models have QPOs (less visible at the orientation chosen for the {\tt MAD\_thick-jet} model)
    \citep{2012MNRAS.423.3083M,2013ApJ...774L..22S}. Note that
    differences between conjugate baselines arise from polarization
    alone and vary greatly between models.  Both SANE and MAD models
    give high enough $\breve{m}$ amplitude and asymmetry for
    well-chosen baseline orientations, but the MAD models are
    consistent for a broader set of baseline orientations.
    \label{fig:mbreve-vs-t}}

\end{figure*}

In this section, we consider how non-zero baselines at $230$GHz with
linear polarization provide additional constraints on the models
beyond zero baseline frequency-dependent observations.  The simplest
EHT observation is to only consider a single baseline consisting of a
specific antenna pair at various points in time. Each physical baseline has two directions but 
provides a single measurement in intensity and fractional circular
polarization ($\tilde{I}(u,v) = \tilde{I}^\ast(-u,-v)$ and
$\breve{v}(u,v) = \breve{v}^\ast(-u,-v)$) because their corresponding
images are real. However, the baseline provides two independent linear polarization measurements $\breve{m}(\pm u, \pm v)$ because the linear
polarization image $Q+iU$ is complex.

Fig.~\ref{fig:mbreve-vs-t} shows light curves for $\breve{m}$, which
take into account the time dependence of the emissivity from the GRMHD
simulation and the evolution of the flow during radiative transfer (no
fast-light approximation). 
The MAD models show data from $20212r_g/c-22212r_g/c$, {\tt
SANE\_quadrupole-disk} shows $4000r_g/c-4560r_g/c$, and {\tt
SANE\_dipole-jet} $2800r_g/c-3200r_g/c$.
We consider both fixed and changing
baseline orientations (due to the rotation of the Earth) computed for an EHT campaign, where the fixed
case just chooses the starting point from the case with changing baseline
positions.  As a reference point, the EHT data show up to
$|\breve{m}|\sim 70\%$ on one baseline point $\{u, v\}$ while at the same time showing 
$|\breve{m}|\sim 30\%$ for the conjugate baseline point $\{-u, -v\}$.  For angles EAST of
NORTH, model {\tt MAD\_thick-disk} used $0^\circ$, {\tt MAD\_thick-jet} used
$+45^\circ$, {\tt SANE\_quadrupole-disk} used $-45^\circ$, and {\tt
  SANE\_dipole-jet} used $-23^\circ$ for the relative baseline
orientation.

All models are highly dynamic with a tendency for larger variation at
larger baseline lengths. Both the amplitude of variations in 
$|\breve{m}|$ and the differences between opposite
baselines are more consistent with observations for the MAD models
than for the SANE models.  The MRI type disk in model {\tt
  SANE\_quadrupole} yields less variability and smaller
$\breve{m}$ even on these baselines that were optimally chosen to give
agreement with EHT observations, but all models can potentially
at some moments in time reproduce the amplitude and asymmetry of $\breve{m}$ seen in EHT data.
As shown in \citet{SM13}, the MAD models have quasi-periodic
oscillations (QPOs) clearly apparent in linear polarization, which may
also help distinguish between MAD and SANE models.  Longer and more
frequent EHT observations could provide further insight by using more
rigorous statistical comparisons.

Our synthetic data show dynamical activity of the source size and
correlated flux. A local minimum in the correlated flux density
observed on day 80 in the 2013 EHT campaign could be due to such
natural variability in the underlying accretion flow.  The level of
variability in all models may be sufficient to produce dips in the
correlated flux as observed on some days. Comparing or finding
agreement in such temporal features is an interesting and important
avenue which we will pursue in more detail in the future.

The absolute position angle orientation of the simulated images is a
priori unknown.  As shown in the prior figures, $\breve{m}$ is more
clearly anisotropic in the visibility plane than $\tilde{I}$ (that is
fairly Gaussian), so polarization on sufficiently long baselines is
more sensitive to the absolute position angle orientation.  The extra
information from polarization can help constrain models that have to
match both the magnitude and asymmetry in $\breve{m}$ vs. time for
both conjugate points in a baseline.

\begin{figure*}[ht]

  \centering
  \includegraphics[width=0.49\textwidth]{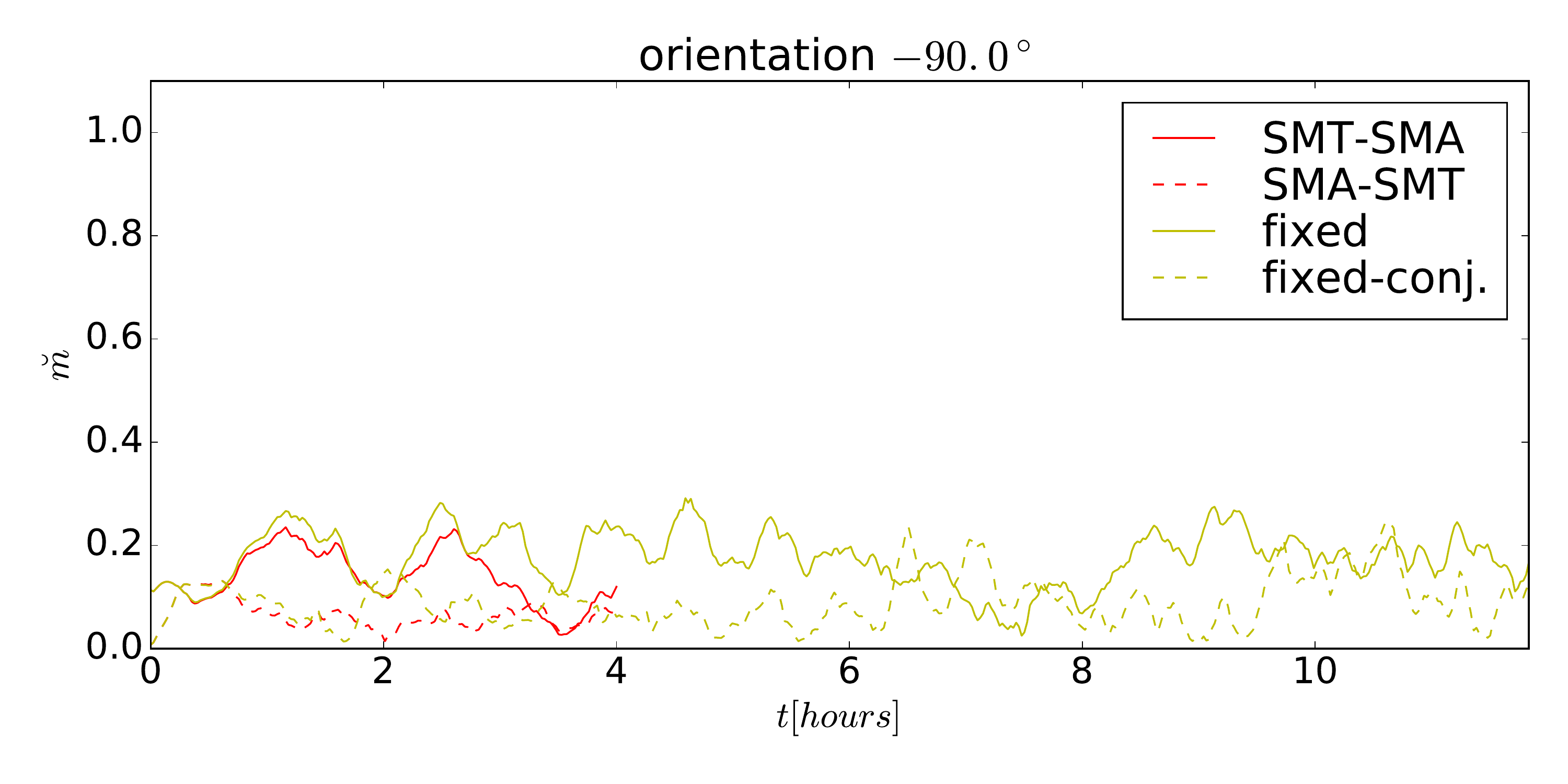}     
  \includegraphics[width=0.49\textwidth]{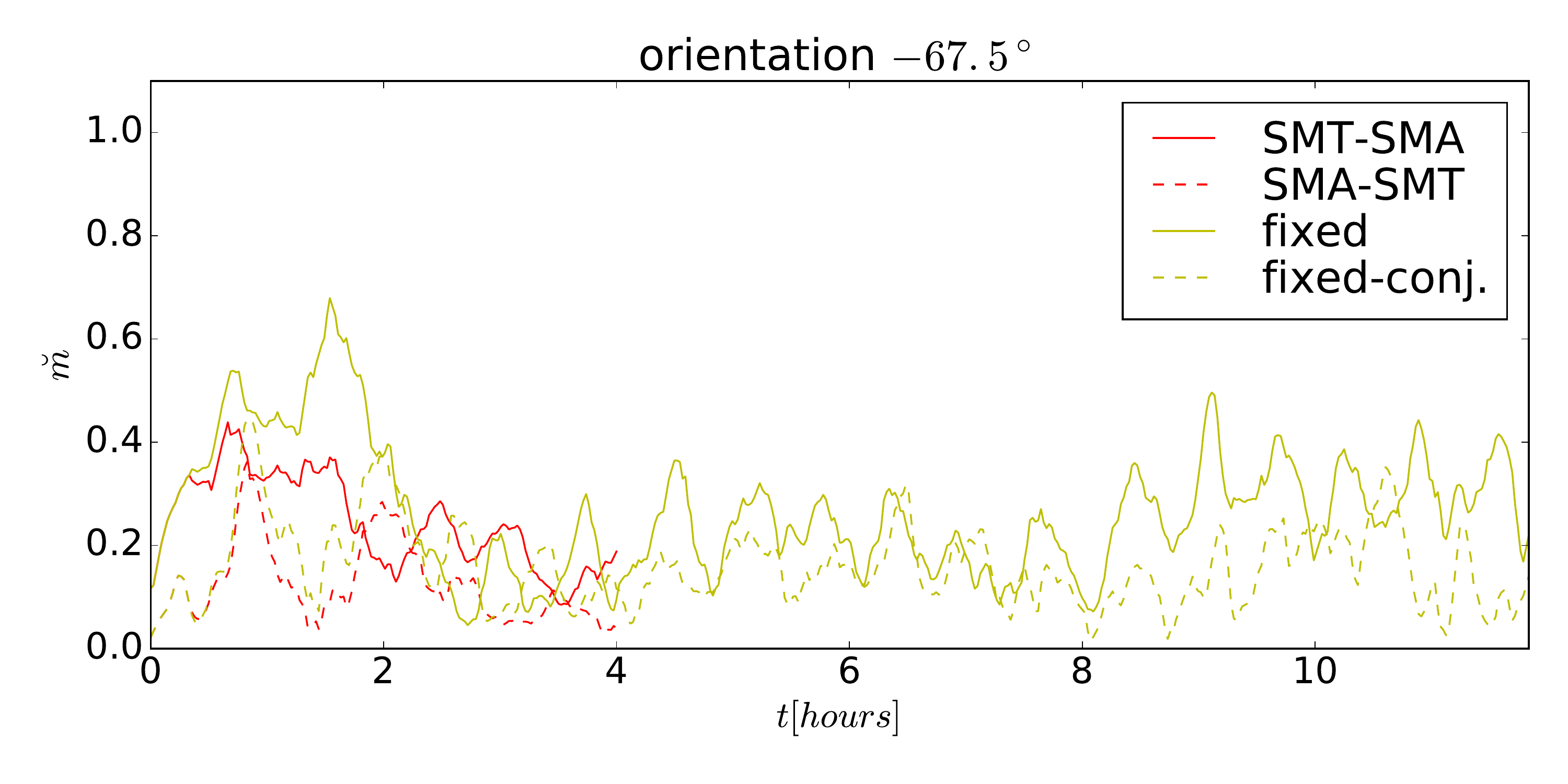} \\  
  \includegraphics[width=0.49\textwidth]{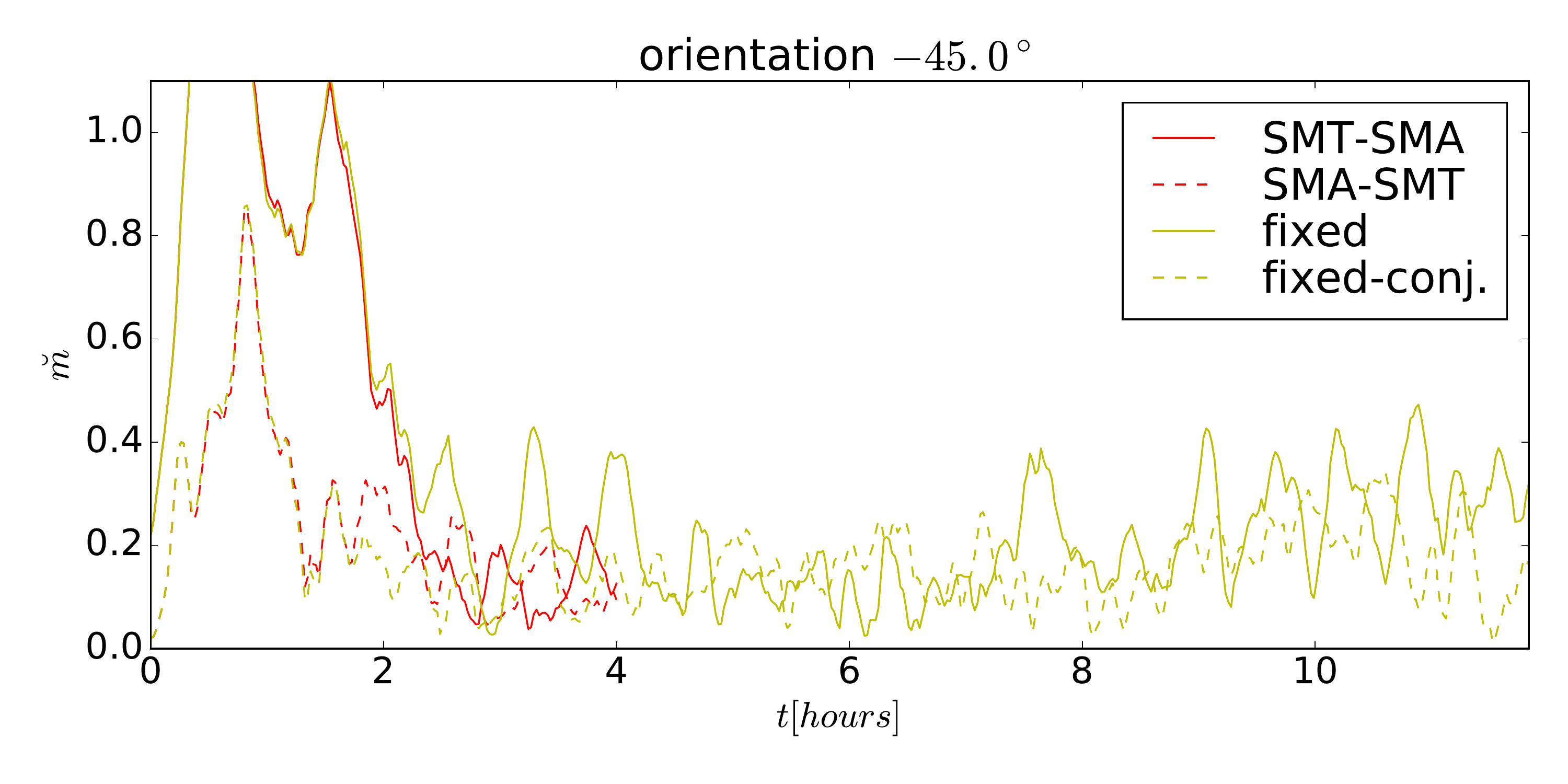}     
  \includegraphics[width=0.49\textwidth]{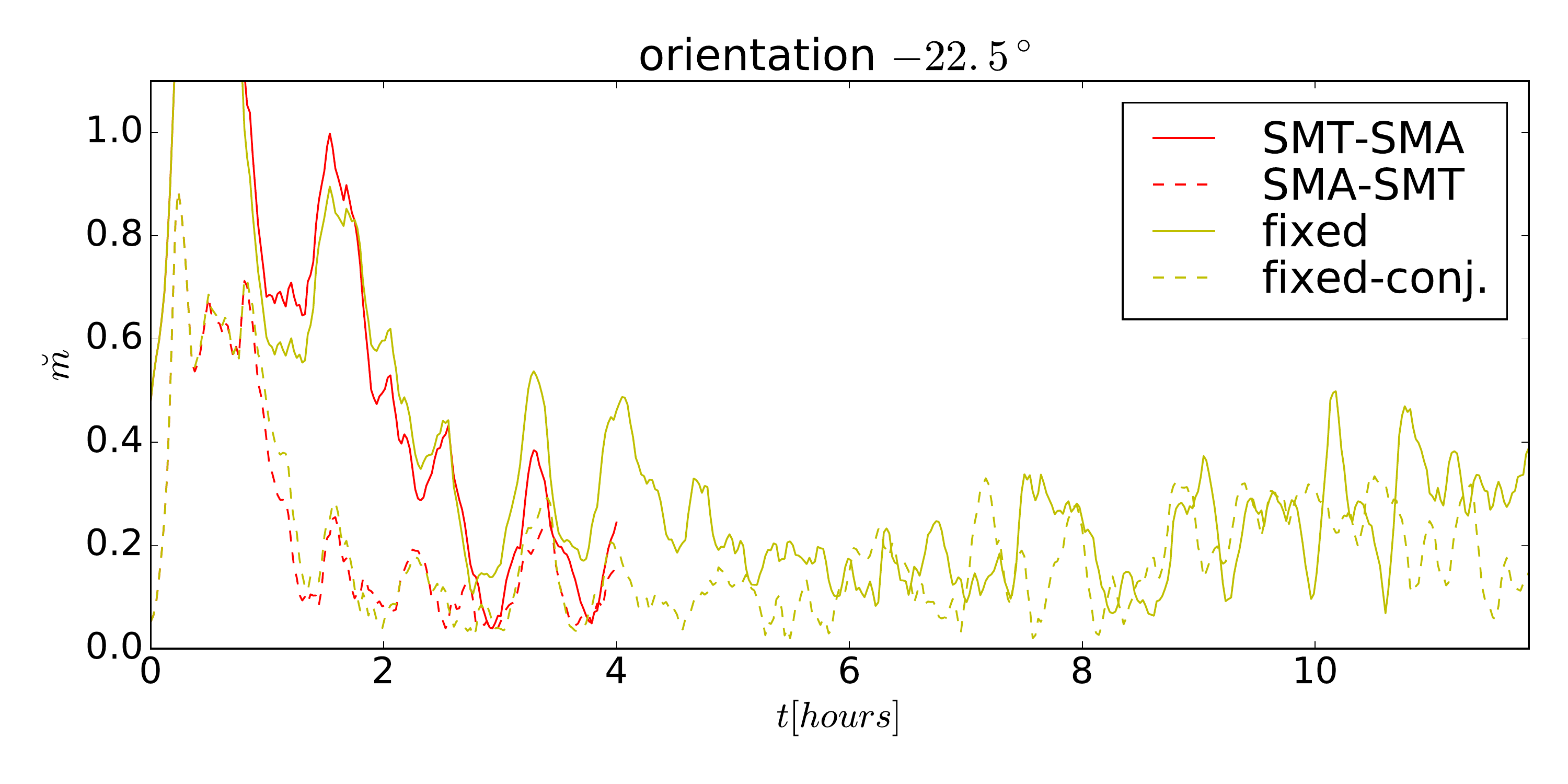} \\  
  \includegraphics[width=0.49\textwidth]{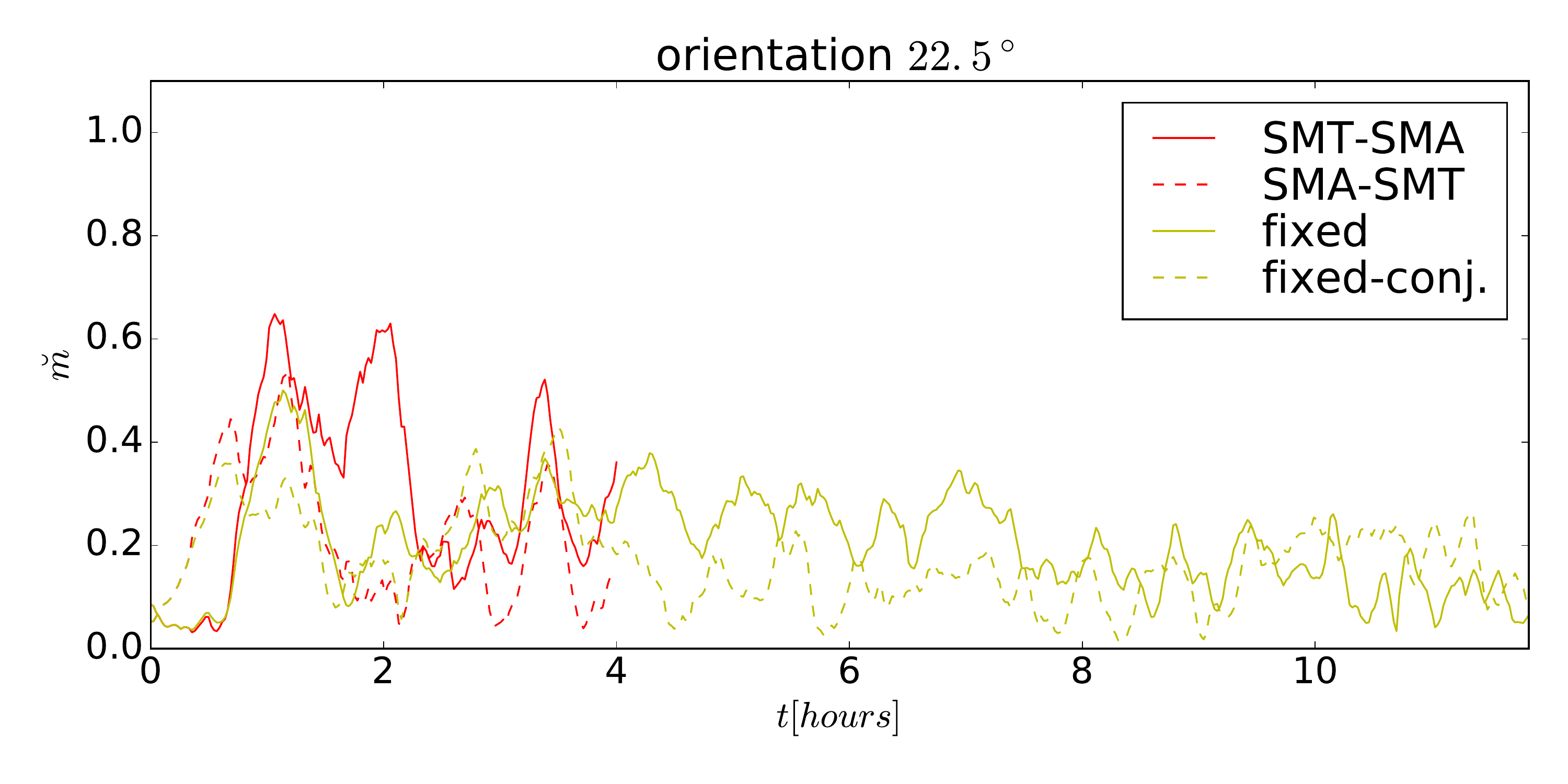}    
  \includegraphics[width=0.49\textwidth]{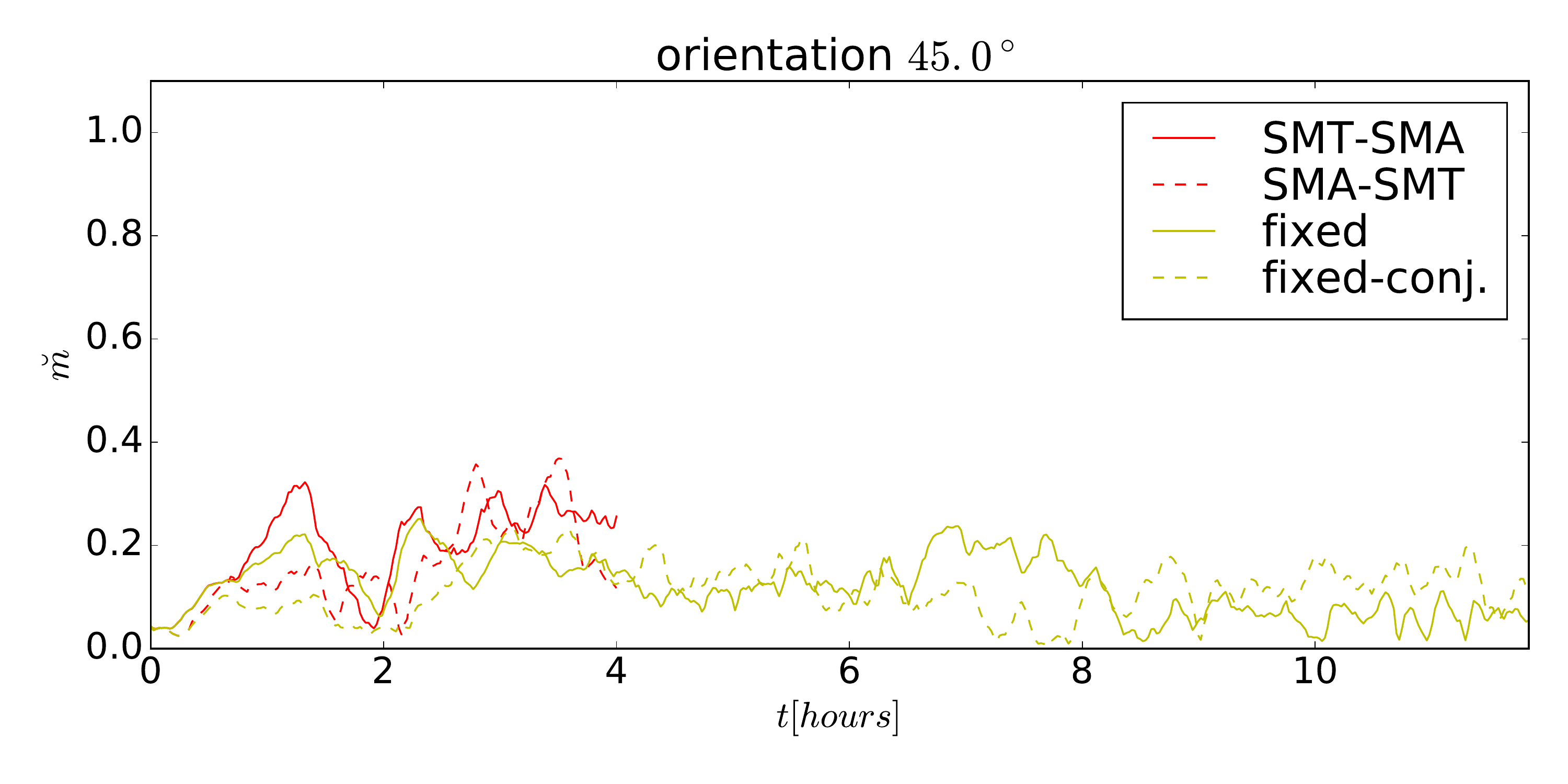} \\ 
  \includegraphics[width=0.49\textwidth]{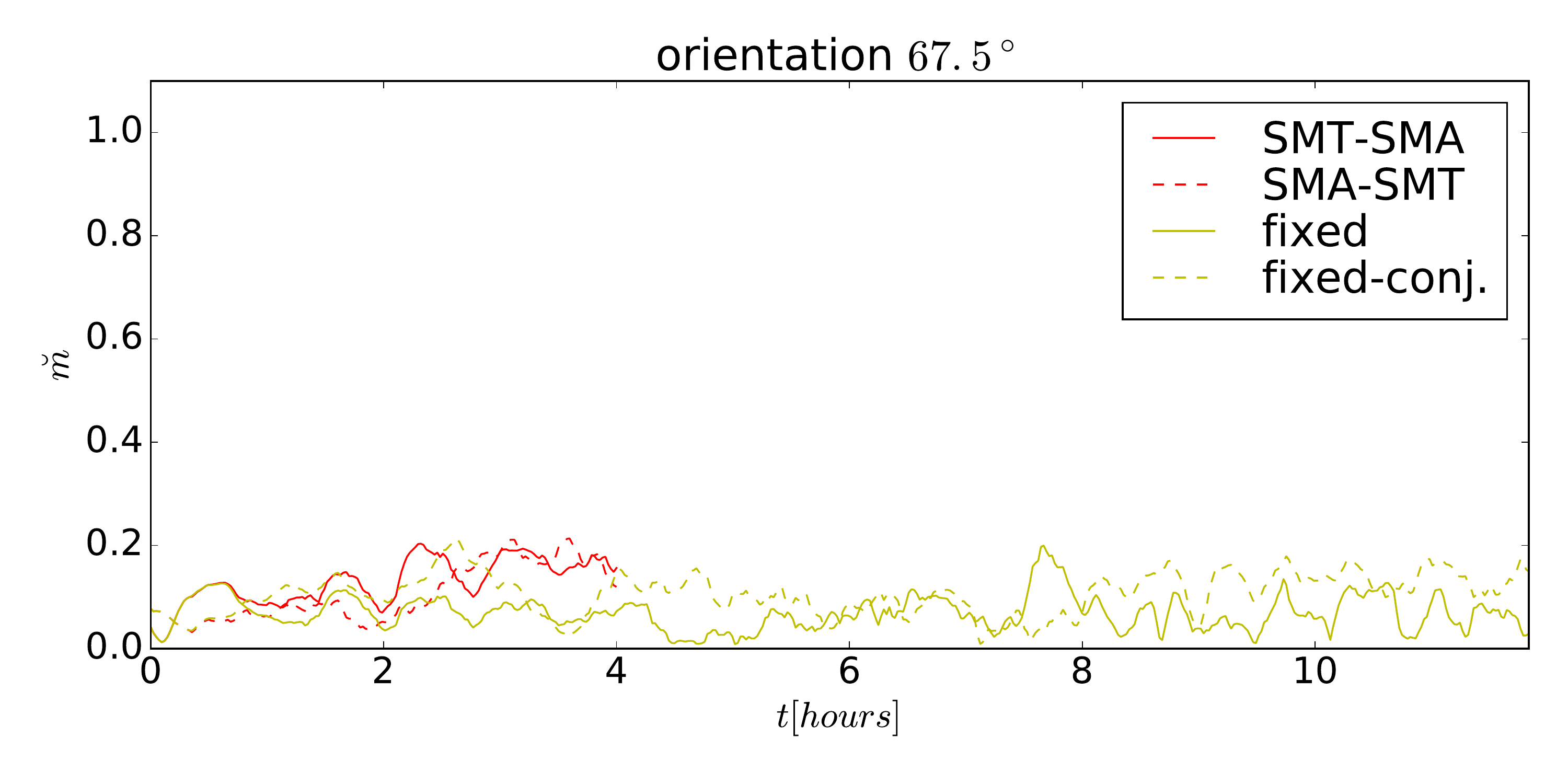}    
  \includegraphics[width=0.49\textwidth]{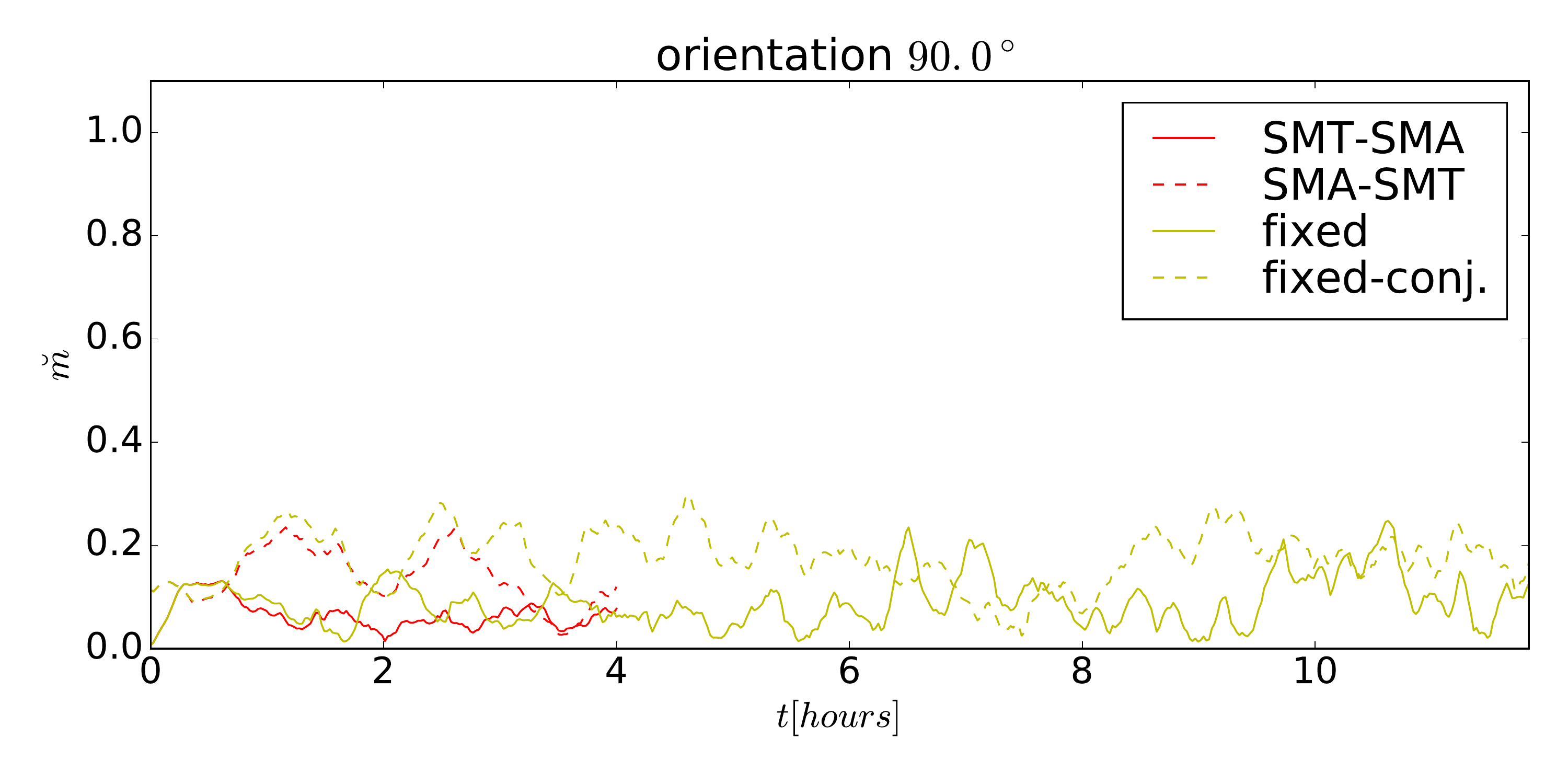}    
  
  \caption{Same as Fig.\ref{fig:mbreve-vs-t} for {\tt MAD\_thick-disk}
    model but for 8 different orientations EAST of NORTH given by $\in
    [-90,-67.5,-45,-22.5,22.5,45,67.5,90]$ (from left to right, top
    to bottom).  For a given baseline orientation, the fixed baseline
    case does not move the baseline position vs.\ time.  For the
    SMT-SMA and SMA-SMT baseline positions, the baselines move with
    time consistent with the EHT.  The agreement with EHT data for
    $\breve{m}$ vs.\ time is not possible with any baseline
    orientation.  The EHT data reaches up to $\breve{m}\sim 70\%$ and
    the SMT-SMA baselines show significant asymmetry in the magnitude
    (down to $\breve{m}\sim 30\%$ for the other side of the baseline).
    Only baseline orientations that are $-45^\circ$ to $+45^\circ$ show
    sufficient amplitude and asymmetry, allowing one to constrain how
    the position angle of the model data relative to the EHT
    baselines.  Such a constraint is not possible with intensity only
    (assuming an anisotropic scattering kernel does not help resolve
    the position angle). 
    \label{fig:mbreve-vs-t-orientations}}
\end{figure*}

Fig.~\ref{fig:mbreve-vs-t-orientations} shows light curves from the
{\tt MAD\_thick-disk} model for 8 different image orientations in the
uv-plane at an angle EAST of NORTH by $\in
[-90,-67.5,-45,-22.5,22.5,45,67.5,90]$. Only if the SMT-SMA baseline
is oriented between $\pm 45^\circ$ in the visibility plane would the
magnitude and asymmetry from the simulations match the EHT data.
Orientations too close to the $u$ (WEST-EAST) axis lead to no large
values of $\breve{m}$ and weak asymmetry.  For other models, like {\tt
  SANE\_dipole-jet}, $\breve{m}$ is high only in small patches in the
uv plane on these EHT baseline lengths.  This comparison alone does
not allow us to exclude the {\tt SANE\_dipole-jet} model, but more EHT
baselines (i.e., beyond EHT 2013) might enable such exclusions.

\subsection{Intensity and Polarization vs.\ Baseline Length}

\begin{figure*}

  \centering
 \includegraphics[width=0.45\textwidth]{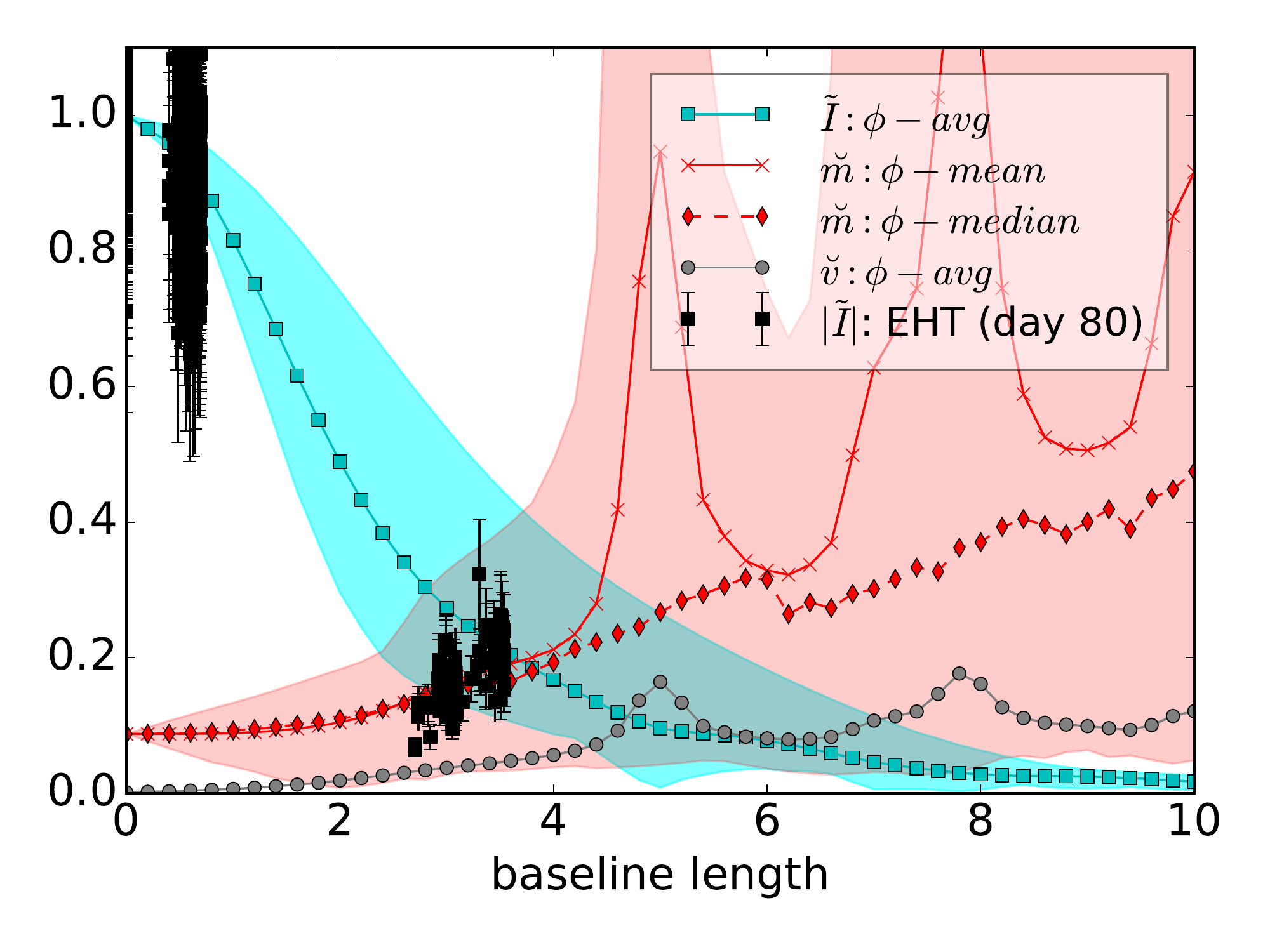}
  \includegraphics[width=0.45\textwidth]{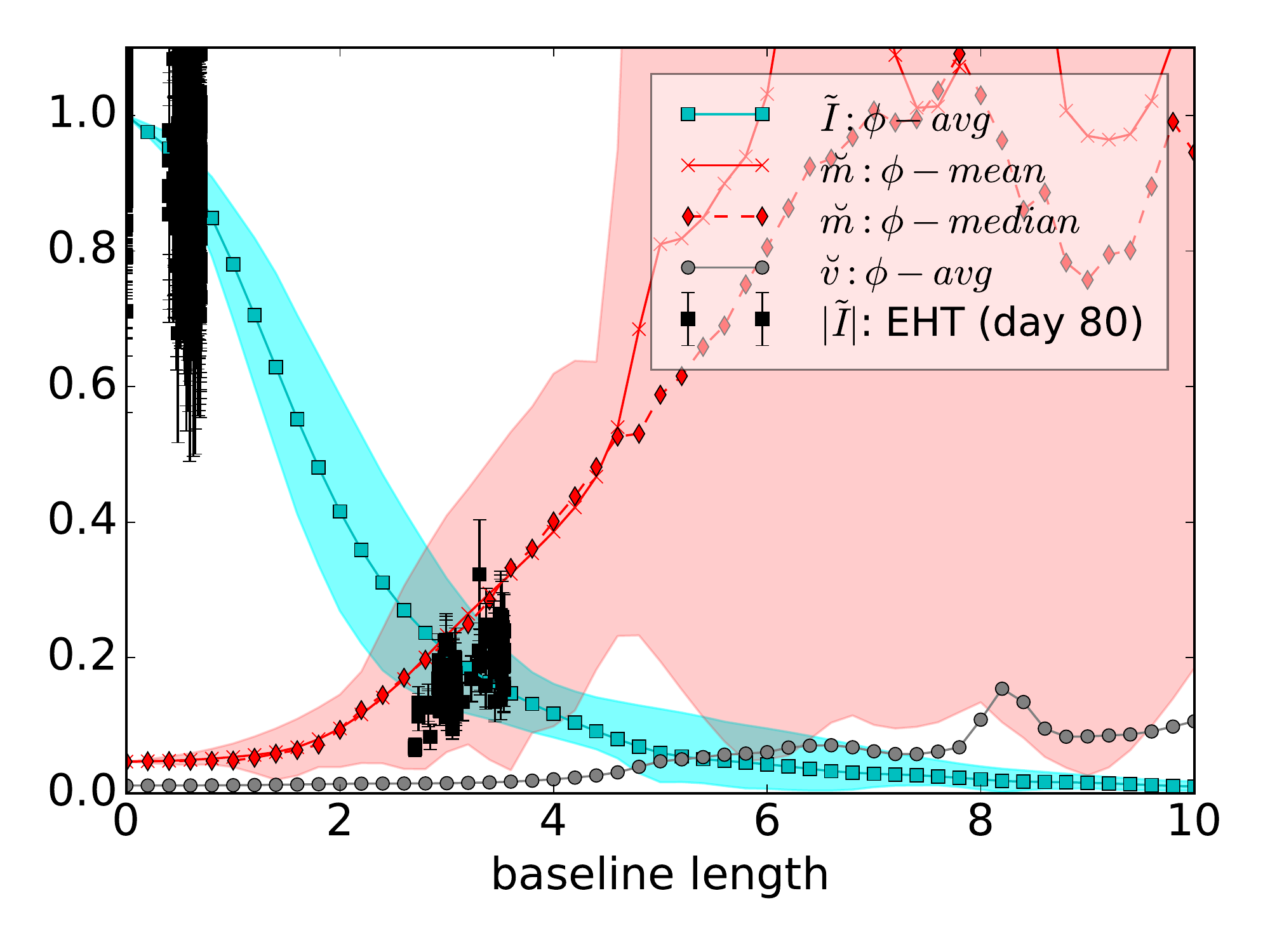}
  \caption{Model {\tt MAD\_thick-disk} (left panel) and {\tt
      MAD\_thick-jet} (right panel) showing $|\tilde{I}|$,
    $|\breve{m}|$, and $|\breve{v}|$ as a function of baseline
    length. Shaded regions span the variation along all baseline
    orientations.  Both models produce high polarization on longer
    baselines. Sharp $\breve{m}$ and $\breve{v}$ peaks are associated
    with baseline lengths with where $\tilde{I}$ falls close to zero.
    \label{fig:thickdisk7-Fuv-image} }
\vspace{.15in}
\end{figure*}

\begin{figure*}

  \centering
  
  \includegraphics[width=0.45\textwidth]{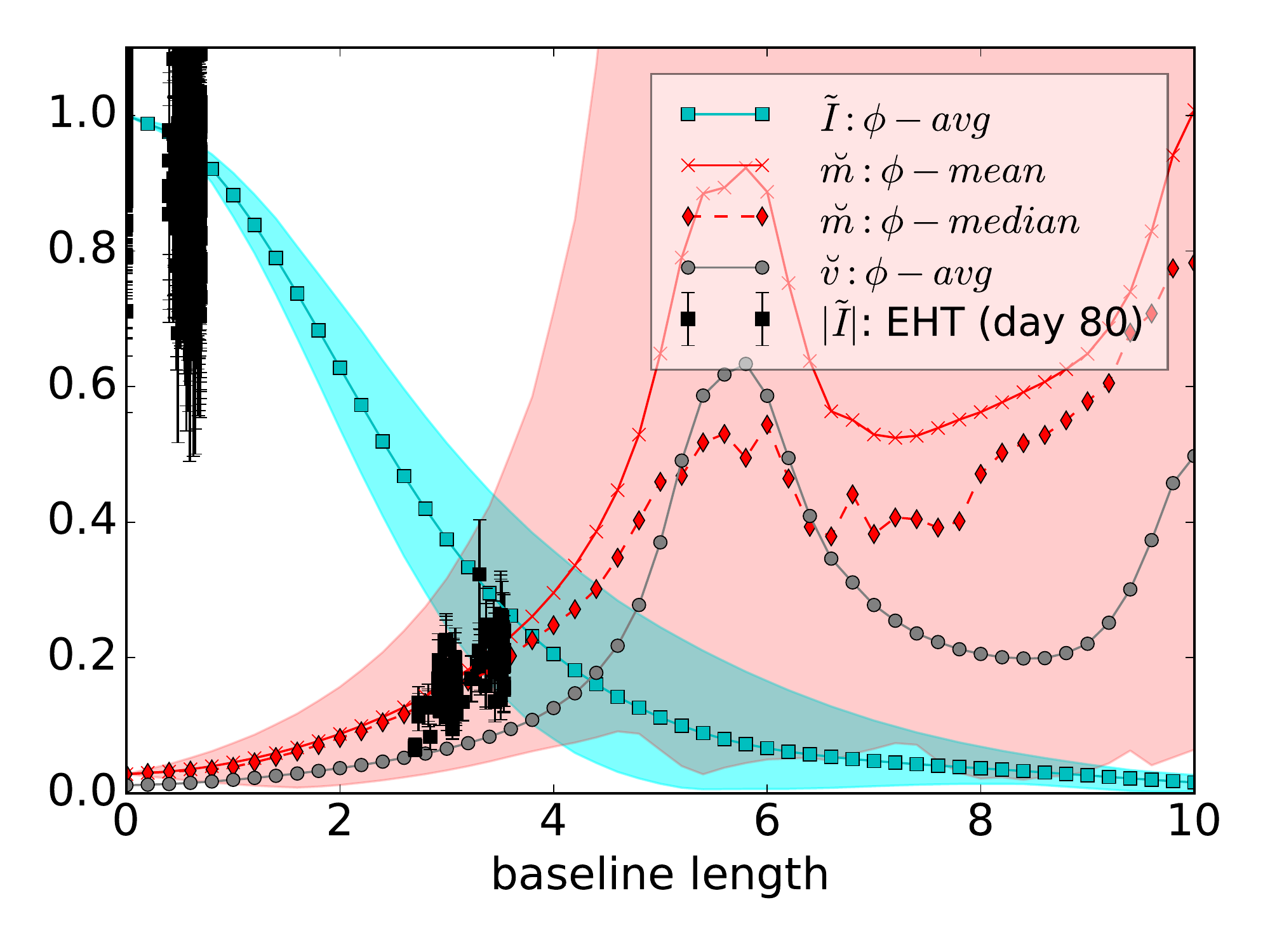}
  \includegraphics[width=0.45\textwidth]{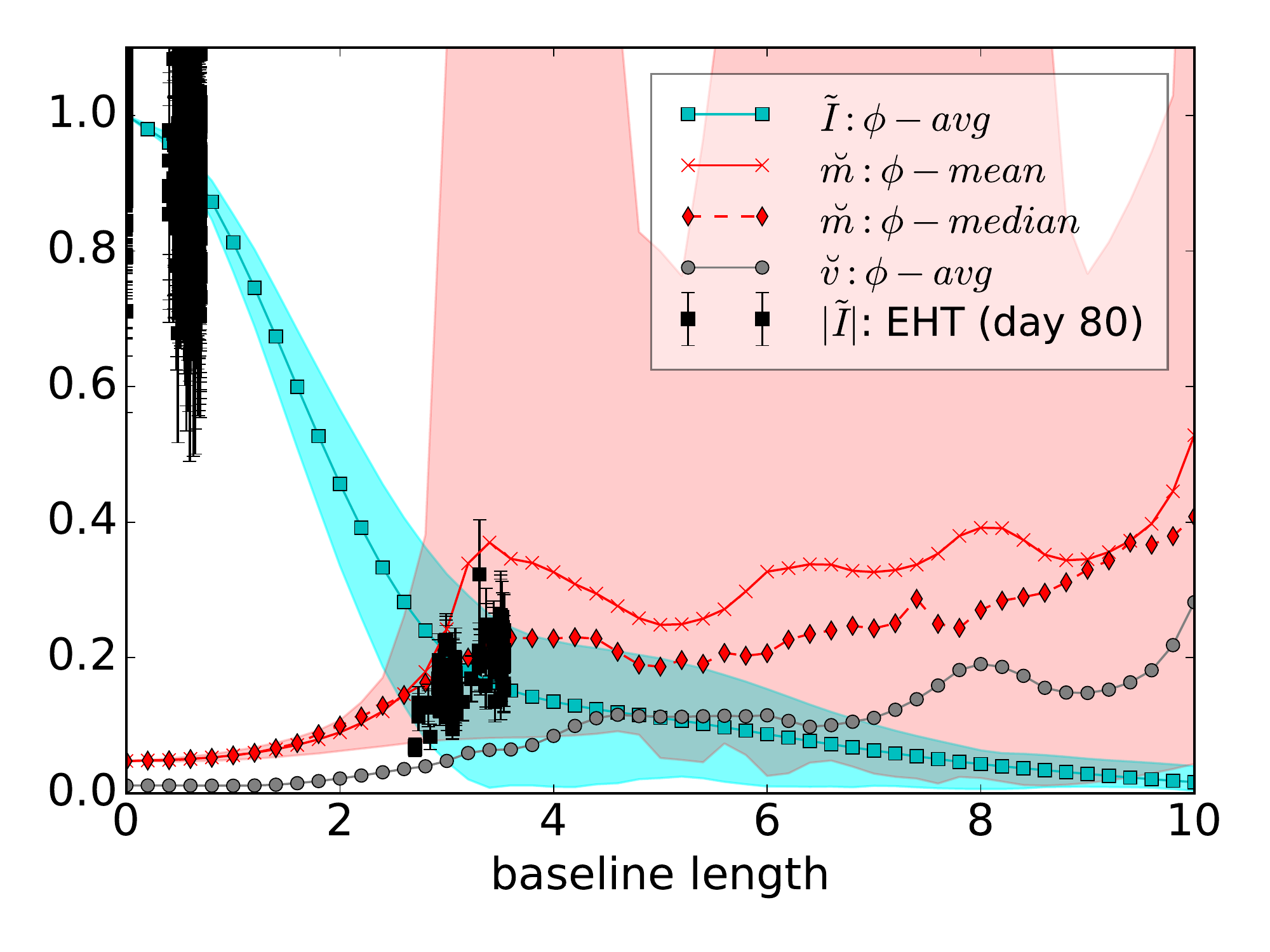}
  \caption{Same as Fig.~\ref{fig:thickdisk7-Fuv-image} for models {\tt
      SANE\_quadrupole-disk} (left panel) and {\tt SANE\_dipole-jet}
    (right panel). The {\tt SANE\_quadrupole-disk} produces large
    scale polarization features comparable to MAD models.  The {\tt
      SANE\_dipole-jet} model shows a partial shadow feature with a
    broad crescent, which leads to a larger image size and more
    consistency with the observations on shorter baseline
    lengths. \label{fig:quadrupole-Fuv-image}}
\end{figure*}

Figs.~\ref{fig:thickdisk7-Fuv-image} and
~\ref{fig:quadrupole-Fuv-image} show the total
intensity and polarization fractions as a function of baseline length. The polarization features are
significantly more distinguishing than the unpolarized emission.

The MAD models {\tt MAD\_thick-jet} and {\tt MAD\_thick-disk} readily produce the
observed polarization, despite emission from the jet region being
suppressed in the {\tt MAD\_thick-disk} model.  All models have fractional
linear polarization in the image domain that reach comparable levels
to recent EHT measurements.

It is apparent from the images that the total intensity features can
be similar for the MAD and {\tt SANE\_quadrupole-disk} models, but the
polarization fields are clearly distinct. This is promising for the
prospect of disentangling different magnetic field structures and
constraining models in a way not possible with only intensity.

\subsection{Dependence on Electron Temperature and Mass-Loading Prescriptions}

\begin{figure*}

  \centering
  \includegraphics[width=0.49\textwidth]{{{thickdisk7-jet-magn40-shotimag93.75th245f230fn5550case78169140_151_polarization-ticks}}}
  \includegraphics[width=0.49\textwidth]{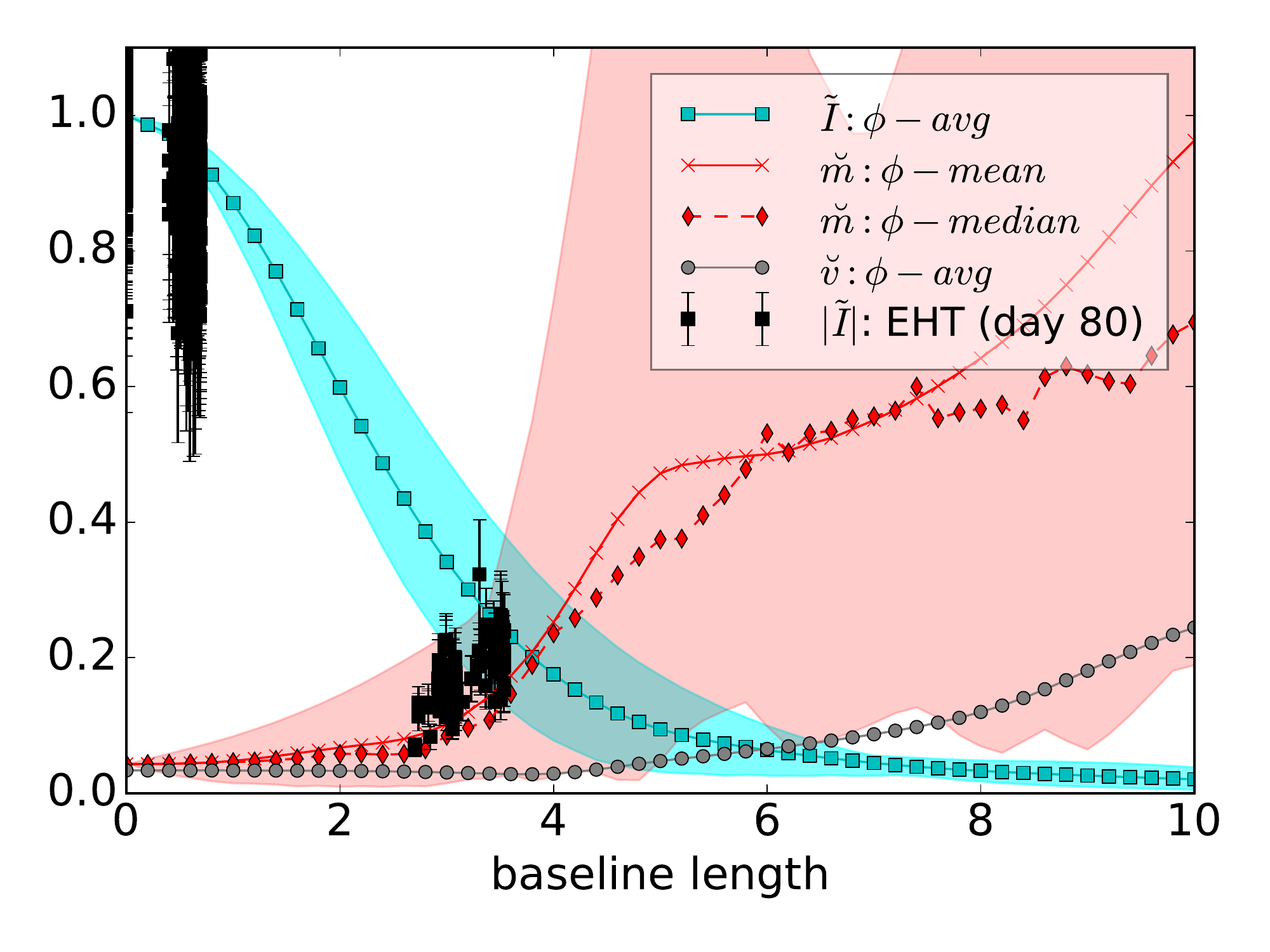}
  \caption{Image with overlaid polarization ticks and visibility domain quantities vs.\ baseline length for a model similar to {\tt
      MAD\_thick-jet}, but with $\sigma_{T_{\rm e}}=40$ corresponding
    to enhanced jet emission.  Dashed circles indicate the expected
    shadow size for a non-spinning (cyan) and maximally spinning
    (grey) BH.  For this data set the density scale and hence the
    accretion rate have been adjusted to be consistent with the observed
    flux density at 230GHz. The emission size is smaller and $|\breve{m}|$ is
    generally lower than for the fiducial {\tt MAD\_thick-jet}.  The BH shadow
    feature is more filled-in due to the enhanced jet emission.  A
    more exhaustive parameter search might find cases where this
    jet-enhanced emission model produces a less compact emission, as
    is required by EHT measurements of $|\tilde{I}|$. \label{fig:thickdisk7-jet-magn40}}
\end{figure*}

The specific electron temperature prescription is uncertain, so we
consider variations away from our default model for $\sigma_{T_{\rm
    e}}$.  Fig.~\ref{fig:thickdisk7-jet-magn40} shows the same plots
as before for the {\tt MAD\_thick-jet} model, but for $\sigma_{T_{\rm
    e}}=40$.  In this case, the BZ-funnel jet (not just the
funnel-wall jet) has its emission boosted.  This occurs because the BZ
funnel has $b^2/\rho\sim 30$--$50$ near the BH, and so changing
$\sigma_{T_{\rm e}}$ from $1$ to $40$ leads to much more of the
material in the funnel having a higher temperature given by $T_{\rm
  e,jet}=35$.

\begin{figure*}
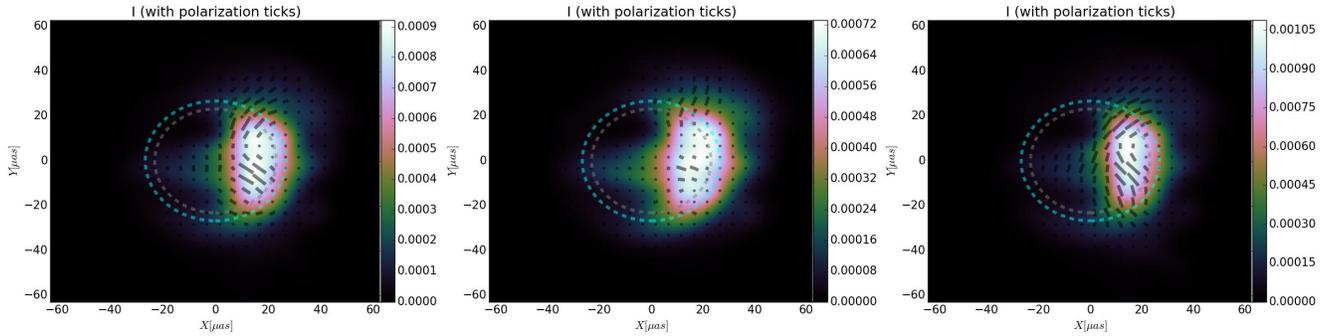

  \centering
  \includegraphics[width=0.97\textwidth]{{{shotimag93.75th170f230fn2140case4442140-4442143_151_polarization-ticks}}}
  \caption{Progression of image plane intensity and overlaid
    polarization ticks for models like {\tt SANE\_quadrupole-disk}
    model (left panel) with modified $\sigma_{T_{\rm e}}=40$ (middle
    panel) consistent with Figs.~\ref{fig:Tejnu-magn40}, and modified
    $T_{\rm e.jet}=100$ (right panel) consistent with
    Figs.~\ref{fig:Tejnu-te100}.  In both modified cases the density
    scale and thus accretion rates was adjusted by $\sim 30\%$ to
    roughly fit the observed zero baseline flux. Differences in
    polarization ticks are dramatic, but changes are minor in
    intensity despite the electron prescription forcing more radiation
    to be emitted from the coronal regions.  This shows how
    polarization could be more sensitive to the electron temperature
    prescription than intensity.
    \label{fig:quadrupole-dep-magn-te-image}}

\end{figure*}

Fig.~\ref{fig:quadrupole-dep-magn-te-image} shows the {\tt
  SANE\_quadrupole-disk} model with different electron temperature
prescriptions, including our default choice, $\sigma_{T_{\rm e}}=40$,
and $T_{\rm e,jet}=100$.  These change the image size slightly, but
the $\sigma_{T_{\rm e}}=40$ has a quite different polarization 
pattern.  This suggests that polarization might be more sensitive to
changes in the electron heating physics or whether a BZ-driven funnel
jet contains hot electrons.

\begin{figure*}
  \centering
  \includegraphics[width=0.97\textwidth]{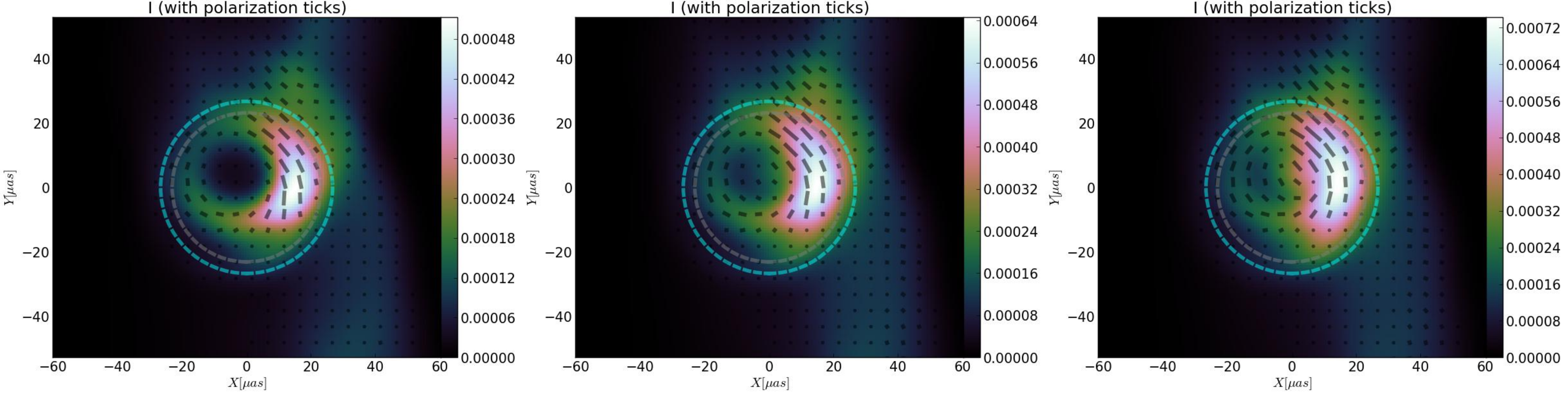}
  \caption{Progression of image plane intensity plots with overlaid
    polarization ticks for models like {\tt MAD\_thick-jet}, but with
    $\sigma_{\rho}=10$ (left panel), $\sigma_{\rho}=35$ (middle
    panel), and $\sigma_{\rho}=100$ (right panel, consistent with our
    default model shown in lower panels of
    Fig.~\ref{fig:16panels-thickdisk7}) with $\rho_{\rm jet}=0$ to
    force the material with $b^2/\rho>\sigma_{\rho}$ to have a lowered
    density value.  The left panel with $\sigma_{\rho}=10$ is
    consistent with removing all numerically-injected floor material,
    leaving only self-consistently evolved material that follows from
    mass conservation.  This removes a significant amount of emission
    from the BZ-driven funnel jet, creating a hole in the emission
    that mimics a BH shadow feature.  The middle and right panels show
    the progression toward our default {\tt MAD\_thick-jet} model
    that includes the BZ-driven funnel jet material (whose emissivity
    is also controlled by the electron temperature
    prescription). These panels show how BZ-driven funnel and
    funnel-wall jets might be distinguishable.
     \label{fig:thickdisk7-sigmajet}}
\end{figure*}

The mass-loading mechanism for BZ-driven funnel jets remains
uncertain, so we vary the mass-loading via changes in $\sigma_{\rho}$
and consider how it changes the intensity and polarization in the
image plane.  Fig.~\ref{fig:thickdisk7-sigmajet} shows how emission
from the BZ-driven funnel jet normally fills-in the BH shadow region
in the {\tt MAD\_thick-jet} model as was shown in
Fig.~\ref{fig:16panels-thickdisk7}.  The BZ-jet funnel material violates mass
conservation, because matter is injected (indirectly motivated by photon annihilation or pair production cascades) in order to
maintain code stability.  By removing the funnel density
material, only self-consistently evolved matter that
obeys mass conservation is left that exists in the
funnel-wall jet, corona, and disk.  The self-consistent material has no funnel emission, so the image plane recovers a BH shadow feature that was
previously obscured by BZ-driven funnel jet emission.  So,
the image plane (and corresponding Fourier transform in the visibility
plane) could potentially distinguish between funnel-wall jet emission
and BZ-driven funnel jet emission.  These differences in the appearance of the shadow might allow one to test
jet theories, and it also gives hope that a BH shadow could be more
easily detected in \sgra, because \sgra~likely has a very low density
in the funnel
\citep{2011ApJ...735....9M,2005ApJ...631..456L,2015ApJ...809...97B}.

These results highlight the need for a better understanding of
electron heating of accretion flows and mass-loading of jets in these
systems. Further progress and more realistic electron physics
\citep{2015MNRAS.454.1848R} and collisionless effects on the proton
temperature \citep{2015arXiv151104445F} will become essential to realistically model
the accretion flow in \sgra\ as EHT data improves.
However, better GRMHD schemes will be required to avoid artificial
numerical heating in magnetized and/or supersonic regions (as near the
funnel-wall or the BH) \citep{2007MNRAS.379..469T}, which feeds into
the collisionless physics terms.  Additional physics or mechanisms are
needed to understand how the BZ-driven funnel jet is mass-loaded and
whether that material emits in systems like \sgra.

These broad range of jet vs.\ disk-dominated electron heating
prescriptions also show that the disk, funnel-wall jet, and BZ-driven
jet could in principle radiate by arbitrary amounts.  This means one
cannot exclude the dynamical presence of a BZ-driven jet based upon
the emission, because the BZ-driven jet may not be dissipating, may
not contain hot electrons, or may contain too few electrons (while
still sustaining force-free or MHD conditions).

\subsection{Implications for Detecting the BH Shadow}\label{shadow}

The BH shadow might be delineated in the image plane by a bright
photon ring or by a crescent feature whose bright Doppler boosted side
is completed by a dimmer Doppler de-boosted side with a null (the
shadow itself) in intensity between \citep{2000ApJ...528L..13F}.
However, several issues can make the shadow appear more or less
detectable for any spin and can potentially introduce features that
are unrelated to the shadow but have a similar appearance. We now
discuss how choices in the modeling, radiative transfer, and dynamics
can affect detectability of the shadow.

The shadow and surrounding ring are most prominent and isotropic at
small viewing inclinations (closer to face-on) because the Doppler
de-boosting is less significant. However, fits to observational data
using GRMHD simulation models tend to favor inclination angles of
$i\sim 45^\circ$ \citep{2010ApJ...717.1092D} or even higher depending
upon spin and electron temperature prescriptions
\citep{2014AA...570A...7M}.  Among our models, nearly edge-on ($i\sim
90^\circ$) models are preferred for disks ({\tt MAD\_thick-disk} and
{\tt SANE\_quadrupole-disk}), while somewhat more tilted angles
($i\sim 130^\circ$) are preferred for the jet models ({\tt
  MAD\_thick-jet} and {\tt SANE\_dipole-jet}). Unlike other works,
these inclinations are influenced by fitting polarimetric data, which
requires more specific inclination angles than intensity due to
cancellation of polarization if the inclination is too face-on. These
high inclinations could make it challenging to detect the faint
Doppler de-boosted side bounding the shadow.

Our simulations also demonstrate that jet emission can also pose a
challenge to detecting a BH shadow feature \citep[see
  also][]{2015ApJ...799....1C}.  For some inclination angles in our
models, most of the $230{\rm GHz}$ emission arises from a jet that
either completely obscures the expected BH shadow or dissects it into
smaller patches (see Fig.~\ref{fig:16panels-thickdisk7}).  Splitting a
shadow feature in smaller patches pushes interferometric signatures of
it (such as notches in $\tilde{I}$ or peaks in $\breve{m}$) to longer
baselines (see Fig.~\ref{fig:thickdisk7-Fuv-image}).

One can also generally see from the images that detecting a shadow
feature can be more difficult depending upon the underlying dynamical
model.  A shadow-like feature can be present that is smaller than the
expected BH shadow, and non-trivial features in the accretion flow can
obscure part of the shadow.  The MAD models show a less clear-cut BH
shadow feature (see Figs.~\ref{fig:16panels-thickdisk7} and
~\ref{fig:thickdisk7-Fuv-image}) than the {\tt SANE\_dipole-jet} model
(see Figs.~\ref{fig:16panels-quadrupole} and
~\ref{fig:quadrupole-Fuv-image}).

\begin{figure*}[t]
  \centering
  \includegraphics[width=0.99\textwidth]{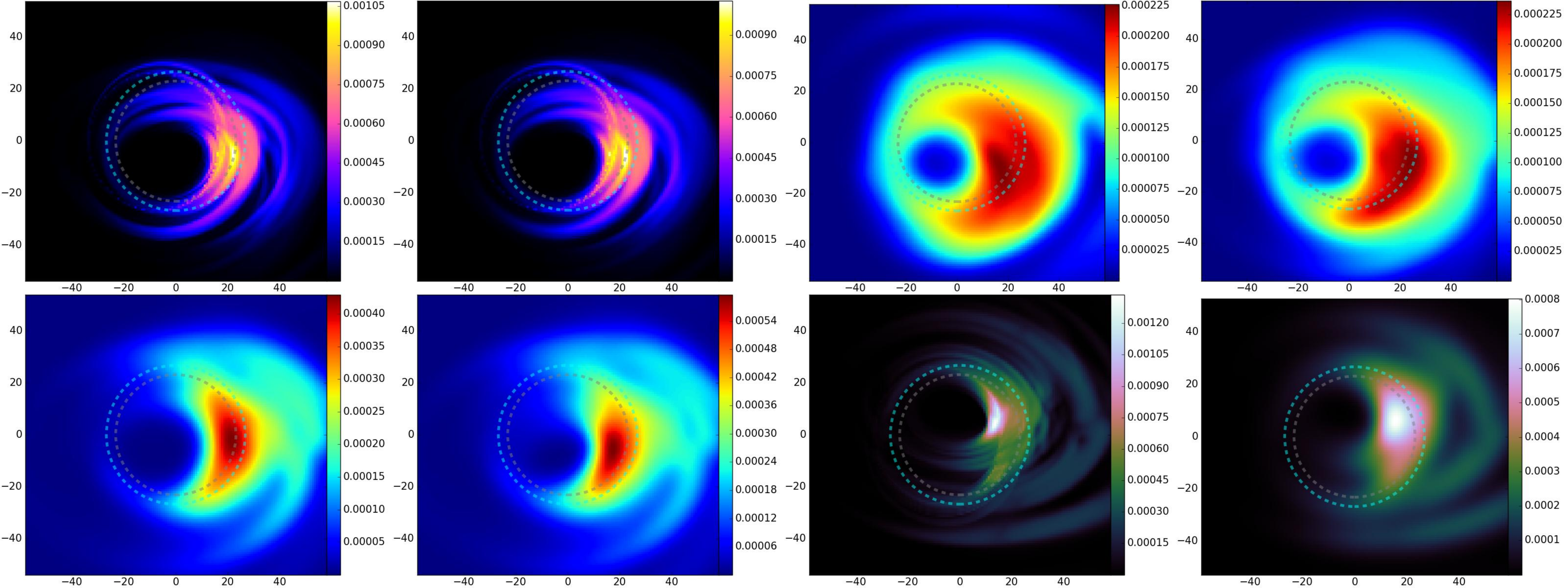} 
  \caption{Progression of BH shadow image plane intensity plots for
    {\tt SANE\_dipole-jet} model.  From top-left to bottom-right, the
    progression shows: 1) single observer time at same time,
    same inclination $i=50^\circ$, same temperature prescription of $T_i/T_e=3$,
    similar mass accretion rate, and similar colormap compared to \citet{2010ApJ...717.1092D} ; 2)
    time-averaged over 10-minute interval using 8 snapshots (shows
    little difference) ; 3) colormap, inclination of $i=45^\circ$, and
    temperature prescription of fixed $T_i/T_e=10$ like for model D in
    \citet{2009ApJ...706..497M}; 4) our default time chosen instead
    (little difference); 5) our temperature prescription with $T_{\rm
      e,jet}=T_{\rm e,gas}$ (i.e., no isothermal jet); 6) our final
    temperature prescription with isothermal jet; 7) our best-fit for
    all parameters (e.g.  inclination and density); and 8) including
    scattering.  In each case, if required, the density was slightly
    changed to reproduce the $230$GHz zero baseline flux.  This
    progression shows how color map choices, physical choices
    (inclination, temperature prescriptions), and scattering (at half
    the actual interstellar scattering, roughly what is possible by
    de-blurring after accounting for noise, see
    \citealt{2014ApJ...795..134F,2015arXiv151208543L,2015ApJ...799....1C}) affect whether the BH shadow
    appears detectable, suggesting that more quantitative measures of BH
    shadow detectability will be important to consider
    \citep{2015ApJ...814..115P}.
    \label{fig:shadow-progression}}
\end{figure*}

The shadows from other radiative transfer GRMHD models range from
having a fairly low level of de-boosted emission
\citep{2010ApJ...717.1092D} to having a fairly bright de-boosted side
\citep{2014AA...570A...7M}.  These differences in GRMHD model results
for the radiative transfer are controlled by differences in electron
temperature prescriptions, mass accretion rates, and dynamical model
differences.

We can highlight how some model choices affect the appearance of the
BH shadow.  Fig.~\ref{fig:shadow-progression} shows a progression of
total-intensity images for the {\tt SANE\_dipole-jet} model. We start
with a model that is similar to the model MBD described in
\citet{2010ApJ...717.1092D} by using the same dynamical simulation,
same inclination, same temperature prescription, and similar mass
accretion rate. Although the radiative transfer codes are independent,
they produce similar images (as expected), which exhibit a relatively
pronounced shadow feature. We then introduce changes that eventually
lead to our fiducial parameters for this model (see
Table~\ref{tab:models}). In addition to being sensitive to the viewing
inclination, temperature prescription, and assumptions about
scattering mitigation, Fig.~\ref{fig:shadow-progression} shows that
even the type of color map used can tend to imply the shadow is more
detectable, because some color maps highlight low-level features so
they can be seen visually (e.g. the so-called ``jet'' color map has
this feature).  As Fig.~\ref{fig:shadow-progression} also
demonstrates, simply analyzing raw unscattered images sets unrealistic
expectations for how the shadow may appear in EHT images.  Compared to
prior work, the final model shown in Fig.~\ref{fig:shadow-progression}
reflects minor changes in the radiative transfer and realistic
scattering limitations that generate a broad crescent feature that is
both one-sided and partially filled-in.

Another origin of differences in simulation model results for the BH
shadow could be due to the choice of initial conditions and run-time.
Many prior simulations have initial conditions of a torus that lead to
relatively thin disks with height-to-radius ratio of $H/R\lesssim
0.2$, while long-term GRMHD simulations evolved with a mass supply at
large radii show radiatively inefficient accretion flows (RIAFs) tend
towards $H/R\sim 0.4$ \citep{2012MNRAS.426.3241N} or even thicker at
$H/R\sim 1$ \citep{2012MNRAS.423.3083M}.  GRMHD simulations that start
with an initial torus too close (pressure maximum within $100r_g$ or
inflow equilibrium within $30r_g$) to the BH remain controlled by
those initial conditions, because the accretion process feeds off of
the vertically thin torus material that has insufficient time to
heat-up and become thick before reaching the BH.  A thinner disk near
the BH more readily produces a narrow sharp crescent (our SANE models
happen to be run with tori with pressure maximum at $r\sim 10r_g$),
while geometrically thick RIAFs with coronae tend to have a broad
fuzzy crescent (our MAD models).

By comparison, analytical models can sometimes show a sharper photon
ring and crescent-like feature
\citep{2006MNRAS.367..905B,2009ApJ...697...45B}, because they tend to
only include disk emission and no corona or jet emission that would
broaden the crescent in a way dependent upon the temperature
prescription.  For example, the vertical structure assumed in
\citet{2009ApJ...697...45B} is that of a Gaussian with $H/R\sim 1$ for
the disk, while hot coronae or jets have an extended column of gas
that may not change the density scale-height much but change the
emission profile to be more vertically extended.  However, even the
crescent feature in \citet{2009ApJ...697...45B} is incomplete and
shows little Doppler de-boosted emission one might need in order to
clearly measure a shadow size.  Further, analytical RIAF solutions
like the advection-dominated accretion flow (ADAF) model have $H/R\sim
0.6$ or smaller for realistic prescriptions of the effective adiabatic
index and any winds present
\citep{1999ApJ...516..399Q,1999ApJ...520..298Q}, which would
presumably lead to a change in the BH shadow feature.

In summary, the filling-in of the BH shadow by corona or jet emission
(as considered for GRMHD simulation-based radiative transfer models),
and the broadening of the expected crescent feature by accretion flow
structure and scattering may make it difficult to unambiguously detect
the BH shadow or to extract information about the space-time from its
size and shape
\citep{2014ApJ...784....7B,2015ApJ...814..115P,2015arXiv151202640J}
before the plasma physics and dynamical properties of the disk and jet
are constrained. Improved techniques to detect the shadow will be
important in order to handle the diverse array of possible accretion
flow physics and to best account for the scattering. In addition, the
BH shadow's appearance in polarization and at higher frequencies
($349$GHz), where the blurring from scattering is $2.3$ times weaker,
will help increase its detectability.

\subsection{349{\rm GHz} \& CP}

Extending the capabilities of the EHT to higher frequencies increases the 
angular resolution, reduces blurring due to interstellar scattering,
and probes the emission structure at lower optical depth. The image-integrated (total)
flux density is comparable at 230 and 345~GHz \cite{Bower2015}.

\begin{figure*}[t]
  \centering
  \includegraphics[width=0.99\textwidth]{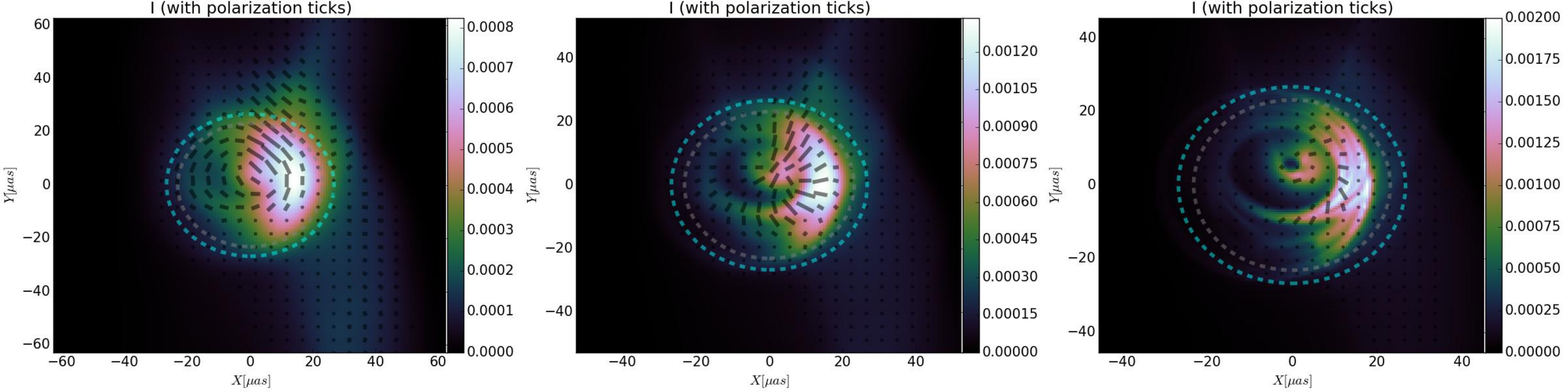} 
  \caption{Images at $230{\rm GHz}$ (left panel), $349{\rm GHz}$
    (middle panel), and $674{\rm GHz}$ for the {\tt MAD\_thick-jet}
    model, which contains significant jet and coronal emission that
    fills-in the shadow at $230$GHz.  As is similar for all our models,
    at higher frequencies, any emission in the central region begins
    to become optically thin and develops an intensity null, which is
    bounded by emission from the photon ring and some accretion
    material.  Despite the presence of jet emission even at higher
    frequencies, the development of a null in intensity could be used
    to help detect the BH shadow.\label{fig:shadow-3freqs}}
\end{figure*}

Fig.~\ref{fig:shadow-3freqs} shows the {\tt MAD\_thick-jet} model and how the
shadow feature becomes more distinguishable at higher frequencies.
This revealing of the BH shadow (null in intensity bounded by
emission) occurs because the surrounding emission structure becomes
less opaque and less luminous at higher frequencies. Thus, even for underlying accretion models containing jet and corona emission, 
EHT data at $349{\rm GHz}$ can potentially detect the BH shadow. These
results strongly motivate EHT efforts at $349{\rm GHz}$.

There are no CP data yet available from the EHT, but CP 
appears to differentiate magnetic field configurations, including among 
different MRI-type GRMHD models.  For example, the {\tt
  SANE\_dipole-jet} model shows different circular polarization
patterns in the visibility plane than the {\tt SANE\_quadrupole-disk}
model, especially at long EHT baselines (see
Figs.~\ref{fig:16panels-thickdisk7}
and~\ref{fig:16panels-quadrupole}).  Also, the visibility plane
structure in $\breve{m}$ is quite different than the structure in
$\breve{v}$, suggesting that linear and circular polarization each provide
independent constraints on the underlying accretion flow structure. 
Higher sensitivity at higher frequencies may not be required for
linear or circular polarization, which both increase with frequency.

\section{Future Work}
\label{sec:future}

Validation of radiative transfer codes is crucial to ensure reliable
comparisons are made between models and observations.  Several codes
exist to perform radiative transfer of GRMHD simulations during
post-processing
\citep{2009ApJ...703L.142D,2009ApJ...706..497M,2010ApJ...717.1092D,2014AA...570A...7M,2015ApJ...812..103C,2015ApJ...799....1C}.
In section~\ref{shadow}, we have compared our radiative transfer
results with \citet{2010ApJ...717.1092D} for the same underlying
simulation model and assumptions but using a different radiative
transfer scheme, and we found reasonable qualitative agreement in the
appearance of the shadow and overall emission structure in the image.
In section~\ref{shadow}, we have shown how even slight differences in
color maps and physical assumptions lead to what looks like large
changes in the appearance of the BH shadow.  As more physics (such as
polarization included in this paper, non-thermal particles, etc.) is
considered, it will become crucial to ensure all radiative transfer
codes can achieve the same quantitative and qualitative results for a
suite of tests that include a diverse range of simulation and
analytical models.  In future work, we plan to compare the results of
radiative transfer codes used by various researchers in the field.

In light of how EHT polarimetric observations offer a probe of models
not possible with zero-baseline or non-polarimetric observations, we
plan a more comprehensive investigation of a larger suite of GRMHD
models beyond our models with relatively high spins.  These include
models which vary across a broad range of BH
spins \citep{2012MNRAS.423.3083M} as well as models that vary across a
broad range of tilt angles between the accretion disk angular momentum
axis and BH spin axis \citep{2013Sci...339...49M} (which should occur
in \sgra and M87 on horizon scales,
see \citealt{2013MNRAS.432.2252D,2015arXiv151207969P}).  In order to
better control emission from the funnel region, we will include new
simulations that track the matter numerically-injected into the funnel
to distinguish this material from self-consistently evolved material.
In order to improve our ad hoc electron temperature prescriptions, we
will include new simulations that track the electron and proton
temperatures.

Currently, our fitting procedure does not include minimizing $\chi^2$
over image size or various nuisance parameters that modify the
electron temperature or BZ-driven funnel jet mass-loading.  Also, EVPA
is not part of our zero baseline fitting procedure.  Regarding the
EHT, we only compare observations to simulation results, but we do not
use that as part of our fitting procedure, so in principle SANE models
might do better to match EHT observations in some part of our
parameter space.  We only compare EHT observations to simulations in
linear polarization vs. intensity in the visibility plane, not
focusing on EVPA and not focusing on CP that is not available yet.  We
provide some discussion of CP, because it is currently being measured
by the EHT and some theoretical guidance is useful.  Much more work on
the radial extension of the disk model and any additional Faraday
screens will be required to better fit zero-baseline CP and EVPA
observations vs. frequency or any EHT data.  For the GRMHD simulations
used in this paper and any new simulations, in the future, we will
also consider how zero-baseline EVPA and EHT measurements of CP and
EVPA constrain the models.  We will also consider closure phases and
other similar diagnostics that are independent of visibility
amplitudes and insensitive to single antenna phase
errors \citep{Doeleman:2001:SSA,2011ApJ...738...38B}.  CP, EVPA, and
Faraday rotation can potentially probe the helical orientation of jet
magnetic field \citep{2009ApJ...702L.148C} or probe the degree of
order of the disk magnetic field
\citep{2012ApJ...745..115M}.   The hope is that EHT CP
and EVPA observations at non-zero baselines at $230$GHz or higher
frequencies will further constrain the disk, jet, and plasma
properties in unique ways compared to linear polarization fraction.

\section{Conclusions}
\label{sec:conclusions}

We have performed general relativistic (GR) polarized radiative
transfer calculations on time-dependent three-dimensional GRMHD
simulations to model thermal synchrotron emission from the Galactic
Center source \sgra.  We considered several models of MAD and SANE
types with various kinetic-physics-inspired electron heating
prescriptions that enhance the emission from the disk corona,
funnel-wall jet that hugs the boundary between the disk and jet, and
highly magnetized BZ-driven funnel jet.

We have compared our results to most recent (2013) polarimetry
measurements by the EHT \citep{JohnsonEtAl2015Science} and showed how
polarization in the visibility (Fourier) domain distinguishes and
constrains accretion flow models with different magnetic field
structures.  To identify which model would be favored, we compared a
binned visibility amplitude ($\tilde{I}$) vs.\ binned visibility
fractional linear polarization ($\breve{m}$), which the EHT can
reliably measure \citep{JohnsonEtAl2015Science}.  We also compared the
simulation results for $\breve{m}$ vs.\ time for different baselines
to see if the behavior matched the EHT observations.  We focused on
comparing to linear polarization at various (including zero) baseline
lengths.

Our comparisons between the simulations and observations favor models
with ordered magnetic fields near the event horizon in \sgra.
Specifically, the MAD models are broadly consistent with these most
recent EHT data sets for linear polarization fraction vs. intensity as
well as linear polarization fraction vs. time.  MADs occur when a
supply of magnetic flux at larger distances accretes and accumulates
near the BH.  This leads to an ordered magnetic field threading the
region around the horizon and threading the accretion disk near the
photon orbit.  For electron heating prescriptions that highlight
emission from the disk (i.e., model {\tt MAD\_disk}), the emission
that escapes to the observer primarily occurs from the disk near the
photon orbit that is threaded by ordered poloidal magnetic field.  For
electron heating prescriptions that produce more jet emission (i.e.,
model {\tt MAD\_jet}), both the disk and funnel-wall jet contribute
about equally.  The agreement between the EHT data and MAD models is
robust to fairly substantial changes in inclination angle.

The EHT observations disfavor the standard MRI-type SANE models with
an initial dipolar field (i.e., model {\tt SANE\_dipole-jet}) more
strongly than another MRI-type disk with an initial large-scale
quadrupolar field (i.e., model {\tt SANE\_quadrupole-disk}).  The
former model contains a clear jet but contains mostly disordered
magnetic field in the disk, while the latter model has no jet but
contains ordered field within the disk.  This suggests that, broadly
speaking, those models with ordered magnetic fields threading the
accretion disk are favored by the EHT data.  The MAD models have the
strongest degree of ordered dipolar magnetic field threading the
accretion disk, causing them to be most like the EHT data and so
favored over all other models.  Even stronger constraints on the
magnetic field structure should be possible with CP, EVPA, and higher
($349~{\rm GHz}$) frequencies.

We considered the BH shadow visually in the image plane and how it
changes with different simulations and prescriptions for electron
temperatures and BZ-driven funnel jet mass-loading.  In general, the
shadow feature is not necessarily distinct, e.g., it can be obscured by
coronal emission from collisionless physics-inspired electron heating
prescriptions, jet emission from sufficient jet mass-loading, and
scattering.  We have not performed a detailed analysis of the
detectability of the BH shadow to validate what our visual inspection
suggests.  For some models with strong corona or jet emission at our
fit-preferred inclinations, $349$GHz observations more readily reveal
a BH shadow type null feature than $230$GHz.  The BH shadow's
appearance in polarization should also increase its detectability,
which we will consider in future work.

\acknowledgments

We thank Avery Broderick, Jason Dexter, and Dan Marrone for useful
discussions.  We thank Andrew Chael for providing his scripts to
generate EHT uv-tracks. RG and JCM acknowledge NASA/NSF/TCAN
(NNX14AB46G), NSF/XSEDE/TACC (TG-PHY120005), and NASA/Pleiades
(SMD-14-5451). MJ and SD acknowledge support from the National Science
Foundation (AST-1310896, AST-1211539, and AST-1440254).  MJ and SD
also acknowledge support from the Gordon and Betty Moore Foundation
(GBMF-3561).


\begin{thebibliography}{}
\expandafter\ifx\csname natexlab\endcsname\relax\def\natexlab#1{#1}\fi

\bibitem[{{Akiyama} {et~al.}(2015){Akiyama}, {Lu}, {Fish}, {Doeleman},
  {Broderick}, {Dexter}, {Hada}, {Kino}, {Nagai}, {Honma}, {Johnson}, {Algaba},
  {Asada}, {Brinkerink}, {Blundell}, {Bower}, {Cappallo}, {Crew}, {Dexter},
  {Dzib}, {Freund}, {Friberg}, {Gurwell}, {Ho}, {Inoue}, {Krichbaum},
  {Loinard}, {MacMahon}, {Marrone}, {Moran}, {Nakamura}, {Nagar}, {Ortiz-Leon},
  {Plambeck}, {Pradel}, {Primiani}, {Rogers}, {Roy}, {SooHoo}, {Tavares},
  {Tilanus}, {Titus}, {Wagner}, {Weintroub}, {Yamaguchi}, {Young}, {Zensus}, \&
  {Ziurys}}]{Akiyama2015}
{Akiyama}, K., {Lu}, R.-S., {Fish}, V.~L., {et~al.} 2015, \apj, 807, 150

\bibitem[{{Avara} {et~al.}(2015){Avara}, {McKinney}, \&
  {Reynolds}}]{2015arXiv150805323A}
{Avara}, M.~J., {McKinney}, J.~C., \& {Reynolds}, C.~S. 2015, ArXiv e-prints,
  arXiv:1508.05323

\bibitem[{{Baganoff} {et~al.}(2003){Baganoff}, {Maeda}, {Morris}, {Bautz},
  {Brandt}, {Cui}, {Doty}, {Feigelson}, {Garmire}, {Pravdo}, {Ricker}, \&
  {Townsley}}]{2003ApJ...591..891B}
{Baganoff}, F.~K., {Maeda}, Y., {Morris}, M., {et~al.} 2003, \apj, 591, 891

\bibitem[{{Balbus} \& {Hawley}(1998)}]{MRI1998}
{Balbus}, S.~A., \& {Hawley}, J.~F. 1998, Reviews of Modern Physics, 70, 1

\bibitem[{{Bardeen}(1973)}]{Bardeen1973}
{Bardeen}, J.~M. 1973, in Black Holes (Les Astres Occlus), ed. C.~{Dewitt} \&
  B.~S. {Dewitt} (New York: Gordon and Breach), 215--239

\bibitem[{{Beckwith} {et~al.}(2008){Beckwith}, {Hawley}, \&
  {Krolik}}]{2008ApJ...678.1180B}
{Beckwith}, K., {Hawley}, J.~F., \& {Krolik}, J.~H. 2008, \apj, 678, 1180

\bibitem[{{Blandford} \& {K{\"o}nigl}(1979)}]{1979ApJ...232...34B}
{Blandford}, R.~D., \& {K{\"o}nigl}, A. 1979, \apj, 232, 34

\bibitem[{{Blandford} \& {Znajek}(1977)}]{1977MNRAS.179..433B}
{Blandford}, R.~D., \& {Znajek}, R.~L. 1977, \mnras, 179, 433

\bibitem[{{Bower} {et~al.}(2015){Bower}, {Markoff}, {Dexter}, {Gurwell},
  {Moran}, {Brunthaler}, {Falcke}, {Fragile}, {Maitra}, {Marrone}, {Peck},
  {Rushton}, \& {Wright}}]{Bower2015}
{Bower}, G.~C., {Markoff}, S., {Dexter}, J., {et~al.} 2015, \apj, 802, 69

\bibitem[{{Brinkerink} {et~al.}(2015){Brinkerink}, {Falcke}, {Law}, {Barkats},
  {Bower}, {Brunthaler}, {Gammie}, {Impellizzeri}, {Markoff}, {Menten},
  {Mo{\'s}cibrodzka}, {Peck}, {Rushton}, {Schaaf}, \&
  {Wright}}]{2015arXiv150203423B}
{Brinkerink}, C.~D., {Falcke}, H., {Law}, C.~J., {et~al.} 2015, ArXiv e-prints,
  arXiv:1502.03423

\bibitem[{{Broderick} {et~al.}(2009){Broderick}, {Fish}, {Doeleman}, \&
  {Loeb}}]{2009ApJ...697...45B}
{Broderick}, A.~E., {Fish}, V.~L., {Doeleman}, S.~S., \& {Loeb}, A. 2009, \apj,
  697, 45

\bibitem[{{Broderick} {et~al.}(2011){Broderick}, {Fish}, {Doeleman}, \&
  {Loeb}}]{2011ApJ...738...38B}
---. 2011, \apj, 738, 38

\bibitem[{{Broderick} {et~al.}(2014){Broderick}, {Johannsen}, {Loeb}, \&
  {Psaltis}}]{2014ApJ...784....7B}
{Broderick}, A.~E., {Johannsen}, T., {Loeb}, A., \& {Psaltis}, D. 2014, \apj,
  784, 7

\bibitem[{{Broderick} \& {Loeb}(2006)}]{2006MNRAS.367..905B}
{Broderick}, A.~E., \& {Loeb}, A. 2006, \mnras, 367, 905

\bibitem[{{Broderick} \& {McKinney}(2010)}]{2010ApJ...725..750B}
{Broderick}, A.~E., \& {McKinney}, J.~C. 2010, \apj, 725, 750

\bibitem[{{Broderick} \& {Tchekhovskoy}(2015)}]{2015ApJ...809...97B}
{Broderick}, A.~E., \& {Tchekhovskoy}, A. 2015, \apj, 809, 97

\bibitem[{{Chan} {et~al.}(2015{\natexlab{a}}){Chan}, {Psaltis}, {{\"O}zel},
  {Medeiros}, {Marrone}, {Sa{\c d}owski}, \& {Narayan}}]{2015ApJ...812..103C}
{Chan}, C.-k., {Psaltis}, D., {{\"O}zel}, F., {et~al.} 2015{\natexlab{a}},
  \apj, 812, 103

\bibitem[{{Chan} {et~al.}(2015{\natexlab{b}}){Chan}, {Psaltis}, {{\"O}zel},
  {Narayan}, \& {Sa{\c d}owski}}]{2015ApJ...799....1C}
{Chan}, C.-K., {Psaltis}, D., {{\"O}zel}, F., {Narayan}, R., \& {Sa{\c
  d}owski}, A. 2015{\natexlab{b}}, \apj, 799, 1

\bibitem[{{Contopoulos} {et~al.}(2009){Contopoulos}, {Christodoulou},
  {Kazanas}, \& {Gabuzda}}]{2009ApJ...702L.148C}
{Contopoulos}, I., {Christodoulou}, D.~M., {Kazanas}, D., \& {Gabuzda}, D.~C.
  2009, \apjl, 702, L148

\bibitem[{{Dexter}(2014)}]{2014IAUS..303..298D}
{Dexter}, J. 2014, in IAU Symposium, Vol. 303, IAU Symposium, ed. L.~O.
  {Sjouwerman}, C.~C. {Lang}, \& J.~{Ott}, 298--302

\bibitem[{{Dexter} {et~al.}(2009){Dexter}, {Agol}, \&
  {Fragile}}]{2009ApJ...703L.142D}
{Dexter}, J., {Agol}, E., \& {Fragile}, P.~C. 2009, \apjl, 703, L142

\bibitem[{{Dexter} {et~al.}(2010){Dexter}, {Agol}, {Fragile}, \&
  {McKinney}}]{2010ApJ...717.1092D}
{Dexter}, J., {Agol}, E., {Fragile}, P.~C., \& {McKinney}, J.~C. 2010, \apj,
  717, 1092

\bibitem[{{Dexter} \& {Fragile}(2013)}]{2013MNRAS.432.2252D}
{Dexter}, J., \& {Fragile}, P.~C. 2013, \mnras, 432, 2252

\bibitem[{{Dexter} {et~al.}(2012){Dexter}, {McKinney}, \&
  {Agol}}]{2012MNRAS.421.1517D}
{Dexter}, J., {McKinney}, J.~C., \& {Agol}, E. 2012, \mnras, 421, 1517

\bibitem[{{Dodds-Eden} {et~al.}(2009){Dodds-Eden}, {Porquet}, {Trap},
  {Quataert}, {Haubois}, {Gillessen}, {Grosso}, {Pantin}, {Falcke}, {Rouan},
  {Genzel}, {Hasinger}, {Goldwurm}, {Yusef-Zadeh}, {Clenet}, {Trippe},
  {Lagage}, {Bartko}, {Eisenhauer}, {Ott}, {Paumard}, {Perrin}, {Yuan},
  {Fritz}, \& {Mascetti}}]{2009ApJ...698..676D}
{Dodds-Eden}, K., {Porquet}, D., {Trap}, G., {et~al.} 2009, \apj, 698, 676

\bibitem[{{Doeleman} {et~al.}(2009){Doeleman}, {Agol}, {Backer}, {Baganoff},
  {Bower}, {Broderick}, {Fabian}, {Fish}, {Gammie}, {Ho}, {Honman},
  {Krichbaum}, {Loeb}, {Marrone}, {Reid}, {Rogers}, {Shapiro}, {Strittmatter},
  {Tilanus}, {Weintroub}, {Whitney}, {Wright}, \&
  {Ziurys}}]{2009astro2010S..68D}
{Doeleman}, S., {Agol}, E., {Backer}, D., {et~al.} 2009, in Astronomy, Vol.
  2010, astro2010: The Astronomy and Astrophysics Decadal Survey, 68

\bibitem[{{Doeleman} {et~al.}(2001){Doeleman}, {Shen}, {Rogers}, {Bower},
  {Wright}, {Zhao}, {Backer}, {Crowley}, {Freund}, {Ho}, {Lo}, \&
  {Woody}}]{Doeleman:2001:SSA}
{Doeleman}, S.~S., {Shen}, Z.-Q., {Rogers}, A.~E.~E., {et~al.} 2001, \aj, 121,
  2610

\bibitem[{{Doeleman} {et~al.}(2008){Doeleman}, {Weintroub}, {Rogers},
  {Plambeck}, {Freund}, {Tilanus}, {Friberg}, {Ziurys}, {Moran}, {Corey},
  {Young}, {Smythe}, {Titus}, {Marrone}, {Cappallo}, {Bock}, {Bower},
  {Chamberlin}, {Davis}, {Krichbaum}, {Lamb}, {Maness}, {Niell}, {Roy},
  {Strittmatter}, {Werthimer}, {Whitney}, \& {Woody}}]{2008Natur.455...78D}
{Doeleman}, S.~S., {Weintroub}, J., {Rogers}, A.~E.~E., {et~al.} 2008, \nat,
  455, 78

\bibitem[{{Doeleman} {et~al.}(2012){Doeleman}, {Fish}, {Schenck}, {Beaudoin},
  {Blundell}, {Bower}, {Broderick}, {Chamberlin}, {Freund}, {Friberg},
  {Gurwell}, {Ho}, {Honma}, {Inoue}, {Krichbaum}, {Lamb}, {Loeb}, {Lonsdale},
  {Marrone}, {Moran}, {Oyama}, {Plambeck}, {Primiani}, {Rogers}, {Smythe},
  {SooHoo}, {Strittmatter}, {Tilanus}, {Titus}, {Weintroub}, {Wright}, {Young},
  \& {Ziurys}}]{2012Sci...338..355D}
{Doeleman}, S.~S., {Fish}, V.~L., {Schenck}, D.~E., {et~al.} 2012, Science,
  338, 355

\bibitem[{{Eckart} {et~al.}(2006{\natexlab{a}}){Eckart}, {Sch{\"o}del},
  {Meyer}, {Trippe}, {Ott}, \& {Genzel}}]{2006AA...455....1E}
{Eckart}, A., {Sch{\"o}del}, R., {Meyer}, L., {et~al.} 2006{\natexlab{a}},
  \aap, 455, 1

\bibitem[{{Eckart} {et~al.}(2006{\natexlab{b}}){Eckart}, {Baganoff},
  {Sch{\"o}del}, {Morris}, {Genzel}, {Bower}, {Marrone}, {Moran}, {Viehmann},
  {Bautz}, {Brandt}, {Garmire}, {Ott}, {Trippe}, {Ricker}, {Straubmeier},
  {Roberts}, {Yusef-Zadeh}, {Zhao}, \& {Rao}}]{2006AA...450..535E}
{Eckart}, A., {Baganoff}, F.~K., {Sch{\"o}del}, R., {et~al.}
  2006{\natexlab{b}}, \aap, 450, 535

\bibitem[{{Falcke} {et~al.}(1998){Falcke}, {Goss}, {Matsuo}, {Teuben}, {Zhao},
  \& {Zylka}}]{1998ApJ...499..731F}
{Falcke}, H., {Goss}, W.~M., {Matsuo}, H., {et~al.} 1998, \apj, 499, 731

\bibitem[{{Falcke} \& {Markoff}(2000)}]{2000AA...362..113F}
{Falcke}, H., \& {Markoff}, S. 2000, \aap, 362, 113

\bibitem[{{Falcke} {et~al.}(2000){Falcke}, {Melia}, \&
  {Agol}}]{2000ApJ...528L..13F}
{Falcke}, H., {Melia}, F., \& {Agol}, E. 2000, \apjl, 528, L13

\bibitem[{{Fish} {et~al.}(2013){Fish}, {Doeleman}, {Marrone}, {Lu}, {Wardle},
  \& {EHT Collaboration}}]{2013AAS...22114304F}
{Fish}, V.~L., {Doeleman}, S., {Marrone}, D.~P., {et~al.} 2013, in American
  Astronomical Society Meeting Abstracts, Vol. 221, American Astronomical
  Society Meeting Abstracts \#221, \#143.04

\bibitem[{{Fish} {et~al.}(2009){Fish}, {Doeleman}, {Broderick}, {Loeb}, \&
  {Rogers}}]{2009ApJ...706.1353F}
{Fish}, V.~L., {Doeleman}, S.~S., {Broderick}, A.~E., {Loeb}, A., \& {Rogers},
  A.~E.~E. 2009, \apj, 706, 1353

\bibitem[{{Fish} {et~al.}(2011){Fish}, {Doeleman}, {Beaudoin}, {Blundell},
  {Bolin}, {Bower}, {Chamberlin}, {Freund}, {Friberg}, {Gurwell}, {Honma},
  {Inoue}, {Krichbaum}, {Lamb}, {Marrone}, {Moran}, {Oyama}, {Plambeck},
  {Primiani}, {Rogers}, {Smythe}, {SooHoo}, {Strittmatter}, {Tilanus}, {Titus},
  {Weintroub}, {Wright}, {Woody}, {Young}, \& {Ziurys}}]{2011ApJ...727L..36F}
{Fish}, V.~L., {Doeleman}, S.~S., {Beaudoin}, C., {et~al.} 2011, \apjl, 727,
  L36

\bibitem[{{Fish} {et~al.}(2014){Fish}, {Johnson}, {Lu}, {Doeleman}, {Bouman},
  {Zoran}, {Freeman}, {Psaltis}, {Narayan}, {Pankratius}, {Broderick}, {Gwinn},
  \& {Vertatschitsch}}]{2014ApJ...795..134F}
{Fish}, V.~L., {Johnson}, M.~D., {Lu}, R.-S., {et~al.} 2014, \apj, 795, 134

\bibitem[{{Foucart} {et~al.}(2015){Foucart}, {Chandra}, {Gammie}, \&
  {Quataert}}]{2015arXiv151104445F}
{Foucart}, F., {Chandra}, M., {Gammie}, C.~F., \& {Quataert}, E. 2015, ArXiv
  e-prints, arXiv:1511.04445

\bibitem[{{Gammie} {et~al.}(2003){Gammie}, {McKinney}, \&
  {T{\'o}th}}]{2003ApJ...589..444G}
{Gammie}, C.~F., {McKinney}, J.~C., \& {T{\'o}th}, G. 2003, \apj, 589, 444

\bibitem[{{Gammie} {et~al.}(2004){Gammie}, {Shapiro}, \&
  {McKinney}}]{2004ApJ...602..312G}
{Gammie}, C.~F., {Shapiro}, S.~L., \& {McKinney}, J.~C. 2004, \apj, 602, 312

\bibitem[{{Genzel} {et~al.}(2010){Genzel}, {Eisenhauer}, \&
  {Gillessen}}]{2010RvMP...82.3121G}
{Genzel}, R., {Eisenhauer}, F., \& {Gillessen}, S. 2010, Reviews of Modern
  Physics, 82, 3121

\bibitem[{{Genzel} {et~al.}(2003){Genzel}, {Sch{\"o}del}, {Ott}, {Eckart},
  {Alexander}, {Lacombe}, {Rouan}, \& {Aschenbach}}]{2003Natur.425..934G}
{Genzel}, R., {Sch{\"o}del}, R., {Ott}, T., {et~al.} 2003, \nat, 425, 934

\bibitem[{{Ghez} {et~al.}(2008){Ghez}, {Salim}, {Weinberg}, {Lu}, {Do}, {Dunn},
  {Matthews}, {Morris}, {Yelda}, {Becklin}, {Kremenek}, {Milosavljevic}, \&
  {Naiman}}]{2008ApJ...689.1044G}
{Ghez}, A.~M., {Salim}, S., {Weinberg}, N.~N., {et~al.} 2008, \apj, 689, 1044

\bibitem[{{Globus} \& {Levinson}(2013)}]{2013PhRvD..88h4046G}
{Globus}, N., \& {Levinson}, A. 2013, \prd, 88, 084046

\bibitem[{{Howes} {et~al.}(2008){Howes}, {Dorland}, {Cowley}, {Hammett},
  {Quataert}, {Schekochihin}, \& {Tatsuno}}]{2008PhRvL.100f5004H}
{Howes}, G.~G., {Dorland}, W., {Cowley}, S.~C., {et~al.} 2008, Physical Review
  Letters, 100, 065004

\bibitem[{{Igumenshchev} {et~al.}(2003){Igumenshchev}, {Narayan}, \&
  {Abramowicz}}]{2003ApJ...592.1042I}
{Igumenshchev}, I.~V., {Narayan}, R., \& {Abramowicz}, M.~A. 2003, \apj, 592,
  1042

\bibitem[{{Johannsen} {et~al.}(2015){Johannsen}, {Broderick}, {Plewa},
  {Chatzopoulos}, {Doeleman}, {Eisenhauer}, {Fish}, {Genzel}, {Gerhard}, \&
  {Johnson}}]{2015arXiv151202640J}
{Johannsen}, T., {Broderick}, A.~E., {Plewa}, P.~M., {et~al.} 2015, ArXiv
  e-prints, arXiv:1512.02640

\bibitem[{{Johnson} \& {Quataert}(2007)}]{2007ApJ...660.1273J}
{Johnson}, B.~M., \& {Quataert}, E. 2007, \apj, 660, 1273

\bibitem[{{Johnson} {et~al.}(2014){Johnson}, {Fish}, {Doeleman}, {Broderick},
  {Wardle}, \& {Marrone}}]{Johnson2014}
{Johnson}, M.~D., {Fish}, V.~L., {Doeleman}, S.~S., {et~al.} 2014, \apj, 794,
  150

\bibitem[{{Johnson} \& {Gwinn}(2015)}]{JohnsonGwinn2015}
{Johnson}, M.~D., \& {Gwinn}, C.~R. 2015, \apj, 805, 180

\bibitem[{{Johnson} {et~al.}(2015{\natexlab{a}}){Johnson}, {Loeb}, {Shiokawa},
  {Chael}, \& {Doeleman}}]{2015ApJ...813..132J}
{Johnson}, M.~D., {Loeb}, A., {Shiokawa}, H., {Chael}, A.~A., \& {Doeleman},
  S.~S. 2015{\natexlab{a}}, \apj, 813, 132

\bibitem[{{Johnson} {et~al.}(2015{\natexlab{b}}){Johnson}, {Fish}, {Doeleman},
  {Marrone}, {Plambeck}, {Wardle}, {Akiyama}, {Asada}, {Beaudoin}, {Blackburn},
  {Blundell}, {Bower}, {Brinkerink}, {Broderick}, {Cappallo}, {Chael}, {Crew},
  {Dexter}, {Dexter}, {Freund}, {Friberg}, {Gold}, {Gurwell}, {Ho}, {Honma},
  {Inoue}, {Kosowsky}, {Krichbaum}, {Lamb}, {Loeb}, {Lu}, {MacMahon},
  {McKinney}, {Moran}, {Narayan}, {Primiani}, {Psaltis}, {Rogers}, {Rosenfeld},
  {SooHoo}, {Tilanus}, {Titus}, {Vertatschitsch}, {Weintroub}, {Wright},
  {Young}, {Zensus}, \& {Ziurys}}]{JohnsonEtAl2015Science}
{Johnson}, M.~D., {Fish}, V.~L., {Doeleman}, S.~S., {et~al.}
  2015{\natexlab{b}}, Science, 350, 1242

\bibitem[{{Komissarov} \& {McKinney}(2007)}]{2007MNRAS.377L..49K}
{Komissarov}, S.~S., \& {McKinney}, J.~C. 2007, \mnras, 377, L49

\bibitem[{{Levinson} {et~al.}(2005){Levinson}, {Melrose}, {Judge}, \&
  {Luo}}]{2005ApJ...631..456L}
{Levinson}, A., {Melrose}, D., {Judge}, A., \& {Luo}, Q. 2005, \apj, 631, 456

\bibitem[{{Lu} {et~al.}(2015){Lu}, {Roelofs}, {Fish}, {Shiokawa}, {Doeleman},
  {Gammie}, {Falcke}, {Krichbaum}, \& {Zensus}}]{2015arXiv151208543L}
{Lu}, R.-S., {Roelofs}, F., {Fish}, V.~L., {et~al.} 2015, ArXiv e-prints,
  arXiv:1512.08543

\bibitem[{{Luminet}(1979)}]{Luminet1979}
{Luminet}, J.-P. 1979, \aap, 75, 228

\bibitem[{{Lynn} {et~al.}(2014){Lynn}, {Quataert}, {Chandran}, \&
  {Parrish}}]{2014ApJ...791...71L}
{Lynn}, J.~W., {Quataert}, E., {Chandran}, B.~D.~G., \& {Parrish}, I.~J. 2014,
  \apj, 791, 71

\bibitem[{{Mahadevan} \& {Quataert}(1997)}]{1997ApJ...490..605M}
{Mahadevan}, R., \& {Quataert}, E. 1997, \apj, 490, 605

\bibitem[{{Marrone} {et~al.}(2008){Marrone}, {Baganoff}, {Morris}, {Moran},
  {Ghez}, {Hornstein}, {Dowell}, {Mu{\~n}oz}, {Bautz}, {Ricker}, {Brandt},
  {Garmire}, {Lu}, {Matthews}, {Zhao}, {Rao}, \& {Bower}}]{Marrone2008}
{Marrone}, D.~P., {Baganoff}, F.~K., {Morris}, M.~R., {et~al.} 2008, \apj, 682,
  373

\bibitem[{{McKinney}(2005)}]{2005ApJ...630L...5M}
{McKinney}, J.~C. 2005, \apjl, 630, L5

\bibitem[{{McKinney}(2006)}]{2006MNRAS.368.1561M}
---. 2006, \mnras, 368, 1561

\bibitem[{{McKinney} \& {Blandford}(2009)}]{2009MNRAS.394L.126M}
{McKinney}, J.~C., \& {Blandford}, R.~D. 2009, \mnras, 394, L126

\bibitem[{{McKinney} {et~al.}(2015){McKinney}, {Dai}, \&
  {Avara}}]{2015MNRAS.454L...6M}
{McKinney}, J.~C., {Dai}, L., \& {Avara}, M.~J. 2015, \mnras, 454, L6

\bibitem[{{McKinney} \& {Gammie}(2004)}]{2004ApJ...611..977M}
{McKinney}, J.~C., \& {Gammie}, C.~F. 2004, \apj, 611, 977

\bibitem[{{McKinney} \& {Narayan}(2007)}]{2007MNRAS.375..513M}
{McKinney}, J.~C., \& {Narayan}, R. 2007, \mnras, 375, 513

\bibitem[{{McKinney} {et~al.}(2012){McKinney}, {Tchekhovskoy}, \&
  {Blandford}}]{2012MNRAS.423.3083M}
{McKinney}, J.~C., {Tchekhovskoy}, A., \& {Blandford}, R.~D. 2012, \mnras, 423,
  3083

\bibitem[{{McKinney} {et~al.}(2013){McKinney}, {Tchekhovskoy}, \&
  {Blandford}}]{2013Sci...339...49M}
---. 2013, Science, 339, 49

\bibitem[{{McKinney} {et~al.}(2014){McKinney}, {Tchekhovskoy}, {Sadowski}, \&
  {Narayan}}]{2014MNRAS.441.3177M}
{McKinney}, J.~C., {Tchekhovskoy}, A., {Sadowski}, A., \& {Narayan}, R. 2014,
  \mnras, 441, 3177

\bibitem[{{McKinney} \& {Uzdensky}(2012)}]{mu12}
{McKinney}, J.~C., \& {Uzdensky}, D.~A. 2012, \mnras, 419, 573

\bibitem[{{Mo{\'s}cibrodzka} \& {Falcke}(2013)}]{2013AA...559L...3M}
{Mo{\'s}cibrodzka}, M., \& {Falcke}, H. 2013, \aap, 559, L3

\bibitem[{{Mo{\'s}cibrodzka} {et~al.}(2015){Mo{\'s}cibrodzka}, {Falcke}, \&
  {Shiokawa}}]{2015arXiv151007243M}
{Mo{\'s}cibrodzka}, M., {Falcke}, H., \& {Shiokawa}, H. 2015, ArXiv e-prints,
  arXiv:1510.07243

\bibitem[{{Mo{\'s}cibrodzka} {et~al.}(2014){Mo{\'s}cibrodzka}, {Falcke},
  {Shiokawa}, \& {Gammie}}]{2014AA...570A...7M}
{Mo{\'s}cibrodzka}, M., {Falcke}, H., {Shiokawa}, H., \& {Gammie}, C.~F. 2014,
  \aap, 570, A7

\bibitem[{{Mo{\'s}cibrodzka} {et~al.}(2011){Mo{\'s}cibrodzka}, {Gammie},
  {Dolence}, \& {Shiokawa}}]{2011ApJ...735....9M}
{Mo{\'s}cibrodzka}, M., {Gammie}, C.~F., {Dolence}, J.~C., \& {Shiokawa}, H.
  2011, \apj, 735, 9

\bibitem[{{Mo{\'s}cibrodzka} {et~al.}(2009){Mo{\'s}cibrodzka}, {Gammie},
  {Dolence}, {Shiokawa}, \& {Leung}}]{2009ApJ...706..497M}
{Mo{\'s}cibrodzka}, M., {Gammie}, C.~F., {Dolence}, J.~C., {Shiokawa}, H., \&
  {Leung}, P.~K. 2009, \apj, 706, 497

\bibitem[{{Mo{\'s}cibrodzka} {et~al.}(2012){Mo{\'s}cibrodzka}, {Shiokawa},
  {Gammie}, \& {Dolence}}]{2012ApJ...752L...1M}
{Mo{\'s}cibrodzka}, M., {Shiokawa}, H., {Gammie}, C.~F., \& {Dolence}, J.~C.
  2012, \apjl, 752, L1

\bibitem[{{Mu{\~n}oz} {et~al.}(2012){Mu{\~n}oz}, {Marrone}, {Moran}, \&
  {Rao}}]{2012ApJ...745..115M}
{Mu{\~n}oz}, D.~J., {Marrone}, D.~P., {Moran}, J.~M., \& {Rao}, R. 2012, \apj,
  745, 115

\bibitem[{{Narayan} {et~al.}(2003){Narayan}, {Igumenshchev}, \&
  {Abramowicz}}]{2003PASJ...55L..69N}
{Narayan}, R., {Igumenshchev}, I.~V., \& {Abramowicz}, M.~A. 2003, \pasj, 55,
  L69

\bibitem[{{Narayan} {et~al.}(2012){Narayan}, {S{\c a}dowski}, {Penna}, \&
  {Kulkarni}}]{2012MNRAS.426.3241N}
{Narayan}, R., {S{\c a}dowski}, A., {Penna}, R.~F., \& {Kulkarni}, A.~K. 2012,
  \mnras, 426, 3241

\bibitem[{{Narayan} \& {Yi}(1994)}]{1994ApJ...428L..13N}
{Narayan}, R., \& {Yi}, I. 1994, \apjl, 428, L13

\bibitem[{{{\"O}zel} {et~al.}(2000){{\"O}zel}, {Psaltis}, \&
  {Narayan}}]{Ozel2000}
{{\"O}zel}, F., {Psaltis}, D., \& {Narayan}, R. 2000, \apj, 541, 234

\bibitem[{{Penna} {et~al.}(2010){Penna}, {McKinney}, {Narayan}, {Tchekhovskoy},
  {Shafee}, \& {McClintock}}]{2010MNRAS.408..752P}
{Penna}, R.~F., {McKinney}, J.~C., {Narayan}, R., {et~al.} 2010, \mnras, 408,
  752

\bibitem[{{Philippov} {et~al.}(2015){Philippov}, {Cerutti}, {Tchekhovskoy}, \&
  {Spitkovsky}}]{2015arXiv151001734P}
{Philippov}, A.~A., {Cerutti}, B., {Tchekhovskoy}, A., \& {Spitkovsky}, A.
  2015, ArXiv e-prints, arXiv:1510.01734

\bibitem[{{Polko} \& {McKinney}(2015)}]{2015arXiv151207969P}
{Polko}, P., \& {McKinney}, J.~C. 2015, ArXiv e-prints, arXiv:1512.07969

\bibitem[{{Psaltis} {et~al.}(2015){Psaltis}, {{\"O}zel}, {Chan}, \&
  {Marrone}}]{2015ApJ...814..115P}
{Psaltis}, D., {{\"O}zel}, F., {Chan}, C.-K., \& {Marrone}, D.~P. 2015, \apj,
  814, 115

\bibitem[{{Quataert} \& {Gruzinov}(1999)}]{1999ApJ...520..248Q}
{Quataert}, E., \& {Gruzinov}, A. 1999, \apj, 520, 248

\bibitem[{{Quataert} \& {Gruzinov}(2000)}]{2000ApJ...545..842Q}
---. 2000, \apj, 545, 842

\bibitem[{{Quataert} \& {Narayan}(1999{\natexlab{a}})}]{1999ApJ...516..399Q}
{Quataert}, E., \& {Narayan}, R. 1999{\natexlab{a}}, \apj, 516, 399

\bibitem[{{Quataert} \& {Narayan}(1999{\natexlab{b}})}]{1999ApJ...520..298Q}
---. 1999{\natexlab{b}}, \apj, 520, 298

\bibitem[{{Reid}(2009)}]{2009IJMPD..18..889R}
{Reid}, M.~J. 2009, International Journal of Modern Physics D, 18, 889

\bibitem[{{Ressler} {et~al.}(2015){Ressler}, {Tchekhovskoy}, {Quataert},
  {Chandra}, \& {Gammie}}]{2015MNRAS.454.1848R}
{Ressler}, S.~M., {Tchekhovskoy}, A., {Quataert}, E., {Chandra}, M., \&
  {Gammie}, C.~F. 2015, \mnras, 454, 1848

\bibitem[{{Ricarte} \& {Dexter}(2015)}]{2015MNRAS.446.1973R}
{Ricarte}, A., \& {Dexter}, J. 2015, \mnras, 446, 1973

\bibitem[{{Riquelme} {et~al.}(2012){Riquelme}, {Quataert}, {Sharma}, \&
  {Spitkovsky}}]{2012ApJ...755...50R}
{Riquelme}, M.~A., {Quataert}, E., {Sharma}, P., \& {Spitkovsky}, A. 2012,
  \apj, 755, 50

\bibitem[{{Riquelme} {et~al.}(2015){Riquelme}, {Quataert}, \&
  {Verscharen}}]{2015ApJ...800...27R}
{Riquelme}, M.~A., {Quataert}, E., \& {Verscharen}, D. 2015, \apj, 800, 27

\bibitem[{{Roberts} {et~al.}(1994){Roberts}, {Wardle}, \& {Brown}}]{RWB94}
{Roberts}, D.~H., {Wardle}, J.~F.~C., \& {Brown}, L.~F. 1994, \apj, 427, 718

\bibitem[{{Sch{\"o}del} {et~al.}(2011){Sch{\"o}del}, {Morris}, {Muzic},
  {Alberdi}, {Meyer}, {Eckart}, \& {Gezari}}]{Schodel2011}
{Sch{\"o}del}, R., {Morris}, M.~R., {Muzic}, K., {et~al.} 2011, \aap, 532, A83

\bibitem[{{Shafee} {et~al.}(2008){Shafee}, {McKinney}, {Narayan},
  {Tchekhovskoy}, {Gammie}, \& {McClintock}}]{2008ApJ...687L..25S}
{Shafee}, R., {McKinney}, J.~C., {Narayan}, R., {et~al.} 2008, \apjl, 687, L25

\bibitem[{{Sharma} {et~al.}(2006){Sharma}, {Hammett}, {Quataert}, \&
  {Stone}}]{2006ApJ...637..952S}
{Sharma}, P., {Hammett}, G.~W., {Quataert}, E., \& {Stone}, J.~M. 2006, \apj,
  637, 952

\bibitem[{{Sharma} {et~al.}(2007){Sharma}, {Quataert}, \&
  {Stone}}]{2007ApJ...671.1696S}
{Sharma}, P., {Quataert}, E., \& {Stone}, J.~M. 2007, \apj, 671, 1696

\bibitem[{{Shcherbakov}(2014)}]{2014ascl.soft07007S}
{Shcherbakov}, R.~V. 2014, {ASTRORAY: General relativistic polarized radiative
  transfer code}, Astrophysics Source Code Library, ascl:1407.007

\bibitem[{{Shcherbakov} \& {Huang}(2011)}]{2011MNRAS.410.1052S}
{Shcherbakov}, R.~V., \& {Huang}, L. 2011, \mnras, 410, 1052

\bibitem[{{Shcherbakov} \& {McKinney}(2013{\natexlab{a}})}]{SM13}
{Shcherbakov}, R.~V., \& {McKinney}, J.~C. 2013{\natexlab{a}}, \apjl, 774, L22

\bibitem[{{Shcherbakov} \&
  {McKinney}(2013{\natexlab{b}})}]{2013ApJ...774L..22S}
---. 2013{\natexlab{b}}, \apjl, 774, L22

\bibitem[{{Shcherbakov} {et~al.}(2012{\natexlab{a}}){Shcherbakov}, {Penna}, \&
  {McKinney}}]{2012ApJ...755..133S}
{Shcherbakov}, R.~V., {Penna}, R.~F., \& {McKinney}, J.~C. 2012{\natexlab{a}},
  \apj, 755, 133

\bibitem[{{Shcherbakov} {et~al.}(2012{\natexlab{b}}){Shcherbakov}, {Penna}, \&
  {McKinney}}]{SPM12}
---. 2012{\natexlab{b}}, \apj, 755, 133

\bibitem[{{Takahashi}(2004)}]{2004ApJ...611..996T}
{Takahashi}, R. 2004, \apj, 611, 996

\bibitem[{{Tchekhovskoy} \& {McKinney}(2012)}]{2012MNRAS.423L..55T}
{Tchekhovskoy}, A., \& {McKinney}, J.~C. 2012, \mnras, 423, L55

\bibitem[{{Tchekhovskoy} {et~al.}(2007){Tchekhovskoy}, {McKinney}, \&
  {Narayan}}]{2007MNRAS.379..469T}
{Tchekhovskoy}, A., {McKinney}, J.~C., \& {Narayan}, R. 2007, \mnras, 379, 469

\bibitem[{{Tchekhovskoy} {et~al.}(2010){Tchekhovskoy}, {Narayan}, \&
  {McKinney}}]{2010ApJ...711...50T}
{Tchekhovskoy}, A., {Narayan}, R., \& {McKinney}, J.~C. 2010, \apj, 711, 50

\bibitem[{{Tchekhovskoy} {et~al.}(2011){Tchekhovskoy}, {Narayan}, \&
  {McKinney}}]{2011MNRAS.418L..79T}
---. 2011, \mnras, 418, L79

\bibitem[{{Yuan} {et~al.}(2002){Yuan}, {Markoff}, \&
  {Falcke}}]{2002AA...383..854Y}
{Yuan}, F., {Markoff}, S., \& {Falcke}, H. 2002, \aap, 383, 854

\bibitem[{{Yuan} \& {Narayan}(2014)}]{2014ARAA..52..529Y}
{Yuan}, F., \& {Narayan}, R. 2014, \araa, 52, 529

\bibitem[{{Yuan} {et~al.}(2003){Yuan}, {Quataert}, \&
  {Narayan}}]{2003ApJ...598..301Y}
{Yuan}, F., {Quataert}, E., \& {Narayan}, R. 2003, \apj, 598, 301

\bibitem[{{Yuan} {et~al.}(2004){Yuan}, {Quataert}, \&
  {Narayan}}]{2004ApJ...606..894Y}
---. 2004, \apj, 606, 894

\bibitem[{{Yusef-Zadeh} {et~al.}(2008){Yusef-Zadeh}, {Wardle}, {Heinke},
  {Dowell}, {Roberts}, {Baganoff}, \& {Cotton}}]{Yusef-Zadeh2008}
{Yusef-Zadeh}, F., {Wardle}, M., {Heinke}, C., {et~al.} 2008, \apj, 682, 361

\end{thebibliography}
\end{document}